\documentclass[aps,pra,twocolumn,groupedaddress]{revtex4-1}

\usepackage{amsmath}
\usepackage{amsthm}
\usepackage{amssymb}
\usepackage{graphicx,hyperref,bbm,enumitem,bm}  
\usepackage[pdftex]{color}

\usepackage[toc,page]{appendix}

\newtheorem{theorem}{Theorem}[section]

\theoremstyle{definition}

\def\ave#1{\langle #1 \rangle}
\def\ii{{\rm i}}
\def\sx{\sigma^{\rm x}}
\def\sy{\sigma^{\rm y}}
\def\sz{\sigma^{\rm z}}
\def\tr#1{{\rm tr}{#1}}
\def\1{\mathbbm{1}}
\def\ket#1{{| #1 \rangle}}
\def\bra#1{{\langle #1 |}}
\def\braket#1#2{{\langle #1 | #2 \rangle}}

\def\Ms#1#2{S^{(#1)}_{#2}}
\def\B#1#2{B^{(#1)}_{#2}}
\def\ax{a_{\rm x}}
\def\ay{a_{\rm y}}
\def\az{a_{\rm z}}
\def\ac{a_{\rm c}}
\def\n2{{\lfloor \frac{n}{2} \rfloor}}
\def\vE{{v_{\rm E}}}
\def\rE{{r_{\rm E}}}
\def\rA{\rho_{\rm A}}
\def\nA{{n_{\rm A}}}
\def\nB{{n_{\rm B}}}
\def\av{{\mathbf{a}}}

\def\tit#1{{\em #1},}

\newcommand{\new}[1]{{#1}}

\usepackage{booktabs}
\usepackage{multirow}
\AtBeginDocument{
\heavyrulewidth=.08em
\lightrulewidth=.05em
\cmidrulewidth=.03em
\belowrulesep=.65ex
\belowbottomsep=0pt
\aboverulesep=.4ex
\abovetopsep=0pt
\cmidrulesep=\doublerulesep
\cmidrulekern=.5em
\defaultaddspace=.5em
}

\begin{document}
	
\title{Fastest local entanglement scrambler, multistage thermalization, and a non-Hermitian phantom}
	
	\author{Ja\v s Bensa and Marko \v Znidari\v c}
	\affiliation{Department of Physics, Faculty of Mathematics and Physics, University of Ljubljana, 1000 Ljubljana, Slovenia}
	
	\date{\today}
	
	\begin{abstract}
We study random quantum circuits and their rate of producing bipartite entanglement, specifically with respect to the choice of 2-qubit gates and the order (protocol) in which these are applied. The problem is mapped to a Markovian process and proved that there are large spectral equivalence classes \new{-- different configurations have the same spectrum}. Optimal gates and the protocol that generate entanglement with the fastest \new{theoretically possible rate are} identified. Relaxation towards the asymptotic thermal entanglement proceeds via a series of phase transitions in the local relaxation rate, which is a consequence of non-Hermiticity. In particular, non-Hermiticity can cause the rate to be either faster, or, even more interestingly, slower than predicted by the matrix \new{eigenvalue} gap. This is caused by an \new{exponential in system size} explosion of expansion coefficient sizes resulting in a 'phantom' eigenvalue, and is due to non-orthogonality of non-Hermitian eigenvectors. \new{We numerically demonstrate that the phenomenon occurs also in random circuits with non-optimal generic gates, random U(4) gates, and also without spatial or temporal randomness, suggesting that it could be of wide importance also in other non-Hermitian settings, including correlations.}
	\end{abstract}
	
	\maketitle

\section{Introduction}

Entanglement is one of the key properties that can make quantum systems different than classical ones. This is reflected in quantum information -- large entanglement is a necessary resource to gain advantage over classical computation, and many of the new phases discovered in recent decades can be distinguished by different patterns of entanglement~\cite{book}. Because entanglement and the related concept of quantum information plays such a fundamental role it is also instrumental in the quest to push the boundaries of the present day physics, for instance, trying to figure out essential rules that quantum gravity should obey.

An elementary question is how can one efficiently generate this resource? One procedure that we focus on are the so-called random quantum circuits~\cite{emerson03}, where quantum gates are chosen randomly from a certain set of gates. What set one takes might foremost depend on the available resources; while one wants to generate entanglement as quickly as possible, one must use the resources as efficiently as possible. What is meant by efficient will depend on the context, however, there are some common conditions. Richness of nature emerges due to two ingredients, innate properties of constituent objects (particles) and local interactions between them. Locality, being \new{intimately related} to causality, is rather important and is typically also the costly resource in quantum computation. Local transformations, i.e. 1-site unitary operations, are faster to perform and typically have higher fidelity, while interactions in the form of 2-site gates are expensive. We shall focus on random circuits in which 1-site resources are random (1-qubit unitary from the Haar measure) while the 2-site transformations are held fixed. Such a choice makes sense for two reasons: (i) it follows quantum information cost guidelines, and (ii) in some cases allows exact solvability. We will demonstrate that taking a fixed good entangling 2-qubit gate is actually better than randomly choosing the whole 2-qubit transformation. Optimal random circuits that generate entanglement the fastest should be of interest also in the near-term applications of noisy quantum computers where they have been identified as prime candidates in the quest to demonstrate quantum supremacy~\cite{Google}. \new{Using the optimal circuit in such a quest is of high relevance as it directly affects the execution speed and therefore the attainable fidelity, which can in turn be crucial for an experiment to be on the right side of the supremacy frontier.}

Random circuits have also another use -- they are believed to correctly describe some of the properties of generic quantum systems, like e.g., dynamics of operators~\cite{adam18,vedika18,Frank18,nick18}, and can therefore serve as models of chaotic many-body systems~\cite{Chalker18}. The main advantage over chaotic systems is that their randomness enables analytical simplifications, leading to exact results. Exact results for entanglement evolution have been obtained also for the CFT~\cite{Calabrese05}. Recently the two pictures, that of random circuits and solvable quantum systems, emerged in the form of the so-called dual-unitary circuits, e.g. Ref.~\cite{prosen19}, which are solvable models having some elements of chaotic systems (as well as of integrable ones). \new{As we shall see, extremal random circuits use the same building blocks as dual-unitary circuits. Interesting phenomena we discuss though are not limited neither to extremal nor to dual-unitary circuits.}

Entanglement generation has been as well studied in the context of black hole physics~\cite{Hayden07,Susskind08,Lorenzo20}. A question of intense interest is in particular the maximal possible entanglement generation, with various bounds end explicit results~\cite{Suh14,Mezei16,Mezei17}. The fact that random circuits 'scramble' information well can be used also for decoupling protocols~\cite{Fawzi12} -- a procedure in which any initial (local) correlations between an observer and a system are spread out globally, such that no local measurement on the system can reveal any correlation anymore -- an observer becomes decoupled from a system (under local measurements). Such quantum information decoupling is in fact what is effectively going on during the thermalization process~\cite{thermalization}, and so random circuits can also be thought of as being models of ideal thermalization \new{(i.e., lacking any Hamiltonian-specific features)}. 

We obtain a complete class of random protocols that generate entanglement optimally, and more importantly, identify several new and surprising features that can emerge in non-Hermitian matrices describing many-body systems. Specifically, we (i) find random circuits that produce entanglement in the fastest possible way. Our fastest circuit is significantly faster than the best previous random circuits. (ii) We identify a number of phase transitions in time -- at certain moments the convergence rate to the asymptotic ``thermal'' entanglement of random states suddenly changes. This shows that an ideal thermalization modeled by a random circuit is a two-stage process rather than relaxation with a constant rate. (iii) We find that the convergence rate may not be given by the transfer matrix gap but can instead be either larger or smaller. This does not occur just in some obscure unimportant cases but in almost every case we looked at, \new{specifically in the fastest circuits as well as in generic ones, and also in circuits with random U(4) gates much studied in the past.} This happens in spite of the spectrum itself being rather innocuous, e.g. the 2nd largest eigenvalue $\lambda_2$ is gapped away from both $\lambda_1=1$ and $|\lambda_3|<|\lambda_2|$. \new{We also interestingly observe that in the thermodynamic limit the phenomenon is observed also in a single realization of a random circuits and that, in fact, randomness is not necessary neither in space, nor in time (one can use the same random 1-qubit transformation at all sites and at all times).} The fact that a common 'folk theorem' that the decay is given by the gap does not hold might have important implications in many areas of physics where one deals with non-Hermitian matrices, e.g., dissipative systems, transfer matrices, etc.. \new{Preliminary results show~\cite{tobe} the same phenomenon also in out-of-time-ordered correlations (OTOC).}

One particularly intriguing case is where the decay is $\sim (\frac{1}{2})^t$ rather than $\sim (\frac{1}{4})^t$ suggested by $|\lambda_2|=\frac{1}{4}$. We show that this occurs due to a 'phantom' eigenvalue $\frac{1}{2}$ -- an 'eigenvalue' that is not in the spectrum but is rather just mimicked by exponentially growing expansion coefficients in front of smaller (true) eigenvalues. We are not aware of any similar observations; the closest is perhaps a recent study~\cite{Mori20} of vanishing gaps in Lindblad generators, finding that \new{due to exponentially growing expansion coefficients} the gap does not always give physically relevant relaxation timescale~\cite{PRE15}.

\begingroup
\squeezetable
\begin{table}[t!]
\begin{ruledtabular}
\begin{tabular}{rccrcrcr} 
\multicolumn{1}{l}{Gate} & \phantom{abc} & \multicolumn{2}{c}{Protocol} & \phantom{abc} & Rate $\rE$ & \phantom{abc} & Ref.\\

\cmidrule(r){3-4} 
&& config. & b.c. & & & \\
\midrule

\multicolumn{1}{l}{1-dim.(n.n.):}\\
U(4) && rand. & PBC && $\frac{2}{5}=0.4$&& \cite{PRA08}\\
U(4) && rand. & OBC && $\frac{1}{5}=0.2$&& \cite{PRA08}\\
U(4) && perm. & OBC && $\ln{\frac{3}{2}}\approx 0.40$&& \cite{Zanardi12}\\
U(4) && BW    & PBC && $4\ln{\frac{5}{4}}\approx 0.89$ && \\
U(4) && BW    & OBC && $2\ln{\frac{5}{4}}\approx 0.45$ && \cite{Frank18,AdamPRB19,adam18}\\
\multicolumn{1}{l}{$\infty$-dim.(all-all):}\\
U(4) && rand. &  && $\frac{6}{5}=1.2$ && \cite{PRA08,Viola10}\\
XY,CNOT && rand. &  && $\frac{4}{3}\approx 1.33$ && \cite{PRA08}\\
\end{tabular}
\end{ruledtabular}
\caption{Existing exact results for the purity decay rate $\rE$, defined as $I(t) \sim \exp{(-\rE t)}$, in qubit random circuits (cf. Fig.~\ref{fig:pregled}). Per unit of time $\sim n$ gates are applied. Deterministic nearest-neighbor (n.n.) protocol with a brick-wall (BW) pattern of gates is faster than randomly choosing a n.n. pair on which the gate acts (rand.). Allowing coupling between an arbitrary pair of gates (all-all) is even faster. Results are shown for open (OBC) and periodic boundary conditions (PBC). \new{U(4) denotes a Haar random gate, while the XY gate is also known as the iSWAP or DCNOT gate.}}
\label{tab:oldrates}
\end{table}
\endgroup
Considering conceptual and practical importance of random circuits it is not surprising that they have a long history. Let us make a brief overview of existing results, focusing on the speed with which entanglement is generated. One of the earliest works that studied the convergence properties of random circuits are Refs.~\cite{emerson03,emerson05}. First exact results about the convergence rate towards random states were made possible by mapping~\cite{oliveira07} the average dynamics to a Markovian chain. Using the mapping, the question about the speed of generating entanglement boils down to the question about the gap of a certain transfer matrix. Such a mapping is rather fruitful, not just for the specific question of entanglement generation~\cite{PRA08,Viola10,cwiklinski13,metoda_redukcija,Swingle20}, but also for a nice systematic treatment of any expectation that involves $t$ copies of the propagator $U$ and $t$ copies of $U^\dagger$ -- a so-called unitary $t$-design~\cite{gross07,Harrow09,brandao16,brandao16b,Hunter20}. The simplest entanglement quantifier purity $I=\tr{\rA^2}$ therefore belongs under the realm of a 2-design. The Markovian mapping in particular allows to get exact results for the purity convergence rate (purity ``entanglement speed'') by recognizing that the resulting matrix is equivalent to an (integrable) spin chain~\cite{PRA08}. Recently, Markovian description has been used to describe evolution of purity for all possible bipartitions \new{at once}~\cite{metoda_redukcija}. Exact results are by now available for various protocols containing the 2-qubit random Haar U(4) gate (including quantities beyond purity), e.g., choosing a random nearest-neighbor pair~\cite{PRA08,Viola10}, picking a random permutation~\cite{Zanardi12}, or having a brick-wall pattern~\cite{Frank18,AdamPRB19,adam18}, or for random single-site unitaries and a global phase~\cite{Swingle20}. If one allows the coupling between all pairs of qubits one can also get the exact entanglement speed for some non-random gates, in particular for the important XY gate (as well as for the CNOT)~\cite{PRA08}, see Table~\ref{tab:oldrates}. Exact results are also available in the limit of large local Hilbert space dimension $q$~\cite{Viola10,Zanardi12,Adam17,Chalker18,Frank18,AdamPRB19} (we shall focus on qubits, $q=2$). For numerical results see Refs.~\cite{Znidaric_2007,Braun08,cwiklinski13,Adam17,Hunter20}.

Summarizing these results in Table~\ref{tab:oldrates} we see that using nearest-neighbor (n.n.) qubit gates, the fastest known protocol is the brick-wall (BW) configuration with a U(4) gate for which the entanglement rate defined via a long-time purity decay $I(t) \asymp \exp(-\rE t)$ is $\rE \approx 0.45$~\cite{foot3} per boundary link. We will find a protocol which is more than $3$ times faster and has the maximal possible entanglement speed. The optimal 2-qubit gate will turn out to be the so-called XXZ gate \new{(also known as the pSWAP)}, a special case of which is the XY gate (iSWAP). The same gate has appeared before in the context of random circuits: it has been observed numerically that the XY gate is the fastest of all gates for random n.n., or all-all protocol~\cite{Znidaric_2007} (which though are \new{much slower than our best BW} protocol). The same type of gate has recently emerged in dual-unitary circuits~\cite{prosen19}. Due to their special properties such circuits (see also related concepts~\cite{Arul20} of 2-unitaries~\cite{Karol15} and perfect tensors~\cite{Yoshida15}) allow exact results, for instance of R\' enyi entropies for a non-symmetric bipartition in a disordered kicked Ising model~\cite{prosen19} (this system has maximal entanglement rate \new{provided one counts time in an optimal way}), see also Ref.~\cite{brunoprb}; for a symmetric half-infinite bipartition see Ref.~\cite{sarang19}. One can also calculate the tripartite information~\cite{Bruno20}, or find circuits with maximal butterfly velocity~\cite{Austen20} in the dual-unitary context.

We are going to characterize entanglement generation through the decay of average purity $I(t)=\tr{\rA^2(t)}$. Because we want to understand entanglement generation on a global scale we shall focus on a symmetric bipartition~\cite{foot2} of our system of $n$ qubits into two equal subsystems A and B, each with $\nA=\nB=n/2$ qubits. \new{That is, for a chain geometry of qubits we split the chain in two equal halfs of consecutive qubits.} We shall calculate the average purity behavior by mapping its dynamics to a Markovian process~\cite{oliveira07,PRA08,metoda_redukcija}, reducing the problem to that of properties of a particular transfer matrix $M$. At long times the purity will decay exponentially with a rate $\rE$ determining how fast entanglement is produced. Because the purity entanglement rate $\rE$ will be proportional to the area ${\cal A}$ of the boundary between A and B one often considers the entanglement speed $\vE$ (called also the purity speed, or tsunami velocity~\cite{Mezei16}) defined as
\begin{equation}
I(t) \asymp \exp(-\rE t), \qquad \rE \equiv \vE {\cal A} s_{\infty}.
\label{eq:vE}
\end{equation}
For random circuits the asymptotic state is an infinite temperature state having maximal entropy density $s_{\infty}=\ln{q}$, while ${\cal A}$ is simply the number of 2-qudit gates applied across the boundary between subsystems A and B per unit of time. If one uses natural units of time such that one applies one gate on each bond per unit of time then ${\cal A}$ is just the number of bonds cut by a bipartition, which for a symmetric bipartition \new{we use} is ${\cal A}=1$ for open boundary conditions (OBC) and ${\cal A}=2$ for periodic boundary conditions (PBC). Because one can generate $2$ maximally entangled 2-qudit states per one application of a 2-site gate~\cite{foot1}, the maximal $\vE$ one can achieve is 
\begin{equation}
  \vE \le 2.
  \label{eq:bound}
\end{equation}
In quantum circuits one therefore has the upper bound $\rE \le 4\ln{q}$ for PBC, and $\rE \le 2\ln{q}$ for OBC.

One can also study other quantities measuring entanglement, like the von Neumann entanglement entropy $S(t)=-\tr{(\rA \ln{\rA})}$, or the R\' enyi entropies $S_r(t)=\ln{\rA^r}/(1-r)$, and use them to define corresponding rates and speeds. In general all those entanglement speeds can be different~\cite{AdamPRB19}, an example being a circuit with random U(4) gates. The reason for the difference is variation in entanglement properties of different circuit realizations, for instance, some rare realizations could use very inefficient entanglers from the U(4) set of gates. For the circuits we focus on we expect such effects to be much less pronounced. In particular, the fastest scramblers that we find have the maximal speed allowed by causality and so all R\' enyi entropies should have the same maximal speed (similarly as e.g. in Ref.~\cite{prosen19}). Furthermore, even for non maximal random circuits with fixed 2-qubit gates the variation in entanglement between different realizations should be much smaller than e.g. for the random U(4) case because the entangling 2-qubit gates are the same at every step (even for random U(4) the difference between the speed from the average purity and the average 2nd R\' enyi entropy is likely very small~\cite{AdamPRB19}). The circuits we study will at not too short times produce typical states that have good self-averaging properties, and we expect von Neumann entropy and R\' enyi entropy to behave essentially the same as the logarithm of the average purity, $S(t) \approx S_2(t)=-\ln{I(t)}$. We leave possible differences for future work. In the remainder of the paper we will always discuss only purity rates and speeds.
\begingroup
\squeezetable
\begin{table}[t!]
\begin{ruledtabular}
\begin{tabular}{rccrcrcr} 
\multicolumn{1}{l}{Gate} & \phantom{abc} & \multicolumn{2}{c}{Protocol} & \phantom{abc} & Rate $\rE$ & \phantom{abc} & $-\ln{|\lambda_2|}$ \\

\cmidrule(r){3-4} 
 && config. & b.c. & &  & & [for XY] \\
\midrule

\multicolumn{1}{l}{1-dim.(n.n.):}\\
{\bf XXZ}    && {\bf BW}    & {\bf PBC} && $\bm{\ln{16}\approx 2.77}$ && ${\ln{9}\approx 2.20}$\\
{\bf XXZ}    && {\bf BW}     & {\bf OBC} && $\bm{\ln{4}\approx 1.39}$&& ${\ln{4}\approx 1.39}$\\
XXZ    && S     & PBC && $\ln{4}\approx 1.39$&& $\ln{3}\approx 1.10$ \\
XXZ    && S     & OBC && $\ln{2}\approx 0.69$&& $\ln{4}\approx 1.39$ \\
CNOT  && BW    & PBC && $\approx 0.591$&& \\
CNOT  && BW    & OBC && $\approx 0.284$&& \\
CNOT  && S     & PBC && $\approx 0.570$&& \\
CNOT  && S     & OBC && $\approx 0.296$&& \\
\end{tabular}
\end{ruledtabular}
\caption{New results for the purity decay rate $\rE$ in random circuits (cf. Fig.~\ref{fig:pregled}). The fastest entanglement generation for local couplings is achieved for the XXZ gate in a BW configuration (bold). Such circuits are optimal and much faster than e.g. using random U(4) gates or all-all coupling (Table~\ref{tab:oldrates}), or using the CNOT gate. The staircases configuration S is on the other hand the slowest. The rate is in general not given by the 2nd largest eigenvalue $|\lambda_2|$ of the associated transfer matrix.}
\label{tab:nnrates}
\end{table}
\endgroup

\begin{figure}[t!]
  \includegraphics[width=2.8in]{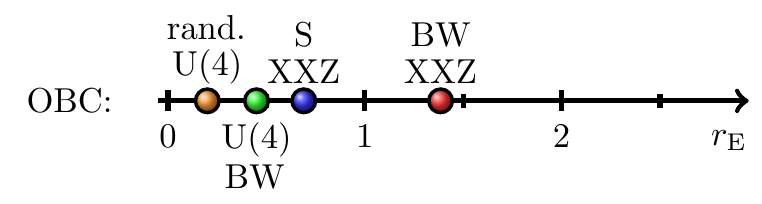}\\
  \includegraphics[width=2.8in]{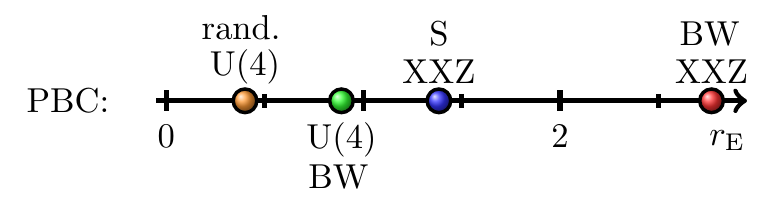}
  \caption{Overview of purity entanglement generation rates $\rE$ for random circuits with different gates and ordering of n.n. gates: random (rand.) vs. brick-wall (BW) vs. staircase (S). See Fig.~\ref{fig:S_in_BW} for BW and S structure.}
  \label{fig:pregled}
\end{figure}

Our results for $\rE$ are summarized in Table~\ref{tab:nnrates} and in Fig.~\ref{fig:pregled}. We can see that the rates are in general not equal to the ones suggested by the 2nd largest eigenvalue $\lambda_2$ of the transfer matrix $M$, and that in the two optimal cases, namely the BW configuration with PBC, or the OBC, the entanglement velocity is $\vE=2$ and therefore saturates the bound (\ref{eq:bound}). The optimal gate is in both cases found to be the XXZ gate for any value of the anizotropy $0\le \az < 1$ in front of the $\sz_j\sz_{j+1}$ coupling (see Eq.~\ref{eq:XXZ} for its definition). \new{In particular}, using the XXZ gate is much faster than using random U(4) gates, as well as for instance the CNOT gate. We do not find any transition with $\az$ in the relevant entanglement rate $\rE$, though we do find it in the leading eigenvalue $\lambda_2$ of the transfer matrix (which in most cases nevertheless does determine the late-time entanglement rate after the phase transition at the thermalization time). For dual-unitary systems with transitions with $\az$ in transfer matrices see Refs.~\cite{BrunoSci,Bruno20}.

\section{Methods}
\label{sec:methods}

\subsection{Random quantum circuits}

We study random quantum circuits in which 2-qubit unitary transformations are drawn from some set of unitary gates. One elementary gate -- a 2-qubit transformation $U_{i,j}$ acting on qubits $i$ and $j$ -- is a product of independent single-qubit random unitaries $V_i$ acting on $i$-th qubit and $V_j$ acting on $j$-th qubit, drawn according to the Haar measure on group U(2), and a 2-site unitary $W_{i,j}$. The whole gate is therefore $U_{i,j}=W_{i,j} V_i V_j$. The 2-site $W_{i,j}$ will be the same for all steps, while the single-site $V_i$ are independent for each step and \new{each} qubit. The formalism that we will use can actually accommodate also the situation where $W_{i,j}$ would be random from U(4) \new{(we are going to briefly comment on that case in Sec.~\ref{sec:gen})}, however, such circuits produce entanglement slower (see Table~\ref{tab:oldrates}) than the ones we study.

\begin{figure}[t!]
	\begin{center}
		\includegraphics[width=80mm]{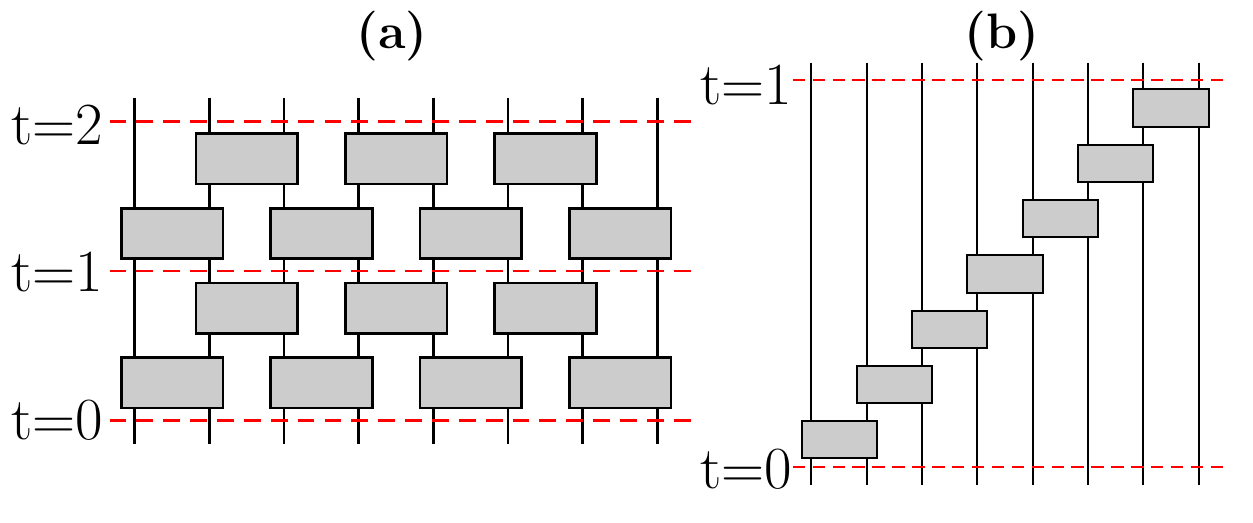}
		\caption{An illustration of different quantum circuit configurations (here for OBC): (a) the brick-wall (BW) configuration, (b) the staircase (S) configuration.}
		\label{fig:S_in_BW}
	\end{center}
\end{figure}

A given random circuit is specified by a fixed 2-site $W$ and a sequence of qubit pairs, called a protocol or configuration, on which $U_{i,j}$ are applied. For instance, one could choose $W$ to be the CNOT gate and a sequence to be a brick-wall pattern (Fig.~\ref{fig:S_in_BW}). We are going to find a gate $W$ and a protocol that generate entanglement the fastest.

In the main body of the paper we shall limit to protocols in one-dimension (1D), where the gates are allowed only between nearest-neighbor (n.n.) qubits with either open (OBC) or periodic boundary conditions (PBC), i.e., qubits are on a line or on a circle. We shall only briefly mention the 2D case of the Sycamore~\cite{Google} quantum processor, and in Appendix~\ref{app:random_couplings} the all-to-all coupling (n.n. protocols are faster than those). Focusing on the 1D setting, the space of quantum circuits over which we need to optimize is still large (infinite), therefore simplification is necessary. First, we shall focus on periodic protocols in which the geometry of applied gates repeats after each period. One period will consist of exactly one gate applied on each allowed bond, that is, for OBC there are $T=n-1$ gates in a period whereas there are $T=n$ for PBC. Optimization therefore needs to be performed over all $T!$ permutations describing different orderings of gates within one period, and over all choices of the gate $W$. Our time $t$ that we use throughout the paper measures the number of periods (see Fig.~\ref{fig:S_in_BW}). In Sec.~\ref{sec:spectral} we shall prove that one does not need to consider all $\sim n!$ permutations but only $\sim n$ of them.

One can also reduce the number of $W$ that need to be checked. Because we have a single-site invariance due to random single-qubit gates it suffices to consider $W$ in the canonical form parameterized by only $3$ parameters (instead of $16$ for general $W \in U(4)$). Namely, every 2-qubit unitary can be written in the following canonical form~\cite{dekompozicija_1,dekompozicija_2_in_simetrije},
\begin{align}
	W_{j,k} &= V^{(1)}_j V^{(2)}_k w_{j,k}(\av) V^{(3)}_j V^{(4)}_k, \nonumber\\
	w_{j,k}(\av)&=\exp{[\mathrm{i}\frac{\pi}{4}(\ax \sx_j \sx_k + \ay \sy_j \sy_k + \az \sz_j \sz_k )]},
	\label{canonical_form}
\end{align}
where $V^{(\alpha)}_{k}$ are one-site unitary operators, $\sigma^{\mathrm{x},\mathrm{y},\mathrm{z}}$ are Pauli matrices and $\av=(\ax,\ay,\az)$ has three real parameters, which can be constrained to $0 \le \az \le \ay \le \ax \le 1$. A detailed explanation on how to calculate this decomposition from an arbitrary unitary two-site gate $W$ can be found in Ref.~\cite{dekompozicija_recept}. Fixed single-site $V_k^{(\alpha)}$ for a specific gate $W$ can be lumped together with random unitaries $V_k$, so that we need to explore only canonical gates $w(\av)$. For instance, the CNOT gate has $\av=(1,0,0)$, the gate that will turn out to be the fastest is the XY gate with $\av=(1,1,0)$ (also called the iSWAP or DCNOT gate). More generally, the optimal set will consist of gates
\begin{equation}
W_{\rm XXZ}=  \exp{[\mathrm{i}\frac{\pi}{4}(\sx_j \sx_k + \sy_j \sy_k + \az \sz_j \sz_k )]},
\label{eq:XXZ}
\end{equation}
i.e. with parameters $\av=(1,1,\az)$ and $\az<1$, that we call the XXZ gate (also known as the pSWAP -- parametric SWAP gate). The SWAP gate reached at $\av=(1,1,1)$ generates no entanglement and therefore has completely different entanglement properties than the XXZ family of gates. For our best protocols one therefore has in the thermodynamic limit (TDL) a discontinuous transition from the maximal rate at $\az<1$ (XXZ gate) to zero rate at \new{$\az=1$} (SWAP gate).

To quantify bipartite entanglement of pure states we shall use purity $I(t)$,
\begin{equation}
  I(t) = {\rm tr}_{\rm A}\rA^2(t),
  \label{eq:I}
\end{equation}
where $\rA(t)={\rm tr}_{\rm B} \ket{\psi(t)}\bra{\psi(t)}$, and $\ket{\psi(t)}=U(t)\ket{\psi(0)}$, with $U(t)=\prod_{(i,j)} U_{i,j}$ a product of all 2-qubit gates upto time $t$ (because we have $T$ gates per unit of time the total number of 2-qubit gates in $U(t)$ is therefore $T\cdot t$). The whole system of $n$ qubits is bipartitioned into a subsystem A with $\nA$ qubits and a complement B with $\nB$ qubits, $n=\nA+\nB$. In explicit numerical demonstrations we shall always use a symmetric half-half bipartition where A is composed of the first $n/2$ consecutive qubits while B is the rest. \new{The results though do not depend on the specifics of the bipartition provided one has an extensive $\nA \propto n$ connected subsystem A.} The initial state $\ket{\psi(0)}$ is always chosen to be fully separable with respect to any bipartition such that $I(0)=1$. Purity will therefore decay with time from its initial $1$ to the asymptotic value being equal to the purity of random states~\cite{Lubkin},
\begin{equation}
  I_\infty = \frac{2^\nA+2^\nB}{1+2^n}.
  \label{eq:I_inf}
\end{equation}
For large $t$ our random circuit namely uniformly samples $U$ from the unique unitarily invariant Haar measure. Observe that this asymptotic purity is not exactly equal to the purity of the infinite-temperature state $\rA \propto \1$ which is $I(\beta=0)=\frac{1}{2^{n/2}}$, whereas $I_\infty=\frac{2}{2^{n/2}}\frac{1}{1+2^{-n}}$ for the symmetric bipartition. For easier understanding we shall often show
\begin{equation}
S_2(t)=-\log_2{I(t)},
\end{equation}
which is a quantity that will grow from $0$ at $t=0$ to its maximal value $S_2(\infty) \approx \frac{n}{2}-1$ (for a half-half bipartition) reached when the state $\ket{\psi(t)}$ converges towards a random state. $S_2(t)$ will reach its saturation $\sim n/2$ at time $t_\infty$ given by
\begin{equation}
  \frac{t_\infty}{\nA}=\frac{\ln{2}}{\rE}=\frac{1}{\vE {\cal A}}.
  \label{eq:tinf}
\end{equation}
For instance, for the BW PBC random circuit the scaled thermalization time will be $t_\infty/\nA=1/4$. To more clearly show the relaxation process \new{towards $I_\infty$} we shall also study
\begin{equation}
  \Delta S_2(t)=-\ln{|I(t)-I_\infty|}.
\label{eq:dS}
\end{equation}
Looking at $\Delta S_2(t)$ instead of just $S_2(t)$ will allow to discuss relaxation (thermalization) on timescales beyond $t_\infty$, which will reveal interesting \new{new} phase transitions. Because the state reached (at non-small times) is akin to a random state one will have good self-averaging properties in the thermodynamic limit so that other measures of entanglement, like the von Neumann entropy of the R\' enyi entropies, will behave similarly as the above $S_2(t)$, see Appendix~\ref{app:S} for an explicit numerical demonstration. Also, because the variance of $I(t)$ between different circuit realizations is small in the TDL one can as well study the average $I(t)$, i.e., averaging over different realizations of single-qubit unitaries $V_j$. Such averaging is crucial to get an analytically tractable setting and we shall describe next the formalism used \new{for that}.

\subsection{Markov chain description}

As mentioned, one can describe~\cite{oliveira07} the average dynamics of $\rA^2$, i.e., of squares of the expansion coefficients of $\rA$, by a Markovian matrix $M_{i,j}$ that describes the averaged action of one 2-qubit gate $U_{i,j}$. Such description has been widely used, e.g. Refs.~\cite{PRA08,Viola10,Harrow09,brandao16,Hunter20}, for random U(4) gates or for Clifford gates (gates that map a product of Pauli matrices to a single product of Pauli matrices, examples being the CNOT or the XY gate). Recently, a different approach has been taken~\cite{metoda_redukcija} in which one \new{directly} describes the evolution of purity for all possible bipartitions rather than of $\rA^2$. It allows describing purity evolution for a protocol with an arbitrary 2-qubit gate $W_{i,j}$. As we shall see, despite being more general and formally different, it will, for the cases for which the old method works (Clifford gates), at the end result in the very same transfer matrix $M_{i,j}$ \new{as obtained in Ref.~\cite{PRA08}}.

One starts by formally defining a Hilbert space with basis $\ket{\mathbf{s}}$, where $\mathbf{s} = (s_1, s_2, \dots, s_n)$ with $s_i \in \{ \uparrow, \downarrow \}$. A given $\mathbf{s}$ encodes a bipartition one looks at, that is, if $s_i=\uparrow$ the $i$-th qubit is \new{in the} subsystem A, otherwise it is in B. A state vector whose evolution we want to describe is then defined as
\begin{equation}
\Phi'(t) = \sum_{\mathbf{s}} I_{\mathbf{s}}(t)\, \ket{\mathbf{s}},  
\end{equation}
where $I_{\mathbf{s}}(t)$ is the purity of a state $\ket{\psi(t)}$ for a given bipartition labeled by $\mathbf{s}$. The state $\Phi'(t)$ therefore implicitly depends on the initial state -- for the most interesting fully separable initial state $\ket{\psi(0)}$ the initial $\Phi'(0)\equiv \Phi'_0 $ is simply $\Phi'_0=(\ket{\uparrow}+\ket{\downarrow})^{\otimes n}$. Averaging over two one-qubit random matrices $V_i$ and $V_j$ it has been shown~\cite{metoda_redukcija} that the vector $\Phi'$ is mapped to $M'_{i,j}\Phi'$ \new{(see also Appendix~\ref{app:derivationM})}. Repeating this step for all $T$ two-qubit unitaries \new{(each involving independent 1-qubit random unitaries)} that are applied per unit of time we get the transformation 
\begin{equation}
  \Phi'(t+1) = M'\Phi'(t), \qquad M'=\prod_{(i,j)}^T M'_{i,j}.
  \label{eq:M'}
\end{equation}
Dynamics of purity -- the expansion coefficients of $\Phi'(t)$ -- is therefore given by iterating a single-step transfer matrix $M'$ describing Markovian average purity evolution. To get purity for the most interesting symmetric half-half bipartition one just has to project $\Phi'(t)$ to the basis state $\ket{\uparrow \ldots \uparrow \downarrow \ldots \downarrow}$. Slightly abusing notation and defining the initial vector $\mathbf{\Phi'_0}=(1,1,\ldots,1)$ and the bipartition basis vector $\mathbf{\Phi'}_{\rm half}$ with components $[\mathbf{\Phi'}_{\rm half}]_k=\delta_{k,2^{n/2}}$, we have the average purity after $t$ steps
\begin{equation}
  I(t)=\mathbf{\Phi'}_{\rm half}\,(M')^t\,\mathbf{\Phi'}_0.
\end{equation}
The $4\times 4$ matrix $M'_{i,j}$ can be calculated for an arbitrary $W_{i,j}$ in its canonical form $w(\av)$ (\ref{canonical_form}). Following the procedure in Ref.~\cite{metoda_redukcija} we obtain \new{(Appendix~\ref{app:derivationM})}
\begin{equation}
	M'_{i,j} =
	\begin{pmatrix}
		1 & 0 & 0 & 0 \\
		h & b_+ & b_- & h \\
		h & b_- & b_+ & h \\
		0 & 0& 0& 1
	\end{pmatrix}, \quad b_\pm = \frac{1}{36} \left( 3\pm 6u+5v\right)
	\label{M'_cd}
\end{equation}
where the basis is ordered as $s_i s_j = \{ \uparrow\uparrow, \uparrow\downarrow, \downarrow\uparrow, \downarrow\downarrow \}$, and $h = \frac{1}{9} \left(3-v\right)$, $u = \cos\left( \pi \ax \right)+\cos\left( \pi \ay \right)+\cos\left( \pi \az \right)$ and $v = \cos\left( \pi \ax \right)\cos\left( \pi \ay \right)+\cos\left( \pi \ax \right)\cos\left( \pi \az \right)+\cos\left( \pi \ay \right)\cos\left( \pi \az \right)$. For instance, for the XY gate we have $u=v=-1$, resulting in
\begin{equation}
	M'_{i,j}({\rm XY}) =
	\frac{1}{9}\begin{pmatrix}
		9 & 0 & 0 & 0 \\
		4 & -2 & 1 & 4 \\
		4 & 1 & -2 & 4 \\
		0 & 0& 0& 9
	\end{pmatrix}.
\end{equation}
\new{
  Not surprisingly, due to local unitary averaging the parameters $h,u,v$ can be related to known 2-qubit gate objects. Specifically, one has $h=2e(W)$, where $e(U)$ is the entangling power~\cite{Paolo00} of unitary $U$, while $v$ itself is equal to the local invariant $G_2$~\cite{Makhlin00,Whaley03}. Note though that knowing the 2-site properties of the gate $W$ does in no way help us understand the entanglement generating properties of the whole circuit. As we will see, the latter is a full many-body property (the interesting effects that we discuss only arise in the TDL) and also crucially depends on the configuration, not just on the choice of a 2-qubit gate. For instance, the entangling power~\cite{footeU} of $W$ is maximal for the XY as well as for the CNOT gate, and continuously decreases with $\az$ for the XXZ gate. Our results on the other hand show that in the many-body setting of random circuits the CNOT gate is for instance more than $4$ times slower than the XY gate (Table~\ref{tab:nnrates}), and that for the XXZ gate the rate does not depend on $\az$.
  }

We observe that the matrix $M'_{i,j}$ is not symmetric. Often it is easier to work with symmetric matrices and it turns out that it is possible to transform $M'$ to a symmetric form. This will make symmetries more transparent and facilitate comparison with previous Markovian descriptions of random circuits. We can obtain the symmetric form with a similarity transformation $M_{i,j} = A_i^{-1} A_j^{-1} M'_{i,j} A_i A_j$, where
\begin{equation}
	A_i =  
	\begin{pmatrix}
		1 & -1/\sqrt{3}\\
		1 & 1/\sqrt{3}
	\end{pmatrix},
	\label{A_i}
\end{equation}
arriving at
\begin{equation}
	M_{i,j} = 
	\begin{pmatrix}
		\frac{1}{36} \left( 33+v \right) & 0 & 0 & \frac{1}{12}\left(3-v\right) \\
		0 & u_+ & u_- & 0 \\
		0 & u_- & u_+ & 0 \\
		\frac{1}{12} \left( 3-v \right) & 0 & 0 & \frac{1}{4}\left(1+v\right)
	\end{pmatrix},
	\label{eq:M_cd}
\end{equation}
where $u_\pm=(3\pm u)/6$. This elementary 2-site \new{matrix} can be equivalently written in terms of Pauli matrices,
\begin{equation}	
	M_{i,j} = d \1  + J_\mathrm{x} \sx_i \sx_j +J_{\mathrm{y}} \sy_i \sy_j+ J_{\mathrm{z}} \sz_i  \sz_j + \frac{h}{2} (\sz_i + \sz_j),
	\label{Mcd_Heisenberg}
\end{equation}
where $d = \left( 39 + 6 u + 5 v\right)/72$, $J_\mathrm{x} = \left(9 - 2u - v\right)/24$, $J_\mathrm{y} = \left(3 - 2u + v\right)/24$, $J_\mathrm{z} = \left(3 - 6u +5v\right)/72$ and $h = \left(3-v\right)/9$. For the XY gate it is equal to
\begin{equation}
	M_{i,j}({\rm XY}) =
	\frac{1}{9}\begin{pmatrix}
		8 & 0 & 0 & 3 \\
		0 & 3 & 6 & 0 \\
		0 & 6 & 3 & 0 \\
		3 & 0& 0& 0
	\end{pmatrix}.
        \label{eq:Mxy}
\end{equation}
\new{Observe that even if one starts with a noninteracting gate like the XY the resulting $M$ is interacting ($J_{\rm z} \neq 0$).}
For Clifford gates, e.g. CNOT and XY, this is in fact the same transfer matrix as the one already derived in Ref.~\cite{PRA08}. The new mapping~\cite{metoda_redukcija} that we use therefore generalizes the Markovian mapping to non-Clifford gates. As a side note, the transfer matrix $M_{i,j}$ for a random U(4) gate~\cite{PRA08} is obtained by formally using $u=0$ and $v=-3/5$ in Eq.~(\ref{Mcd_Heisenberg}).

The similarity transformation $A$ preserves the spectrum -- the spectrum of $M$ is the same as of $M'$, only the vectors have to be transformed. Similarly as in Eq.~(\ref{eq:M'}) we use $M$ to denote a product of $T$ 2-qubit $M_{i,j}$ appearing in one unit of time. We shall use a prime to denote matrices and vectors written in the (unrotated) basis $\ket{\mathbf{s}}$, like $M'_{i,j}$ or $\Phi'(t)$, whereas unprimed objects, like $M_{i,j}$ or $\Phi(t)$ are in the basis transformed by $A$ (where needed, we will denote the basis vectors of this space by $\mathbf{e}_k$).  For instance, the initial vector corresponding to a product initial state $\mathbf{\Phi'}_0$ is transformed to $\mathbf{\Phi}_0=(A^{-1})^{\otimes n}\mathbf{\Phi'}_0$ and has components $\mathbf{\Phi}_0=(1,0,\ldots,0)$, while $\mathbf{\Phi}_{\rm half}=(A^T)^{\otimes n} \mathbf{\Phi'}_{\rm half}$. A boldface notation is used to stress that a vector (components) is written in a given basis, like $\mathbf{\Phi}_0$, while $\Phi_0$ is used for a basis-independent ket notation. The average purity after $t$ periods of our random circuits in this new basis is $I(t)=\mathbf{\Phi}_{\rm half}\, M^t\, \mathbf{\Phi}_0$. For a symmetric bipartition and a product initial state, due to the structure of $\mathbf{\Phi}_0$ and $\mathbf{\Phi}_{\rm half}$, purity is determined by one column of $M^t$, or equivalently by one row of $(M')^t$. Due to a block structure of $M_{i,j}$ (\ref{eq:M_cd}) it is also clear that $M$ conserves the parity $Z:=(-1)^{N_\downarrow}=\prod_j \sz_j$ of a vector $\ket{\Phi(t)}$, i.e., the number of down spins $N_\downarrow$. Because the initial state in the primed basis $\ket{\Phi'_0}$ is invariant under the particle-hole transformation (i.e., spin-flip), $X \ket{\Phi'_0}=\ket{\Phi'_0}$, where $X=\prod_j \sx_j$ (this is because the purity is invariant under exchanging subsystems A$\leftrightarrow$B), the parity of the initial $\ket{\Phi_0}$ is always even (even $N_\downarrow$). Indeed, if $X \ket{\Phi'_0}=\ket{\Phi'_0}$ then $X A^{\otimes n} \ket{\Phi_0}=A^{\otimes n} \ket{\Phi_0}$, leading to $ (A^{-1})^{\otimes n} X A^{\otimes n} \ket{\Phi_0}=\ket{\Phi_0}$. Because $A_j^{-1} \sx_j A_j=\sz_j$, one immediately gets $(-1)^{N_\downarrow} \ket{\Phi_0}=\ket{\Phi_0}$. The relevant parity sector of $M$ describing purity evolution of physical $\ket{\Phi_0}$ is therefore the one with an even number of down spins. We note that in some cases other additional symmetries are present. For instance, for BW protocol with PBC and $n$ divisible by $4$ one also has a reflection symmetry about the site $n/2$ and the translation symmetry by $2$ sites.

Another useful observation is that $M_{i,j}$ has eigenvectors that are independent of gate parameters $u$ and $v$, namely
\begin{align}
\label{eq:v}
\mathbf{v}_{1} &= (3,0,0,1), & \lambda &= 1, \\
\mathbf{v}_{2} &= (0,1,1,0),  & \lambda &= 1, \nonumber \\
\mathbf{v}_{3} &= (0,-1,1,0), & \lambda &= u/3, \nonumber \\
\mathbf{v}_{4} &= (-1,0,0,3), & \lambda &= (3+5v)/18. \nonumber
\end{align}
One consequence of this is that $M$ always has (at least) $2$ eigenvalues equal to $\lambda_1=1$. The corresponding eigenvectors are steady states and are also independent of the gate $W$. This is consistent with the asymptotic convergence to purity of random states. The physically relevant steady-state vector $\mathbf{\Phi'}_\infty$, i.e. the even parity eigenvector of $M'$ with $\lambda_1=1$, has components equal to $I_\infty$ (\ref{eq:I_inf}),
\begin{equation}
  \ket{\Phi'_\infty}=\sum_\mathbf{s} I_\mathbf{s}(\infty) \ket{\mathbf{s}}.
  \label{eq:Phiinf}
  \end{equation}
One can indeed use the above 1-site eigenvectors $\mathbf{v}_{1,2}$ to construct the steady state on $n$ qubits and explicitly verify $I_\infty$ (see Appendix~\ref{app:even_parity}). The steady state (\ref{eq:Phiinf}) can also be written compactly as a matrix product state of rank $2$, see Appendix~\ref{app:MPS}.

\new{Time evolution of the average purity is therefore Markovian (\ref{eq:M'}), i.e., there is no memory. The transition matrix, either $M'$ or $M$ (\ref{M'_cd},\ref{eq:M_cd}), is however not stochastic (rows do not sum to $1$) and therefore does not describe transition probabilities. The probabilistic interpretation is recovered if one works with squares of the expansion coefficients of the state $\rho$ on the original full $16$ dimensional 2-site Pauli basis~\cite{oliveira07}, it is though lost if one uses reduction~\cite{PRA08} down to $4$ nontrivial basis states that results in $M$ (\ref{eq:Mxy}), or directly works with purity evolution~\cite{metoda_redukcija}.}

Note that the transfer matrix for one step $M$ is in general not symmetric, even though the individual gates $M_{i,j}$ are. The spectrum of $M$ is therefore complex and contained within the unit circle. Provided the spectrum has a gap (which is always the case for nontrivial gates) the next largest eigenvalue $|\lambda_2|$ should determine the asymptotic decay of purity, and therefore the growth of entanglement \new{(decay of $I(t)$)}. Asymptotically one should have
\begin{equation}
	|I(t)-I_\infty| \asymp  |\lambda_2|^t,
\end{equation}
where we denote by $\lambda_2$ the largest eigenvalue smaller than $1$ (in absolute value). The corresponding entanglement rate (\ref{eq:vE}) would then be $\rE=-\ln{|\lambda_2|}$.

In the next Section we shall therefore study the gap of $M$ for various gates $W$ and configurations, and in particular prove that there are large classes of configurations with the same spectrum. This will enable us to then find circuits with the largest gap, i.e., the smallest $|\lambda_2|$. However, it will turn out, surprisingly, that the entanglement rate is in fact not given by $|\lambda_2|$, the reason being a non-symmetric nature of $M$. \new{As we will show, one can have two configurations with exactly the same spectrum but with different entanglement rates (for example, taking XXZ gates and comparing S configuration with OBC, and BW with OBC, see Table~\ref{tab:nnrates}). Nevertheless, the spectral analysis will suggest good candidates for optimal circuits. Furthermore, the two theorems on spectral equivalence are completely general and hold for products of any matrices acting on nearest-neighbor sites (for PBC the matrices need to be symmetric) and could therefore be of use in other situations.

  Combining spectral results on $\lambda_2$ from Sec.~\ref{sec:spectral} and numerically calculated true entanglement rates in Sec.~\ref{sec:rates} will result in the main general message of our work: in a many-body setting involving non-Hermitian transfer matrices spectral analysis can give wrong results.
}

\section{Spectral optimization}
\label{sec:spectral}

In this section we will determine the protocol and the gate with the smallest 2nd largest eigenvalue $|\lambda_2|$, i.e., the largest gap of $M$, in a 1D geometry where only local gates between nearest-neighbor sites are allowed. Specifically, if we have $T$ distinct n.n. gates, i.e., $T=n-1$ for OBC and $T=n$ for PBC, we find a sequence (configuration) of those $T$ gates among all $T!$ different permutations as well as among all possible choices of a 2-qubit gate $M_{i,i+1}$ (held fixed for all gates) \new{that has the smallest $|\lambda_2|$}.

As mentioned, we shall treat chains with periodic boundary conditions where also the gate $M_{n,1}$ is allowed, and chains with open boundary conditions. In both cases we shall first prove that among all factorially many $T!$ configurations there are only few with different gaps. The idea of proofs is to show that the spectrum of a product of gates $M_{i,i+1}$ does not change under certain rearrangements (holding the 2-site gate, i.e., $M_{i,j}$, fixed). This will then allow us to focus on equivalence classes defined as arrangement with the same spectrum. We shall use the equivalence notation $A \simeq B$ if the matrices $A$ and $B$ have the same spectrum. It turns out that for OBC there is only one equivalence class, while for PBC there are $\n2$ ($\n2$ is the largest integer smaller or equal to $\frac{n}{2}$). This greatly reduces complexity and allows to numerically find the optimal configuration and the gate.

\subsection{OBC protocols}

For OBC we have $(n-1)!$ possible configurations, however, all are spectrally equivalent. Namely, one can prove the following theorem.
\begin{theorem}\label{thm:OBC}
	Any product of $n-1$ 2-site matrices acting on $n-1$ distinct n.n. sites on a line (OBC protocol) is spectrally equivalent to the canonical staircases configuration (denoted shortly by S)
	\begin{equation}
	\Ms{n-1}{1} = M_{n-1,n} \dots M_{2,3} M_{1,2}.
	\end{equation}
\end{theorem}
The superscript in $\Ms{k}{j}$ indicates the number of consecutive gates in such canonical order and the subscript $j$ the first site on which they act. For the proof which uses the spectral equivalence under cyclic permutations, $ABC \simeq CAB$, \new{see Appendix~\ref{app:obc}}.

\begin{figure}[t]
	\begin{center}
		\includegraphics[width=80mm]{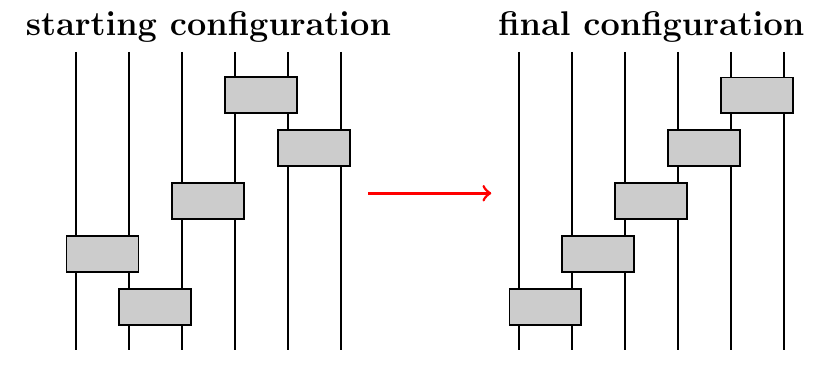}
		\caption{Transforming the initial configuration $M=M_{4,5} M_{5,6} M_{3,4} M_{1,2} M_{2,3}$ to the final S configuration preserves the spectrum of $M$. For OBC any configuration can be transformed to S.}
		\label{fig:M_to_canonic}
	\end{center}
\end{figure}
\new{As a consequence, one can transform any configuration with OBC to the canonical form S (see Fig.~\ref{fig:M_to_canonic} for an illustration) without affecting the spectrum of $M$. For OBC there is therefore a single spectral equivalence class and we can focus only on one representative staircases (S) configuration, so we only need to find the optimal two-qubit gate $M_{i,i+1}$ (\ref{eq:M_cd}). Staircases configurations have been considered before in e.g. the context of operator spreading~\cite{adam18} or complexity~\cite{andrew21}.}

\begin{figure}[h]
	\begin{center}
		\includegraphics[width=95mm]{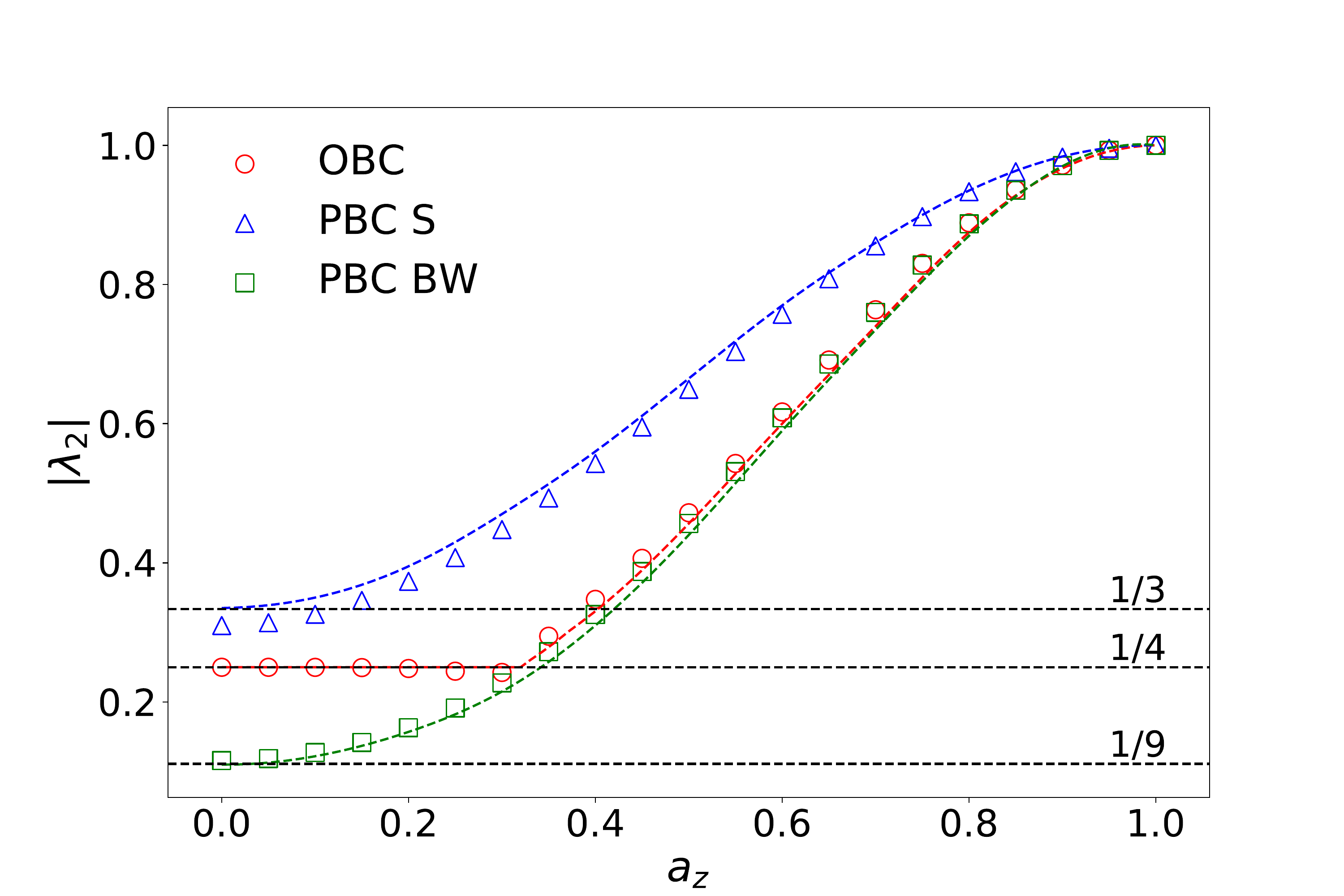}
		\caption{$|\lambda_2|$ for the XXZ gate (\ref{eq:XXZ}). Symbols are numerical data for $n=16$ while dashed curves are results in the TDL (obtained by extrapolation like in Fig.~\ref{fig:OBC_in_PBC_tdl}). For $\az=1$ all three cases converge to a non-entangling SWAP gate with $\lambda_2=1$. For the BW configuration with PBC ('fastest' class $p=\n2$) and for the S configuration with PBC ('slowest' PBC class $p=1$) the dependence is smooth. For the OBC (all configurations are equivalent) the gap does not change until a transition at $\az=\ac\approx 0.32$.}
		\label{fig:OBC_in_PBC_reze_az}
	\end{center}
\end{figure}

We have calculated the 2nd largest eigenvalue $|\lambda_2|$ of $M=\Ms{n-1}{1}$ for all different 2-qubit gates in the canonical form. Results (see Fig.~\ref{fig:OBC_gaps} in Appendix~\ref{app:numerics1}) show that the smallest eigenvalues $|\lambda_2|$ come from the region around $\ax = \ay = 1, \az=0$,  which is the XY gate. Looking more precisely at the $n$ dependence for the XY gate (Fig.~\ref{fig:OBC_in_PBC_tdl}) we find that in the TDL the S configuration with OBC has
\begin{equation}
|\lambda_2| =\frac{1}{4}.
\label{eq:lam_OBC}
\end{equation}
Exploring also the region around the XY gate, we find that the same largest eigenvalue (\ref{eq:lam_OBC}) is achieved also for sufficiently small nonzero values of $\az$ and $\ax=\ay=1$. In Fig.~\ref{fig:OBC_in_PBC_reze_az} we can see that at $\az=\ac \approx 0.32$ the transition happens in $|\lambda_2|$, so that the optimal $|\lambda_2|=\frac{1}{4}$ is obtained for all XXZ gates with $\az \le \ac$.

\subsection{PBC protocols}

\begin{figure}[ht!]
	\begin{center}
		\includegraphics[width=40mm]{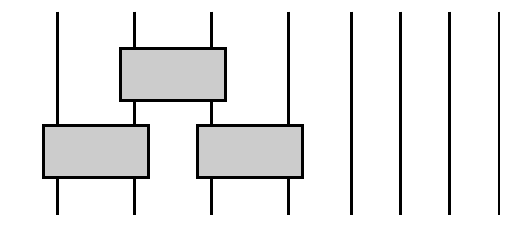}
		\caption{Brick-wall configuration \new{$\B{3}{1}$} with $3$ gates, i.e., $\B{2p-1}{1}$ with $p=2$.}
		\label{fig:OBC_BW}
	\end{center}
\end{figure}

\new{For PBC things are a bit more complicated, resulting in $\n2$ spectral equivalence classes. Each class can be labeled by an integer $p=1,\ldots,\n2$, being the maximal number of consecutive commuting gates in the canonical representative configuration}
\begin{equation}
\Ms{n-2p+1}{2p}\B{2p-1}{1},
\end{equation}
where $\B{2p-1}{1}$ is a brick-wall configuration made out of $2p-1$ gates (see Fig.~\ref{fig:OBC_BW} for an example)
\begin{equation}
\B{2p-1}{1}:=M_{2p-2,2p-1}\cdots M_{4,5}M_{2,3} M_{2p-1,2p}\cdots M_{3,4}M_{1,2}.
\end{equation}
We note that $\B{n}{1}$ denotes the full BW configuration with PBC, i.e. $\B{n}{1}=M_{n,1} \B{2 \lfloor \frac{n}{2} \rfloor - 1}{1}$.

\begin{figure}[t]
	\begin{center}
		\includegraphics[width=85mm]{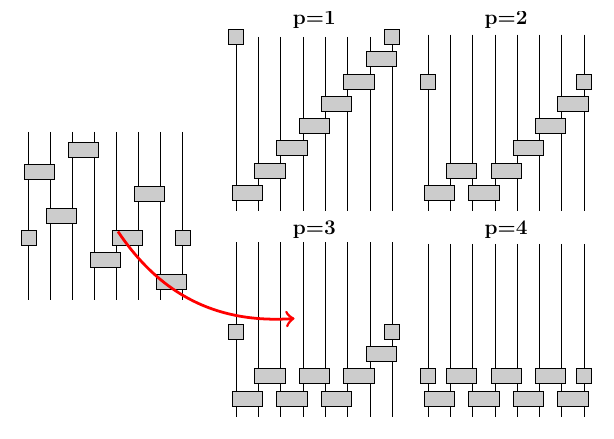}
		\caption{Spectrally equivalent configurations for PBC protocols. In the example with $n=8$ there are just 4 classes labeled by $p=1,\ldots,4$. The specific configuration shown in the left can be transformed to $\Ms{3}{6}\B{5}{1}$ with $p=3$ (red arrow).}
		\label{fig:PBC_equivalent}
	\end{center}
\end{figure}

\new{More precisely, the following theorem holds.}
\begin{theorem}\label{thm:PBC}
	Any product of $n$ \new{symmetric} 2-site matrices acting on n.n. sites on a circle (PBC configuration) is spectrally equivalent to one of $\n2$ canonical configurations of the form (Fig.~\ref{fig:PBC_equivalent})
	\begin{equation}
	\Ms{n-2p+1}{2p}\B{2p-1}{1}.
	\end{equation}
\end{theorem}
The proof that in addition to cyclic permutations uses also that the spectra of $A$ and $A^{\rm T}$ are the same can be found in Appendix~\ref{app:pbc} \new{(observe that our $M_{i,j}$ (\ref{eq:M_cd}) are symmetric matrices)}.

For PBC protocols we therefore have to find the optimal configuration among the $\n2$ different ones as well as the optimal 2-qubit gate. We have numerically computed $|\lambda_2|$ for small sizes $n$, see Appendix~\ref{app:numerics1} and Fig.~\ref{fig:PBC_gaps_n10}. One observes that fixing a 2-qubit gate (canonical parameters $\ax,\ay,\az$) and $n$ the gap always monotonically increases as we change the equivalence class from $p=1$ to $p=\n2$. Configurations in the class $p=1$ have in the TDL the smallest gap, those in $p=\n2$ the largest. Focusing on the region around the XY gate we see that in the TDL and for $p=1$ as well as for $p=\n2$ the fastest gate is obtained for $\az=0, \ax=\ay=1$ (Figs.~\ref{fig:PBC_p1} and~\ref{fig:PBC_pn2} \new{in Appendix~\ref{app:numerics1}}), which is the XY gate. Analyzing the dependence on $n$ (Fig.~\ref{fig:OBC_in_PBC_tdl} \new{in Appendix~\ref{app:numerics1}}) we get that in the TDL the eigenvalue is
\begin{equation}
|\lambda_2|=\frac{1}{9}, \qquad \hbox{\rm BW with PBC}
\label{eq:BW_PBC}
\end{equation}
for the fastest class $p=\n2$, and 
\begin{equation}
|\lambda_2|=\frac{1}{3}, \qquad \hbox{\rm S with PBC}
\end{equation}
for the slowest $p=1$ that we simply call S PBC. Contrary to the OBC case the smallest eigenvalue $|\lambda_2|$ is achieved only at a single point $\ax=\ay=1$, $\az=0$. For the XXZ gate the gap smoothly decreases as one moves away from $\az=0$, see Fig.~\ref{fig:OBC_in_PBC_reze_az}.

Among all protocols, OBC and PBC, the one with the largest gap, and therefore expected fastest entanglement generation, is the brick-wall (BW) with PBC and the XY gate for which $|\lambda_2|=1/9$. While we used numerics to obtain this value we can in fact rigorously show, see Appendix~\ref{app:9}, that $\lambda=1/9$ is in the spectrum of $M$ for $p=\n2$, which implies an exact relation $|\lambda_2| \ge 1/9$. \new{Proving analytically that $|\lambda_2|$ is $1/3, 1/4, 1/9$ for the XY gate and PBC S, OBC, and PBC BW, respectively (Fig.~\ref{fig:OBC_in_PBC_reze_az}) is an interesting open problem.}

\new{Note that the same eigenvalues $\lambda_2=1/3,1/4,1/9$ are also the leading non-trivial eigenvalues of the Markovian matrix propagating OTOCs~\cite{tobe}.}

\section{Entanglement generation speed}
\label{sec:rates}

\begingroup
\squeezetable
\begin{table}[t!]
  \begin{ruledtabular}
  \begin{tabular}{rcc}
\multicolumn{1}{l}{Configuration} & \multicolumn{2}{c}{Gate} \\
\cmidrule(r){2-3} & XY, $\az=0$ & XXZ, $\az \in (0,1)$ \\
\midrule
BW OBC  & $\lambda_2$   & $\lambda_2$ [$\az \in (0,\ac)$], faster\\
BW PBC  & faster   & faster\\
S OBC  & phantom   & phantom, faster\\
S PBC  & faster   & faster\\
\end{tabular}
\end{ruledtabular}
\new{
  \caption{The origin of entanglement rate: ``phantom'' denotes a phantom eigenvalue resulting in a slower entanglement generation than given by $|\lambda_2|$, ``faster'' denotes cases with faster generation than given by $|\lambda_2|$, and $\lambda_2$ the one case where the rate is actually determined by $|\lambda_2|$.}
         \label{tab:fantom}
  }
\end{table}
\endgroup
In the previous section we have identified gates and configurations that have the smallest $|\lambda_2|$ and are therefore expected to have the fastest asymptotic entanglement generation rate $\rE$. Comparing the spectral gap for the S configuration between PBC and OBC (Table~\ref{tab:nnrates}) we see that the gap is larger for the OBC than for the PBC. If the gap would be the end of the story that would mean that the entanglement generation rate would be larger for the OBC than for the PBC \new{which would be odd as} one would expect exactly the opposite. For symmetric bipartition one has 2 cuts for the PBC and only 1 for the OBC and therefore the rate should be twice as large for PBC than for OBC. Faster rate for the OBC would mean that applying less gates across the cut would produce more entanglement. \new{We will see in this section that numerically analyzing the true entanglement rate will rectify this odd spectral situation -- for OBC the true rate will be slower than suggested by $|\lambda_2|$, for PBC it will be faster than $|\lambda_2|$, such that at the end the ratio of the two is recovered at the expected value 2.}

\new{For} all extremal random circuits, either the fastest (BW), or the slowest in its class (the S configuration), the relevant entanglement rate $\rE$ on times smaller than the thermalization time $t_\infty$ (\ref{eq:tinf}) is, surprisingly, not necessarily given by $-\ln{|\lambda_2|}$ \new{despite $|\lambda_2|$ being gapped away from other eigenvalues}. In other words, the time $t_\infty$ at which $S_2$ reaches its random-state saturation value $\sim n/2$ is not equal to $t_\infty = n/(-2\log_2{|\lambda_2|})$. \new{It can be either larger due to a phantom eigenvalue -- a phenomenon where exponentially large expansion coefficients mimic a fake eigenvalue larger than $|\lambda_2|$, or smaller due to again exponentially large coefficients in front of eigenvalues that are smaller than $|\lambda_2|$.} As we shall see, the culprit lies in the non-Hermiticity of the transfer matrix $M$. \new{A quick overview of our results for extremal circuits that we focus on is in Table~\ref{tab:fantom}.}

\new{Non-Hermiticity is important for the following reason.} The spectral decomposition of a non-Hermitian operator $M$ has the form (assuming diagonalizability),
\begin{equation}
  M=\sum_j \lambda_j \ket{R_j}\bra{L_j},\qquad \braket{L_j}{R_k}=\delta_{j,k},
  \label{eq:spectral}
\end{equation}
where $\lambda_j$ are in general complex eigenvalues, while $\ket{R_j}$ and $\ket{L_j}$ are the associated right and left eigenvectors. Importantly, left and right eigenvectors are mutually orthogonal, however, left and right eigenvectors are not orthogonal between themselves. Expanding the initial state $\ket{\Phi_0}$ over a non-orthogonal basis $\ket{R_j}$ we have
\begin{equation}
  \ket{\Phi(t)}=\sum_j c_j \lambda_j^t \ket{R_j},\quad c_j=\braket{L_j}{\Phi_0}.
  \label{eq:x}
\end{equation}
If $M$ would be Hermitian (having orthogonal $\ket{R_j}$) the triangle inequality would bound $|c_j|^2\le \braket{\Phi_0}{\Phi_0}$. \new{For} non-Hermitian $M$ no such bound exists and the expansion coefficient $c_j$ can be arbitrarily big \new{(see e.g. Ref.~\cite{sarang20} for a simple Ising chain in a tilted field with an imaginary component, where the eigenvectors effectively span a subspace of lower dimension and thus the expansion coefficients can get large; see also the Lindblad case in Ref.~\cite{Mori20})}. This is precisely what will happen for our random circuits. Exponentially large (in $n$) coefficients $c_{j>2}$ will dominate over the term $c_2 \lambda_2^t$ for times $t<t_\infty$, causing the entanglement rate $\rE$ to be larger than $-\log{|\lambda_2|}$. In one case things will be even stranger -- the entanglement rate will be smaller than $-\log{|\lambda_2|}$, i.e., like there is a ``phantom'' eigenvalue larger than $|\lambda_2|$. We shall also explain how that comes about.

To calculate the correct entanglement rate we are going to use numerical simulations to get $I(t)$, or closely related $\Delta S_2(t)$ (\ref{eq:dS}). We will use two methods \new{with the goal of simulating as large a system as possible. This will turn out to be important as the multistage thermalization reveals itself clearly only in relatively large systems -- simulating systems with  e.g. just $n=16$ qubits one could have easily missed it.} The first method works by directly iterating the full state $\ket{\Phi(t)}$, obtaining
\begin{equation}
  I(t)-I_\infty= \mathbf{\Phi}_{\rm half}\cdot M^t \cdot (\mathbf{\Phi}_0-\mathbf{\Phi}_\infty),
  \label{eq:dI}
  \end{equation}
where $\mathbf{\Phi}_\infty=(A^{-1})^{\otimes n} \mathbf{\Phi'}_\infty$ (\ref{eq:Phiinf}). We used such iteration to obtain the exact $\Delta S_2(t)$ for $n \le 34$. For larger circuits \new{memory requirements of the exact method become prohibitive} and we used the matrix product state ansatz (MPS) for more \new{memory-efficient} representation of $\ket{\Phi'(t)}$. While the MPS method~\cite{Vidal} allows to study large systems it has its limitation beyond times when $S_2$ becomes \new{too large. Namely, when e.g. $S_2 \sim 50$ the floating point double precision ``noise'' of size $\sim 10^{-15} \approx 2^{-50}$ becomes comparable to $I(t)$, making further simulation hard despite using large MPS bond dimensions upto $1500$.}

\new{The main part of this section is devoted to the brick-wall configuration in section~\ref{sec:BW} and to the staircases in section~\ref{sec:S}. In section~\ref{sec:gen} we briefly demonstrate that the same phenomena are observed also for generic fixed gates as well as for the random U(4) gates. In section~\ref{sec:google} we touch upon the experimentally relevant Sycamore processor that uses a 2D layout of qubits.}

\subsection{Fastest scrambler}
\label{sec:BW}

\begin{figure}[t]
  \centerline{\includegraphics[width=2.4in]{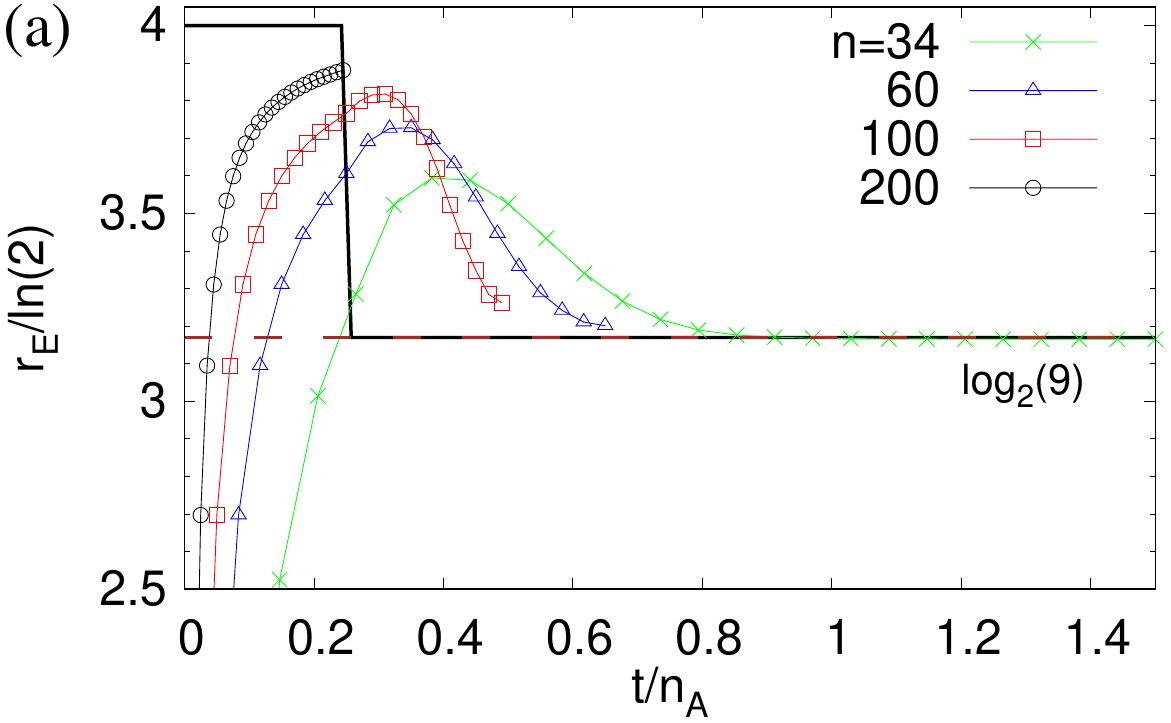}}
  \vskip2mm
\centerline{\includegraphics[width=3.3in]{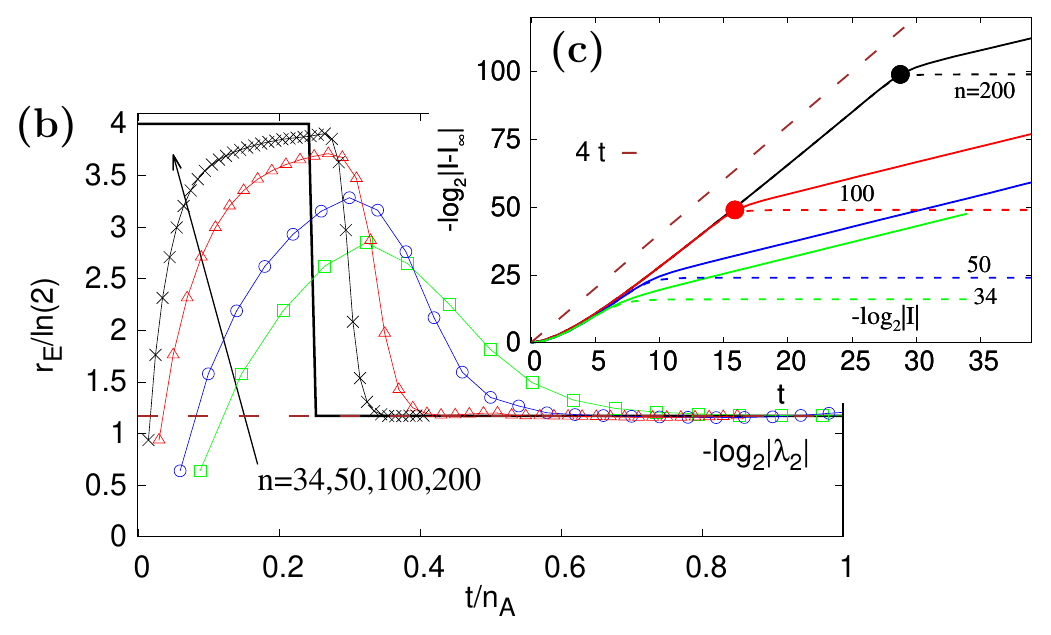}}
\caption{The local entanglement rate $\rE$ for the BW configuration with PBC. (a) is for $\az=0$ (the XY gate), (b) and (c) for $\az=0.5$. The full black line in (a) is a conjectured $\rE$ in the TDL, which has a discontinuous jump from $4$ for $t\le t_{\rm c}=t_{\infty}$ to a smaller value given by $\lambda_2$. In (c) dotted curves are $-\log_2I(t)$, full curves $-\log_2|I(t)-I_\infty|$, while the transition times $t_{\rm c}$ for $n=100$ and $200$ are marked by full circles. The dashed brown line in (a) and (b) agrees with $-\log_2{|\lambda_2|}$ from Fig.~\ref{fig:OBC_in_PBC_reze_az}}
\label{fig:bwpbc}
\end{figure}
\begin{figure}[t]
  \centerline{\includegraphics[width=2.7in]{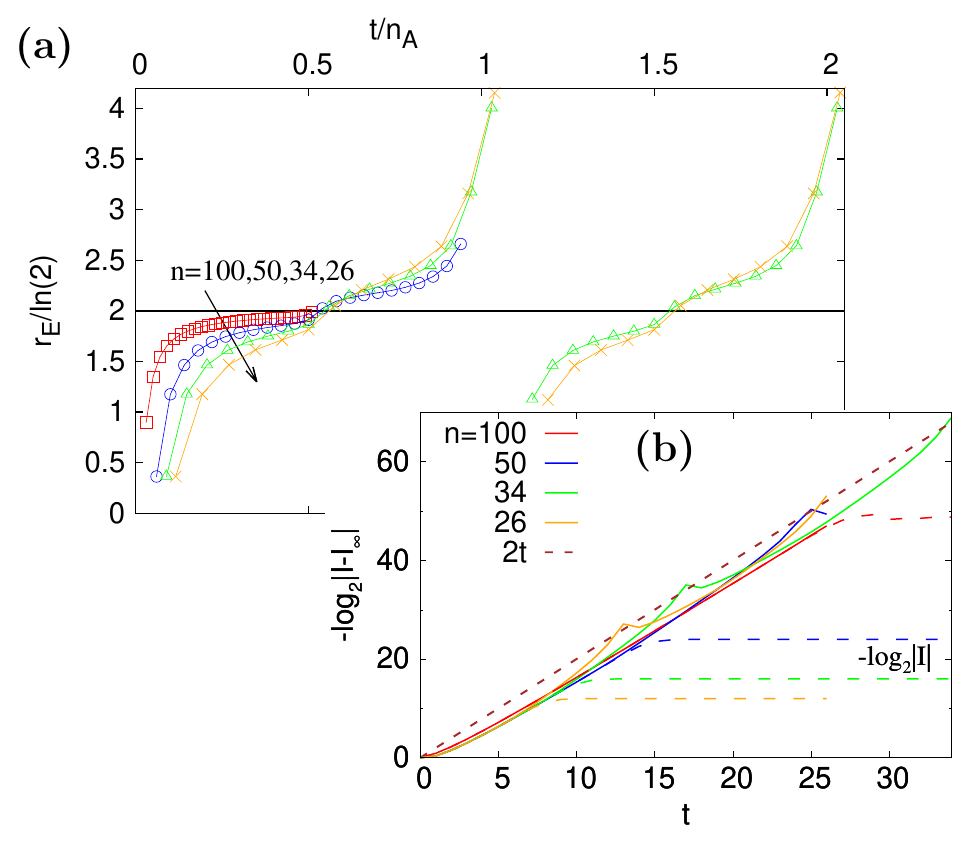}}
  \vskip2mm
\centerline{\includegraphics[width=3.1in]{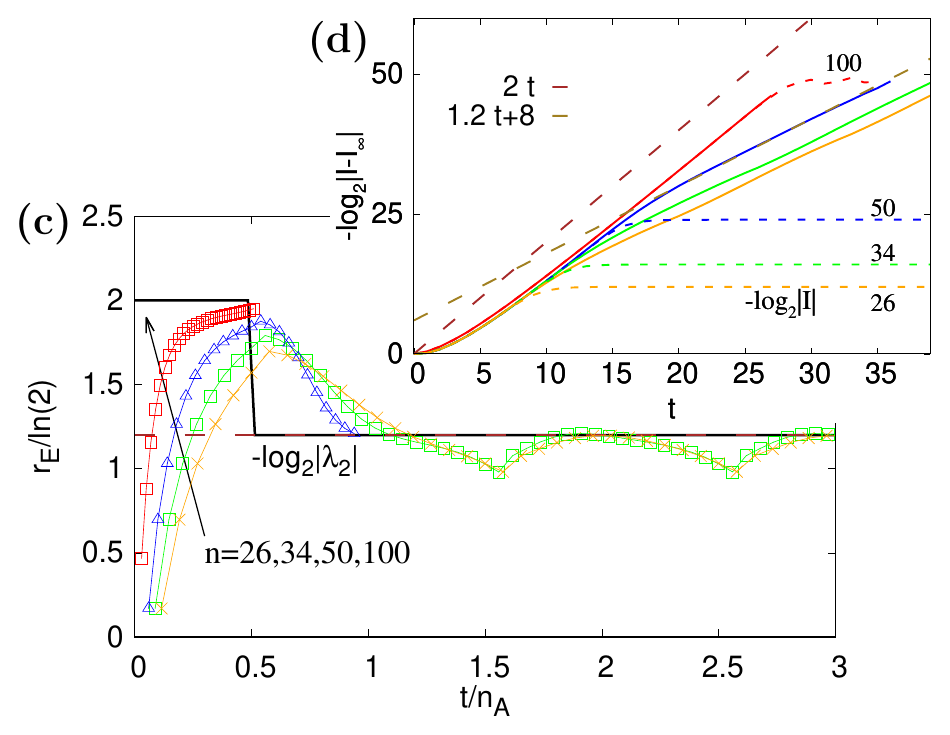}}
\caption{Entanglement rate for the BW and OBC. (a) and (b) is for $\az=0$ (the XY gate), (c) and (d) for $\az=0.5$. For $\az=0$ the local rate is always equal to $\ln{4}$ in the TDL. For $\az=0.5$ the conjectured rate in the TDL (full black line in (c)) is $\ln{4}$ for $t<t_{\rm c}=t_\infty=\nA/2$ and smaller $-\ln{|\lambda_2|}$ for $t>t_{\rm c}$. Dashed brown line in (c) is at $1.2$ and is approximately equal to $-\ln_2{|\lambda_2|}$ read from Fig.~\ref{fig:OBC_in_PBC_reze_az}.
}
\label{fig:bwobc}
\end{figure}

Let us first have a look at the fastest random circuits, namely the BW configuration for which the spectral analysis would predict $\rE=\ln{9}$ for the XY gate and PBC, and $\rE=\ln{4}$ for the XY with OBC. \new{The BW configuration with an appropriate gate type that we will identify will turn out to be the fastest possible entanglement scrambler saturating the theoretical upper bound on the entanglement speed in Eq.~(\ref{eq:bound}).}

\new{In the following we shall show a number of figures of the same type which will demonstrate the asymptotic entanglement rate and a multistage thermalization. Let us describe what those figures will be showing.} The main quantity is the local entanglement rate $\rE$, being equal to the local slope of $\Delta S_2(t)=-\ln{|I(t)-I_\infty|}$. Specifically, from numerical data we shall plot
\begin{equation}
  \frac{\rE(t+\frac{1}{2})}{\ln{2}}=\log_2{\left| \frac{I(t)-I_\infty}{I(t+1)-I_\infty}\right|}.
  \label{eq:localr}
\end{equation}
\new{Logarithms used will be base-2 so that for qubits the theoretical upper bound on the rate is an integer ($2$ or $4$ depending on boundary conditions). We will in addition show plots of purity $-\log_2 I(t)$ as well as purity with subtracted asymptotic saturation value $I_\infty$, i.e., $-\log_2|I(t)-I_\infty|$.}

\begin{figure}[hbt!]
\centerline{
  \includegraphics[width=1.6in]{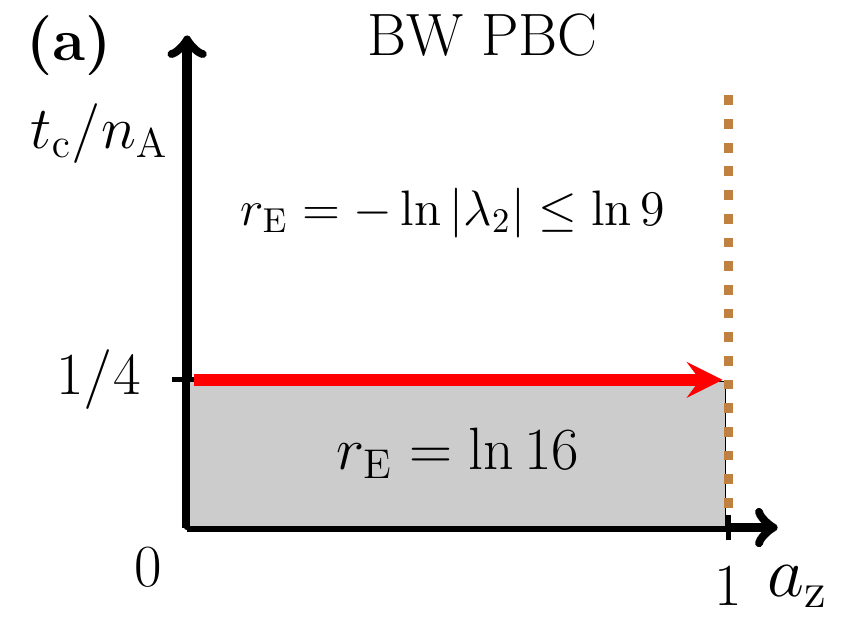}\hskip3mm
  \includegraphics[width=1.6in]{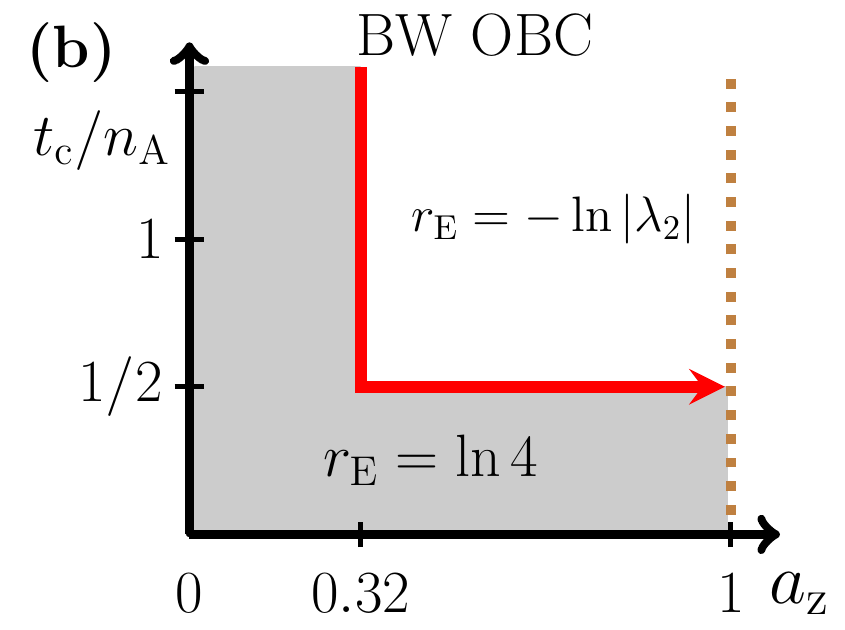}
}
\caption{Phase diagram of the local entanglement rate $\rE$ for BW configuration. (a) For PBC there is one phase transition, for OBC (b) there is again one discontinuous transition for $\az>a_{\rm c}\approx 0.32$. \new{Brown dashed vertical line at $\az=1$ indicates a non-entangling SWAP gate.}}
\label{fig:faznibw}
\end{figure}

In Fig.~\ref{fig:bwpbc} we show results of numerical simulations for the BW configuration with the XXZ gate and PBC. \new{In frame (a) we show the XY gate ($\az=0$) whereas in (b) and (c) a more general XXZ gate with $\az=0.5$.} The behavior is similar for any $\az<1$ ($\az=1$ corresponds to the SWAP gate which does not generate any entanglement). \new{We can see from the purity plot in Fig.~\ref{fig:bwpbc}(c) that there is a kink -- a change in the entanglement growth rate -- at the point when $I(t)$ gets close to $I_\infty$ (dashed saturating curves). The change in $\rE$ is also clearly seen in Fig.~\ref{fig:bwpbc}(b). Regardless of $\az$ the rate in the TDL increases towards $4$ for times smaller than the saturation time $t_\infty$ (note that we rescale the time axis by $\nA=n/2$). The convergence with $n$ is however rather slow.} If one would look only at small $n$ one could be mislead into thinking that the rate  is given by $\rE=-\ln{|\lambda_2|}$ ($n=34$ in Fig.~\ref{fig:bwpbc}(a)), however, at larger $n$ the initial bump gets higher and moves to smaller times. \new{The rate is therefore not determined by $\lambda_2$ and equal to $-\ln|\lambda_2|$, but is instead larger.} For instance, for $\az=0$ we have $\lambda_2=\frac{1}{9}$, while the rate converges with $n$ towards $\ln{16}$ (Fig.~\ref{fig:bwpbc}(a)). This is perhaps even more clear for $\az=0.5$ in Fig.~\ref{fig:bwpbc}(b). Based on the data we \new{therefore} conjecture the following scenario in the TDL, holding for any $\az<1$: the rate is $\rE=\ln{16}$ for $t\le t_\infty=\frac{1}{4} \nA$, at which point there is a discontinuous transition in the rate to a smaller value determined by the 2nd largest eigenvalue $|\lambda_2|$. For any zero or nonzero $\az<1$ there is one discontinuous phase transition in the local rate at the critical time $t_{\rm c}=\frac{1}{4}\nA$. This is reflected in a phase diagram shown in Fig.~\ref{fig:faznibw}(a). The relaxation process has two phases: in the first faster phase, whose entanglement rate is in the TDL independent of $\az$ (even if we are very close to a non-entangling SWAP gate), one approaches a random state, in the 2nd, that starts at $t_{\rm c}=t_\infty=\nA/(2{\cal A})$ when the state is already close to random, the relaxation is slower and determined by $-\ln{|\lambda_2|}$. The physically relevant entanglement generation rate that determines how long it takes to generate all $\sim \frac{n}{2}$ bits of entanglement is therefore not determined by the matrix gap, rather, it is equal to $\ln{16}$.

How can that be? By studying the spectrum of $M$ (in the even sector) and the relevant overlaps with $\mathbf{\Phi}_0$ and $\mathbf{\Phi}_{\rm half}$ (\ref{eq:dI}) that appear in the spectral expansion
\begin{equation}
  I(t)=I_\infty+\sum_{j=2}^\infty d_j \lambda_j^t ,\quad d_j=\braket{\Phi_{\rm half}}{R_j} \braket{L_j}{\Phi_0}
  \label{eq:Isum}
\end{equation}
we can identify the leading terms in the sum. Focusing on $\az=0$, there is exactly one with $|\lambda_2|\approx \frac{1}{9}$ and then there are $\sim (n-4)/4$ with $|\lambda_j|\approx \frac{1}{16}$. The weight $d_j$ of terms with $\lambda_j \sim 1/16$ grows with $n$, albeit first for small $n$ rather slowly (see e.g. slow convergence with $n$ in Fig.~\ref{fig:bwpbc}(a)). Nevertheless, for large $n$ the terms $d_3 (\frac{1}{16})^t$ overwhelm the ``leading'' $d_2 (\frac{1}{9})^t$; this happens roughly when the ratio $d_3/d_2 \sim (16/9)^{n/8} \approx 1.07^n$ gets large.

\begin{figure}[t!]
\centerline{\includegraphics[width=3.0in]{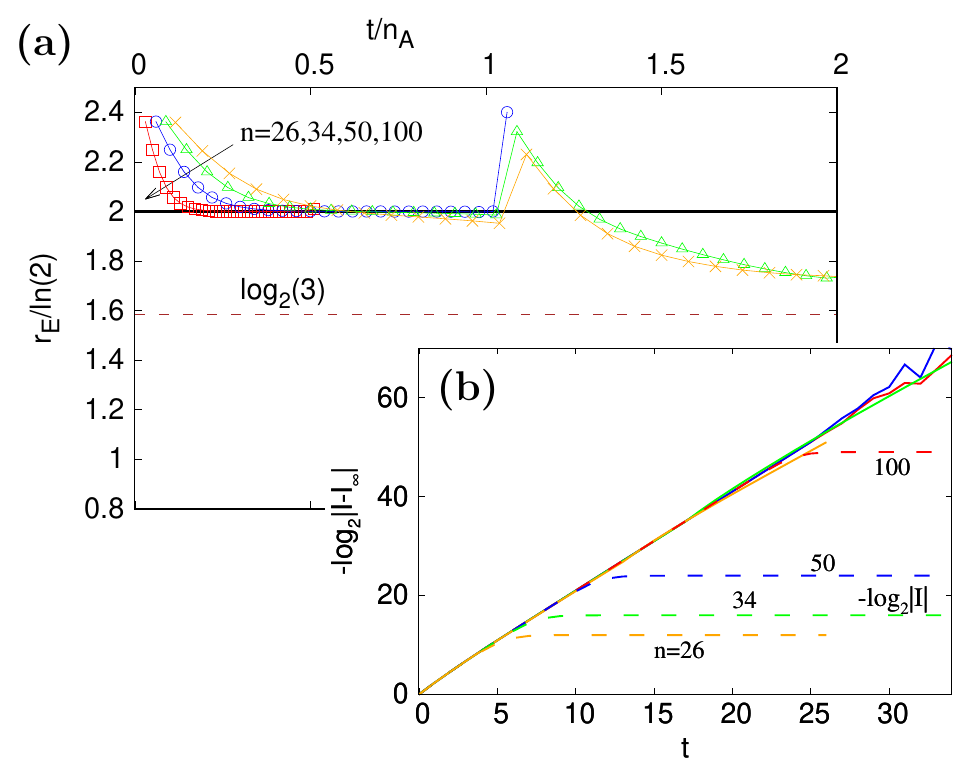}}
\centerline{\includegraphics[width=3.1in]{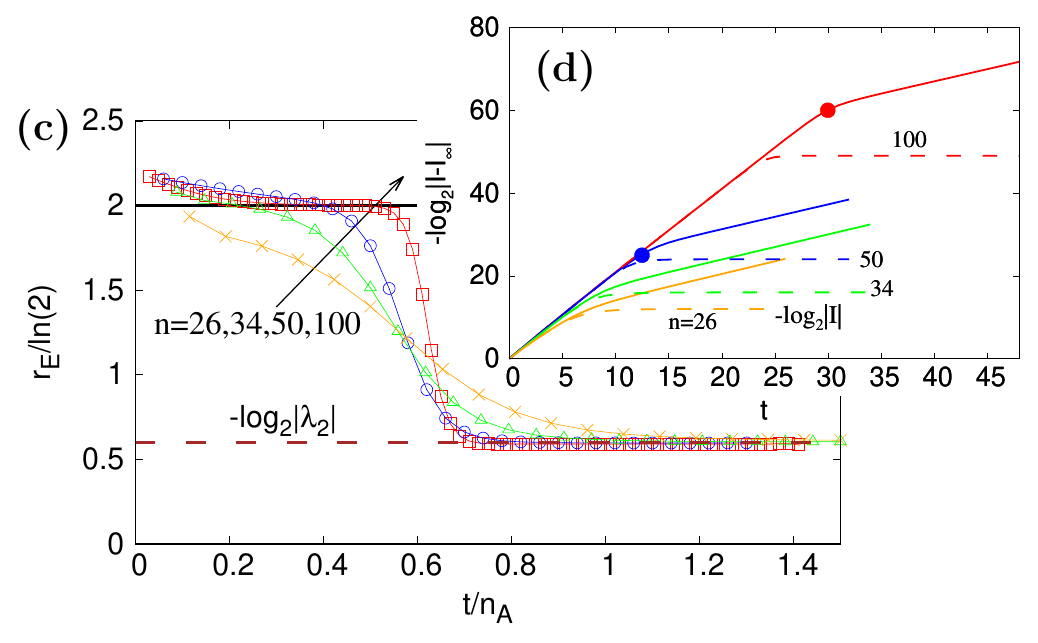}}
\caption{Entanglement rate for the $S$ configuration with PBC and $\az=0$, (a) and (b), and $\az=0.5$, frames (c) and (d). At $\az=0$ the rate is always $\rE=2\ln{2}$ and is larger than $-\ln{|\lambda_2|}$. At nonzero $\az$ there is a phase transition from $\rE=\ln{4}$ for $t_{\rm c}$ to $-\ln{|\lambda_2|}$ at lager times. The transition appears to happen at larger $t/\nA$ than $0.5$ (here around $t_{\rm c}/\nA \approx 0.6$). Brown dashed line at $0.60$ is in-line with $|\lambda_2| \approx 0.66$ from Fig.~\ref{fig:OBC_in_PBC_reze_az}.}
\label{fig:crtapbc}
\end{figure}
Data for the BW with OBC is shown in Fig.~\ref{fig:bwobc}. For $\az=0$ the rate is in the TDL $\rE=2\ln{2}$ for any $t$. This is in-line with the fact that in this case $|\lambda_2|=\frac{1}{4}$ and therefore one does have $\rE=-\ln{|\lambda_2|}$. Out of the 4 cases studied, BW and S with OBC or PBC, this is the only one where the gap does determine the entanglement rate. Apparent singularities in the rate at integer $t/\nA$ visible in Fig.~\ref{fig:bwobc}(a) are due to a sign change in the $I(t)-I_\infty$, which is also visible as sharp kinks in $-\log_2{|I(t)-I_\infty|}$ plotted in frame (b). For nonzero $\az$ things though change. As we have seen in Fig.~\ref{fig:OBC_in_PBC_reze_az}, for $\az<\ac$ the gap for the OBC does not change with $\az$, while for $\az>\ac\approx 0.32$ it starts to \new{decrease}. This is in turn also reflected in the local entanglement rate $\rE$. Upto the thermalization time $t_\infty$ the rate stays at $2\ln{2}$, and is therefore not equal to $-\ln{|\lambda_2|}$ for $\az>\ac$, while at $t_{\rm c}=t_\infty=\nA/2$ the rate jumps to $-\ln{|\lambda_2|}$ (Fig.~\ref{fig:bwobc}(c)). This change in the rate is continuous at $\az=\ac$, but discontinuous for $\az>\ac$. Correspondingly, $\Delta S_2(t)$ (\ref{eq:dS}) exhibits a kink at $t_{\rm c}$ (Fig.~\ref{fig:bwobc}(d)). Worth noting is that while in the scaled time $t/\nA$ the rate will in the TDL be $\ln{4}$ already at $t/\nA \to 0$ (Fig.~\ref{fig:bwobc}(c)), in real time $t$ there is a 'delay' of about $\Delta t\approx 15$ before the rate becomes $\approx \ln{4}$ (slower initial growth in Fig.~\ref{fig:bwobc}(d)). The phase diagrams showing dependence of $\rE$ on $\az$ for BW with OBC is shown in Fig.~\ref{fig:faznibw}(b).

\new{BW configuration with the XXZ gate is the fastest entanglement scrambler. Considering that $\lambda_2$ does not give the correct rate one could wonder whether some other gate that did not look optimal according to $\lambda_2$ in Section~\ref{sec:spectral} could be even better? The answer is no because the rates $4$ and $2$ for PBC and OBC, respectively, saturate the bound (\ref{eq:bound}). There are no better scramblers. What could in principle happen, but we think is unlikely, is that for some other canonical gate with parameters $\mathbf{a}$ different from XXZ one would get the same maximal rate. For instance, for $\mathbf{a}=(0.9,0.8,0.5)$ with BW PBC one again has faster rate than given by $|\lambda_2|$, though at $\approx 2.77$ it is smaller than $4$, see Section~\ref{sec:gen} for more details.}

\subsection{Slow scramblers}
\label{sec:S}

Here we discuss the staircases (S) configuration (Fig.~\ref{fig:S_in_BW}(b)) which is also an extremal configuration in the spectral sense. Namely, for PBC its $|\lambda_2|$ is the largest among all configurations with PBC (spectral equivalence class $p=1$). \new{Note that we do not discuss here the slowest scramblers in the absolute sense because we will still use the fastest gate within the class $p=1$ (XXZ). For instance (Table~\ref{tab:nnrates}), XXZ with S and OBC is still faster than CNOT (``slow gate'') with BW and PBC (fastest configuration). We expect the same behavior that we reveal here for $p=1$ to hold in the TDL also for any finite $p$, i.e., for a spectral class that consists of a finite BW section and an extensive S part.}

\begin{figure}[t!]
\centerline{
\includegraphics[width=1.6in]{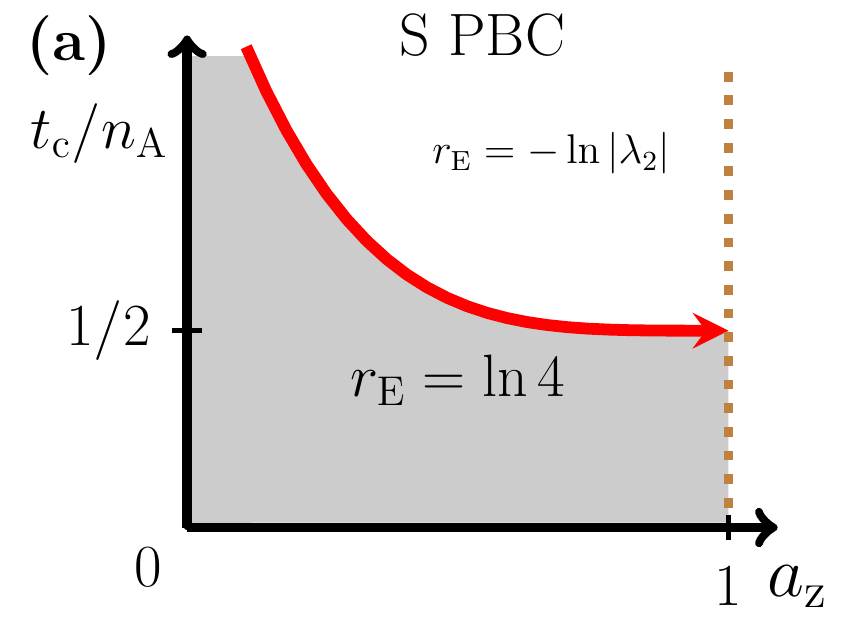}
  \includegraphics[width=1.6in]{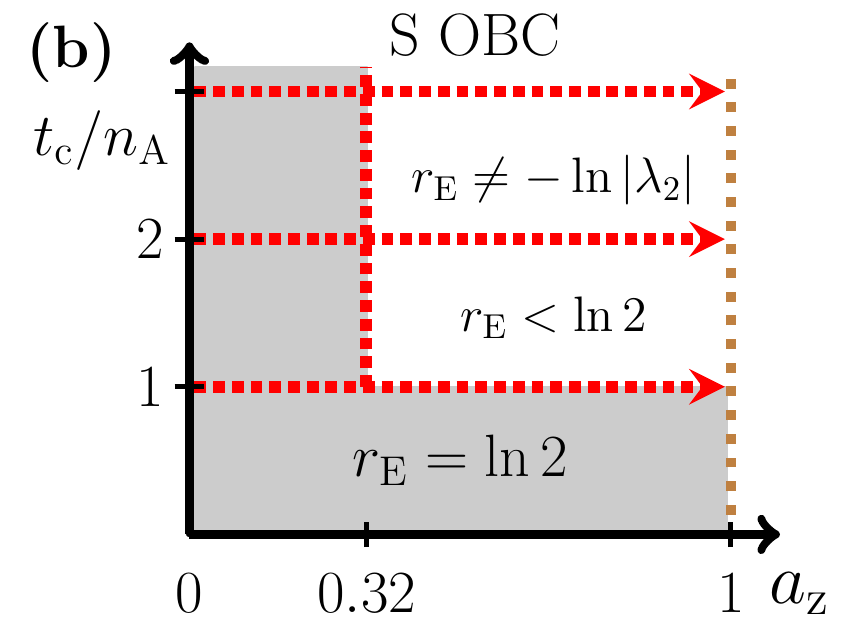}
}
\caption{Phase diagram of $\rE$ for the staircase configuration.}
\label{fig:fazniS}
\end{figure}

Let us first have a look at the S configuration with PBC. Results are shown in Fig.~\ref{fig:crtapbc}. We can see that, similarly as for the BW with PBC, the entanglement rate $\rE$ is larger than $-\ln{|\lambda_2|}$. Specifically, for $\az=0$ one has $-\ln{|\lambda_2|}=\ln{3}$ while the rate is $\rE=\ln{4}$ (Fig.~\ref{fig:crtapbc}(a,b)). \new{While the numerical results are not super clear, perhaps for} $\az=0$ the rate seems to converge in the TDL to $\rE=\ln{4}$ also for times $t>t_{\infty}$ so that there is no phase transition. For nonzero $\az$ one on the other hand has a phase transition in the rate from the initial $\rE=\ln{4}$ to smaller rate given by the spectral gap, Fig.~\ref{fig:crtapbc}(c). For $\az=0.5$ the phase transition seems to happen at a distinctively larger time $t_{\rm c}$ than the thermalization time $t_{\infty}$ (see Fig.~\ref{fig:crtapbc}(d)), while at larger $\az$, for instance $\az=0.7$ (data not shown), it is very close to $t_{\infty}=\nA/2$. As opposed to the BW configuration, here the convergence with $n$ towards the TDL rate $\ln{4}$ is much faster so that there is almost no delay in purity starting to follow the asymptotics (Figs.~\ref{fig:crtapbc}(b,d)). A conjectured phase diagram is sketched in Fig.~\ref{fig:fazniS}(a).

\begin{figure}[t!]
\centerline{\includegraphics[width=2.4in]{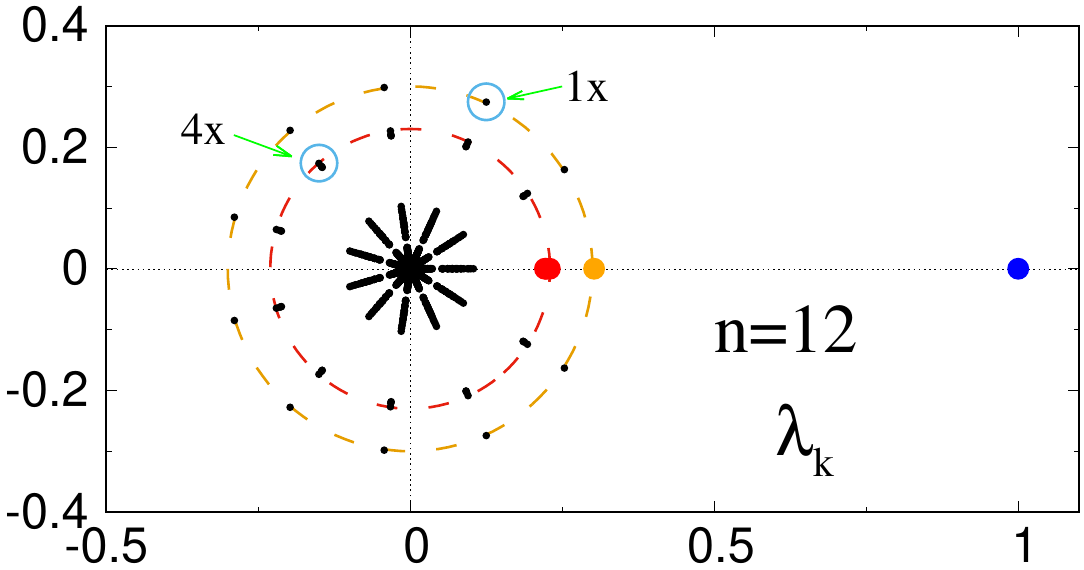}}
\caption{Spectrum of $M$ for S and PBC, $\az=0$ and $n=12$. There are $n-1$ eigenvalues with $|\lambda_j|\to \frac{1}{3}$ (black dots on the \new{larger} (orange) circle) and $(n-1)(n-4)/2$ with $|\lambda_j|\to \frac{1}{4}$ (\new{smaller} red circle; each visible point is composed of $(n-4)/2$ close eigenvalues). The leading contribution to $I(t)$ comes from one eigenvalue with $\lambda_k =\frac{1}{3}$ (orange point) and $(n-6)/2$ eigenvalues with $\lambda_k =\frac{1}{4}$ (almost degenerate 3 red points).}
\label{fig:spek-crtapbc}
\end{figure}
\begin{figure*}[t!]
  \centerline{\includegraphics[width=2.8in]{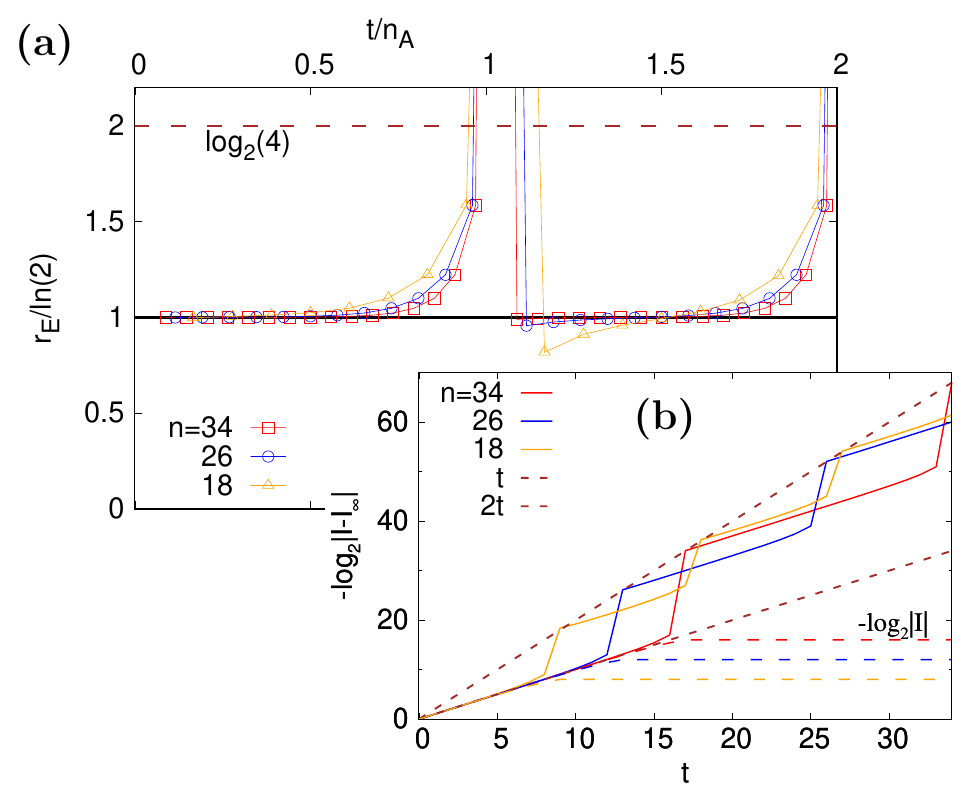}\includegraphics[width=2.8in]{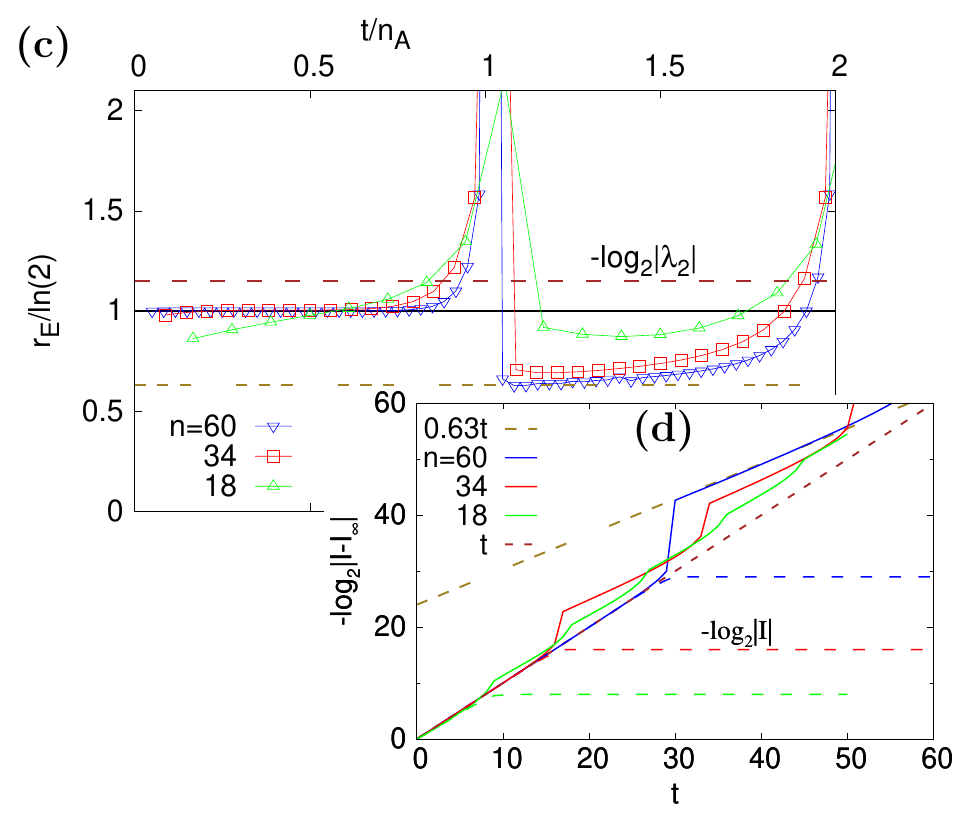}}
  \caption{Entanglement rate for the S configuration with OBC and $\az=0$ (a,b), and $\az=0.5$ (c,d). For $\az=0$ the rate is always $\rE=\ln{2}$ but with discontinuous jumps in $I(t)-I_\infty$ at integer $t/\nA$, see (b). For $\az=0.5$ the rate is again $\ln{2}$ for $t<t_{\rm c}=\nA$, but then discontinuously jumps in the TDL to a smaller value ($\rE \approx 0.63\ln{2}$ at the shown $\az=0.5$ in (c,d), and is not equal to $-\ln{|\lambda_2|}$).}
\label{fig:crtaobc}
\end{figure*}
\begin{figure*}[hbt!]
  \centerline{\includegraphics[width=\textwidth]{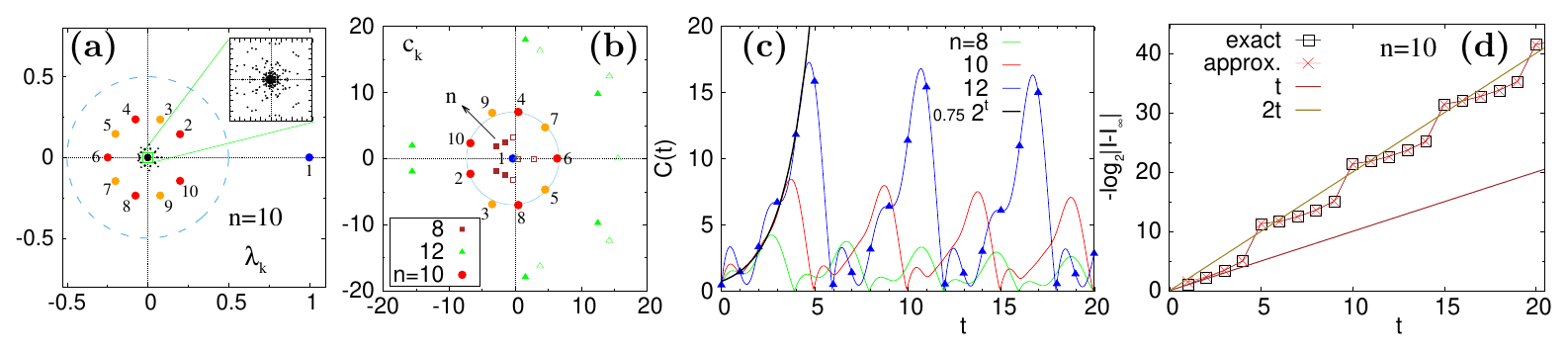}}
 \caption{Slower convergence than $-\ln{|\lambda_2|}$ for the S with OBC and the XY gate due to a phantom eigenvalue (dashed blue circle in (a)). (a) Complex spectrum of $M$ for $n=10$ (even sector with $512$ eigenvalues). (b) Expansion coefficients $c_k$ (\ref{eq:x}). Numbered points correspond to respective eigenvalues in (a); 5 full red circles ($2,4,6,8,10$) are $n/2$ leading terms contributing to $I-I_\infty$. Size of $|c_k|$ grows exponentially with $n$. (c) The sum $C(t)=4^t|\sum_{k=2}^{n/2} d_k \lambda_k^t|$ (\ref{eq:Isum}) of corresponding terms from (b) (blue triangles are at physical integer $t$). (d) Comparison of the exact purity and the approximation with $n/2$ terms from (c).}
\label{fig:spek-crtaobc}
\end{figure*}
The explanation for such larger entanglement rate is similar as for the BW configuration with PBC, though with some different details. \new{The even parity spectrum of $M$ for $n=12$} is shown in Fig.~\ref{fig:spek-crtapbc}. Besides the steady state giving $I_\infty$ (blue point) one has $n-1$ nondegenerate eigenvalues ($\lambda_2$) distributed around the circle with radius $\frac{1}{3}$ (in the TDL, Fig.~\ref{fig:OBC_in_PBC_reze_az}), and $(n-1)(n-4)/2$ eigenvalues around the circle of smaller radius $\frac{1}{4}$ (those are grouped into $n-1$ in the TDL degenerate clusters each having $(n-4)/2$ eigenvalues, e.g. $4$ for the $n=12$ example in Fig.~\ref{fig:spek-crtapbc}). In the TDL though not all contribute in the purity expansion (\ref{eq:Isum}). It turns out that only those on the real axis contribute, that is one with $\lambda_2=\frac{1}{3}$ (orange point), and $(n-6)/2$ with $\lambda_j=\frac{1}{4}$ (red point). The relative weight $d_j/d_2$ of those with $\frac{1}{4}$ grows with $n$ and that is how the faster entanglement rate $\rE=\ln{4}$ emerges in the TDL.

Configuration S with OBC is even more interesting. In Fig.~\ref{fig:crtaobc} we show numerical results for $\rE$ (\ref{eq:localr}). The rate is $\rE=\ln{2}$ and is therefore smaller than $\ln{4}$ suggested by the gap. For $\az=0$ such rate persists for any finite time $t/\nA$, however, looking at $I(t)-I_\infty$ (Fig.~\ref{fig:crtaobc}(b)) we note that there are discontinuous jumps at integer $t/\nA$, the size of which increases with $n$. At such times purity $I(t)$ closes in on its asymptotic value $I_\infty$ by a finite amount (that increases with $n$) in a single time step. For nonzero $\az=0.5$ the situation is similar, with the difference being that the rate at $t>t_{\rm c}=\nA$ jumps to smaller value than $\ln{2}$ (Fig.~\ref{fig:crtaobc}(c)). In this case though the rate after $t_{\rm c}$ is not simply equal to $-\ln{|\lambda_2|}$ as in other cases studied.

Explanation of how one can get slower rate $\rE=\ln{2}$ despite all \new{nontrivial} eigenvalues of $M$ being $|\lambda_2|\le \frac{1}{4}$ is very interesting and is illustrated in Fig.~\ref{fig:spek-crtaobc}. In the frame (a) we can see that there are $n-1$ eigenvalues with $|\lambda_j|=\frac{1}{4}$, of which however only $n/2$ have nonzero overlap with the initial vector (red points in (b)). The size $|c_j|$ of those coefficients grows exponentially with $n$ and they come in conjugate pairs. The resulting expression for purity is therefore of the form $I(t)-I_\infty \approx \sum_{k=2}^{n/2+1} |d_k|\cdot|\lambda_k|^t 2\cos{(\varphi_k t +\alpha_k)}$, where $\varphi_k$ is the phase of $\lambda_k$ and $\alpha_k$ the phase of $d_k$. Coefficients $d_k$, Eq.~(\ref{eq:Isum}), are such that the corresponding sum including \new{cosine} terms mimics exponential growth $\sim 2^t$ upto time $t=n/2$, when it discontinuously jumps to $0$ in a single unit of time (frame (c)). The end result is that their growth $2^t$ partially cancels $(\frac{1}{4})^t$ from the actual eigenvalues, resulting in a decay $(\frac{1}{2})^t$ as there would be a phantom eigenvalue of size $\frac{1}{2}$. Again, this is possible only due to non-Hermitian $M$ whose eigenvectors are not orthogonal. The size of discontinuous jumps in (c) grows exponentially with $n$ and is responsible for jumps we observed in Fig.~\ref{fig:crtaobc}(b). 

In Fig.~\ref{fig:spek-crtaobc}(d) we can transparently see how all this is reflected in $I(t)$: discontinuous jumps are such that on times larger than $t_\infty=n/2$ one actually does get an overall  $\Delta S_2(t) \asymp t \ln{4}$ as predicted from $\lambda_2$, however, in the TDL the time $t_\infty$ is pushed to infinity and the physically relevant entanglement rate towards the random-state entanglement is instead given by the phantom eigenvalue $\frac{1}{2}$. Note that even though the local entanglement rate is for non-integer $t/\nA$ always $\ln{2}$, due to discontinuities in $I(t)$ at integer $t/\nA$ the overall growth of $\Delta S_2(t)$ for $\az=0$ is given by $\ln{4}$ and not the local rate (Fig.~\ref{fig:spek-crtaobc}(d)). For $\az > \ac\approx 0.32$, when $|\lambda_2|$ moves, the local rate at $t>t_{\rm c}=\nA$ also decreases, see Fig.~\ref{fig:crtaobc}(c). It is though not equal to $-\ln|\lambda_2|$ (the rate decreases from $\ln{2}$ whereas the eigenvalue decreases from $\ln{4}$). \new{Note that $\rE=\ln{2}$ for all $\az<1$ regardless of $|\lambda_2|$ increasing with $\az$ (Fig.~\ref{fig:OBC_in_PBC_reze_az}).} This is all reflected in the phase diagram in Fig.~\ref{fig:fazniS}(b).

\new{We remark that one could in principle get a slower effective decay than $|\lambda_2|^t$ for special initial states that would have very small overlap with the corresponding eigenvector; in the asymmetric simple exclusion process that exhibits a cutoff~\cite{cutoffASEP} such are states with an extensive empty or occupied sections. This is however not what is happening in our case; we get such phantom eigenvalue decay for the relevant initial state, i.e., generic bipartition (provided $\nA$ is extensive). The origin of slower decay is in our case different. It is the left and right eigenvectors corresponding to $|\lambda_j|\approx 1/4$ (red points in Fig.~\ref{fig:spek-crtaobc}(a)) that must be special as they mimic exponential function (Fig.~\ref{fig:spek-crtaobc}(c)).}

We have seen that in most circuits that we studied the relevant entanglement rate $\rE$ is not given by the largest nontrivial eigenvalue of $M$. The ratio of $\ln{|\lambda_2|}$ between the PBC and the OBC (Table~\ref{tab:nnrates}) for a fixed $W$ is not equal to $2$ neither for the BW configuration, nor for the $S$, and in particular changes with the $\az$ of the XXZ gate. However, the correct rates $\rE$ \new{we identified in this section} are always in the expected ratio of $2:1$ for the two boundary conditions. Rates that are different than $-\ln{|\lambda_2|}$ are made possible due to non-symmetric nature of $M$ (despite individual gates $M_{i,j}$ being symmetric matrices) that results in expansion coefficients growing with $n$. \new{Another observation is that the ratio of $\rE$ between the BW and S configurations for fixed $W$ and boundary conditions is always $2$ for the XXZ gates, which however is not the case for all 2-qubit gates\cite{footBWS}.}

It is worth noting that in general the exponential growth of coefficients \new{in our case} can not be traced to degeneracies. For instance, taking a simple non-symmetric $2\times 2$ matrix
\begin{equation}
  \begin{pmatrix}
  1 & 1-\epsilon\\
  0 & 1-\epsilon
  \end{pmatrix},
\end{equation}
the eigenvalues are $\lambda_1=1$ and $\lambda_2=1-\epsilon$, with the associated right eigenvectors $\mathbf{x}_1=(1,0)$ and $\mathbf{x}_2=(\epsilon-1,\epsilon)$. In the limit $\epsilon \to 0$ the two eigenvalues become degenerate (an 'exceptional point'; for $\epsilon=0$ one has a non-diagonalizable Jordan canonical form) and the right eigenvectors are almost co-linear. Expanding $\mathbf{y}=(a,1)$ over such a basis will result in large expansion coefficients of the order of the inverse eigenvalue separation that scales as $1/|\lambda_1-\lambda_2| \sim \frac{1}{\epsilon}$, $\mathbf{y}=(a-1+\frac{1}{\epsilon})\mathbf{x}_1+\frac{1}{\epsilon}\mathbf{x}_2$. While in some cases one indeed has $\sim n$ fold degeneracy (S with PBC, Fig.~\ref{fig:spek-crtapbc}), in others one does not (S with OBC, Fig.~\ref{fig:spek-crtaobc}(a)). So it does not seem that the phenomena we found can be simply ascribed to degeneracies.

\new{The phantom eigenvalue phenomenon and more generally of multistage relaxation in which the relaxation rate exhibits a sudden transition has been identified in random circuits. Considering that random circuits are often used as models of chaotic systems one naturally asks if the randomness explicitly present in space and time due to random 1-qubit unitaries is actually necessary. A detailed study of this interesting question goes beyond the present work, however we have numerically checked (see Appendix~\ref{app:flukt}) that in the TDL one gets the same multistage thermalization even in a single circuit realization and without randomness in both space and time, i.e., one can use the same 1-qubit unitary on all qubits and at all time steps.}

\new{
  \subsection{Generic gates}
\label{sec:gen}

\begin{figure}[t]
\centerline{\includegraphics[width=3.2in]{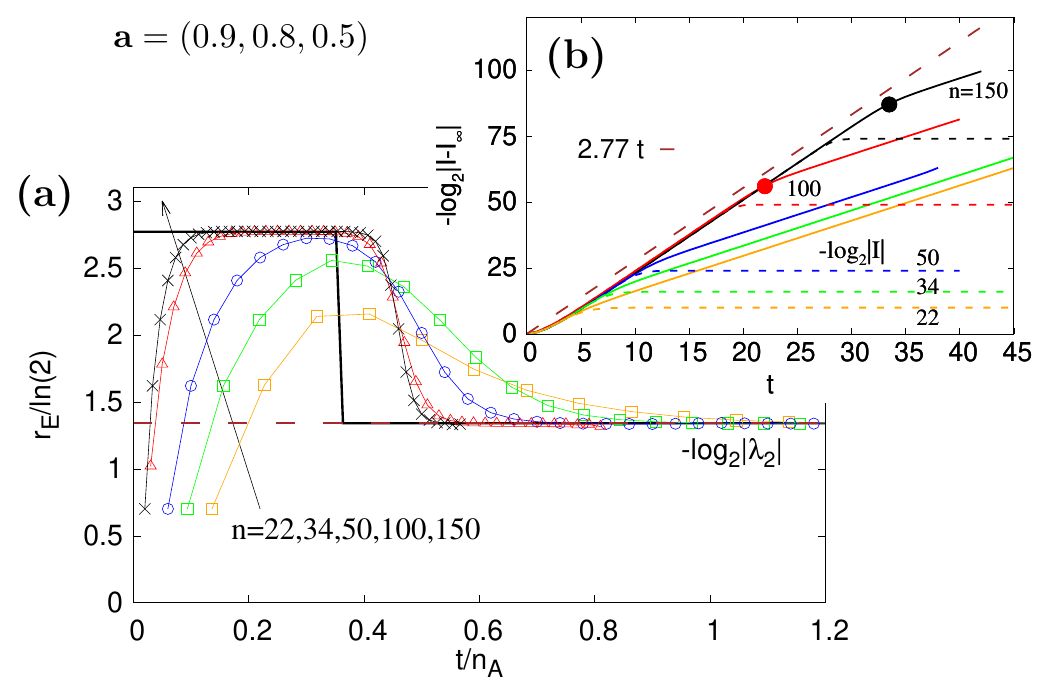}}
\caption{\new{Entanglement rate for the BW configuration with PBC and a 2-qubit gate with parameters $\mathbf{a}=(0.9,0.8,0.5)$. One has a multistage thermalization with rate $\rE/\ln{2} \approx 2.77$ that is non-maximal (less than for XXZ gates) but still faster than the one calculated from $|\lambda_2|$ (dashed brown line at $\approx 1.34$ in (a)).}}
\label{fig:ranA}
\end{figure}
So-far we have focused on the fastest 2-qubit gates in its spectral equivalence class $p=1$ and $p=n/2$, which turned out to be the XXZ-type gates (\ref{eq:XXZ}). Random circuits with XXZ gates are dual-unitary and so one naturally asks if the observed multistage thermalization and phantom eigenvalues are perhaps limited to (some) dual-unitary circuits? The answer is no. All the phenomena we discussed can occur in circuits that are not dual-unitary and this is what we shall briefly demonstrate in this section. Full treatment goes beyond the present paper so we will only show two examples.

The first is a case of the BW configuration with PBC and a generic 2-qubit gate. We pick a completely anisotropic non dual-unitary gate with canonical parameters $\mathbf{a}=(0.9,0.8,0.5)$. In Fig.~\ref{fig:ranA} we can see that one has exactly the same phenomenology as for the XXZ gate (compare with Fig.~\ref{fig:bwpbc}). Namely, the initial rate is in the thermodynamic limit larger than $-\ln{|\lambda_2|}$ and therefore one has a multistage thermalization. This local rate though is smaller than for the XXZ gates, where it is $\rE/\ln{2}=4$. This is the reason why we conjecture that the XXZ gates are the only ones having the maximal entanglement production rate.

\begin{figure}[t]
\centerline{\includegraphics[width=3.1in]{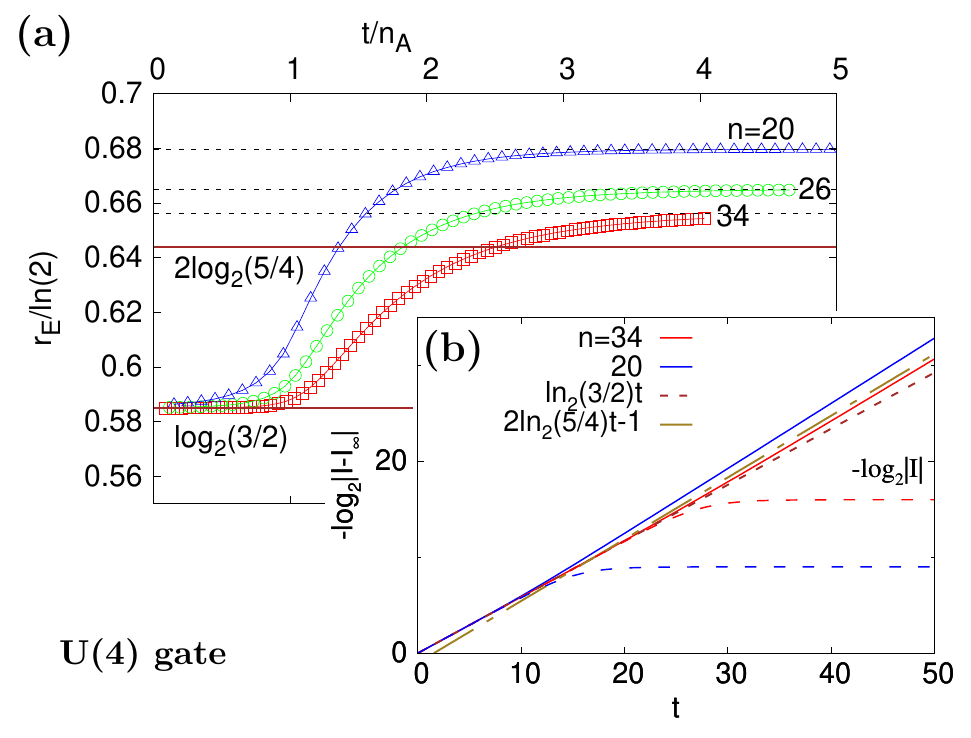}}
\caption{\new{Entanglement rate for the S configuration with OBC and i.i.d. Haar random 2-qubit gates from U(4). In the TDL one has $|\lambda_2|=(4/5)^2$, the entanglement rate though is due to a phantom eigenvalue smaller and equal to $\rE=\ln{(3/2)}$. Three dashed lines in (a) are $\rE=-\log_2{|\lambda_2|}$ using conjectured finite-size expression $|\lambda_2|=[\frac{4}{5}\cos{(\pi/n)}]^2$.}}
\label{fig:u4}
\end{figure}
The second non dual-unitary example that we show is a much studied case of completely random 2-qubit gates distributed according to the Haar measure on U(4). For previous results on the rates for different configurations see Table~\ref{tab:oldrates}. We shall focus on the S configuration with OBC. Their dynamics can again be described by a Markovian matrix~\cite{oliveira07} with an elementary $4\times 4$ two-site matrix~\cite{PRA08}. Due to spectral equivalence of all configurations with OBC we also know that $\lambda_2$ for the S configuration that we study is the same as for the BW configuration. In the TDL it has been calculated (Table~\ref{tab:oldrates}) that for the BW with OBC one has $|\lambda_2|=(4/5)^2$. Therefore, if the rate would be determined by $\lambda_2$ one should have asymptotic decay $I(t) \asymp (4/5)^{2t}$, i.e. $\rE=2\ln{(5/4)}$. Numerical simulations in Fig.~\ref{fig:u4} however show that this is not the case. Based on the data we conjecture that at any finite $t$ the purity for a half-half bipartition is in the TDL in fact exactly equal to
\begin{equation}
  I(t)=\left(\frac{2}{3}\right)^t.
\end{equation}
The decay is slower than one would predict from $\lambda_2$ and one therefore again deals with a phantom eigenvalue like in other staircases circuits that we studied. For the BW configuration U(4) random circuit with open boundary conditions see Ref.~\cite{footBW}.

Because the Markovian matrix $M$ essentially determines evolution of squares of expansion coefficients like those of the density operator, one can expect that the phenomena identified in purity will be present also in other quantities that are quadratic in time-evolved coefficients. One such object are the out-of-time-ordered correlators (OTOC), for which our preliminary results show~\cite{tobe} the same effects.
}

\subsection{Google's Sycamore processor}
\label{sec:google}

\begin{figure}[h]
	\begin{center}
		\includegraphics[width=40mm]{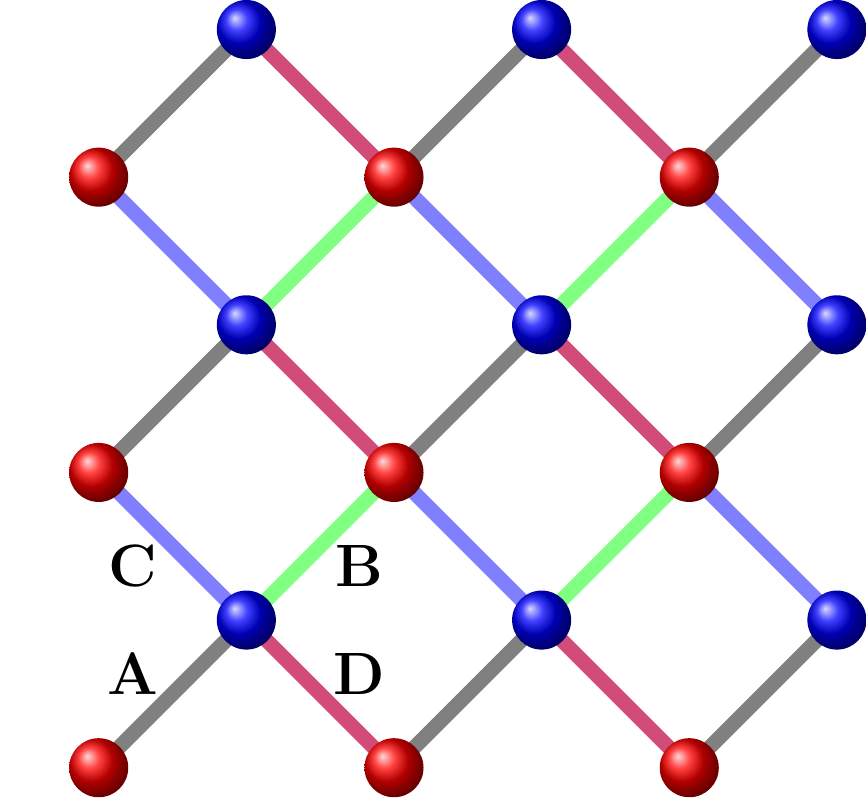}
		\caption{Qubit connectivity used in the Google's quantum processor Sycamore. Dots represent qubits and lines different 2-qubit gates. We shall focus on lattices of $3\times m$ qubits (\new{shown is} $m=6$ giving $n=18$). Gates labeled with the same letter (and color) \new{are} executed simultaneously, see text for details.}
		\label{fig:Google_geometry}
	\end{center}
\end{figure}

\begin{figure}[t]
	\begin{center}
		\includegraphics[width=95mm]{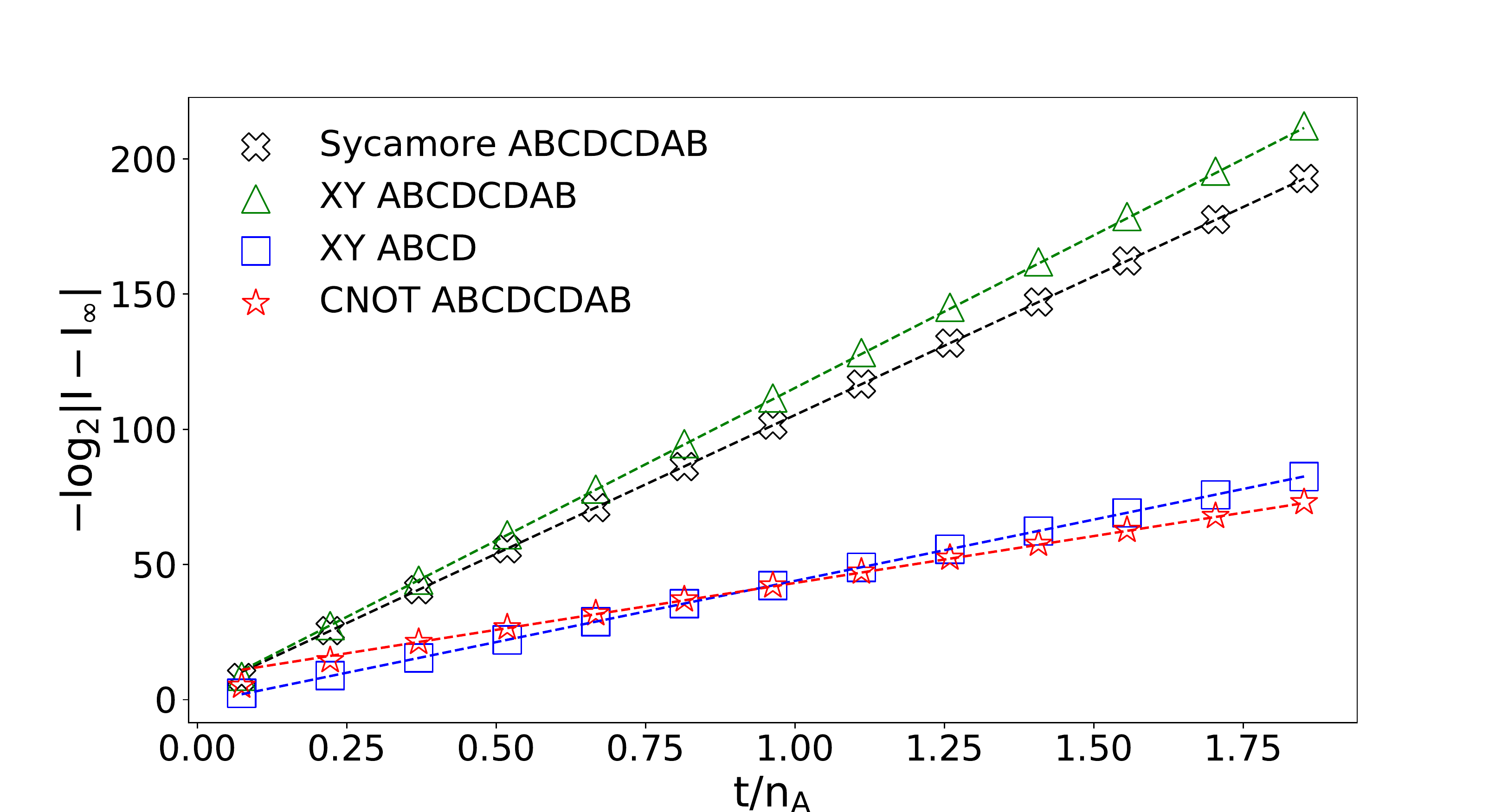}
		\caption{Purity results for Google-like protocol (symbols, $n=27$) and the asymptotic linear growth given by the leading eigenvalue $-\log_2{|\lambda_2|}$ (dashed lines). The protocols examined suggest that purity decay rate agrees with the value predicted from spectral analysis.}
		\label{fig:Google_purity}
	\end{center}
\end{figure}

One of the largest experimental realizations of quantum computation is a recent work by Google AI Quantum performing a random-circuit like computation on 53 qubits~\cite{Google}. Here we analyze the entanglement generation speed of their 2D qubit geometry that is used in the Sycamore quantum processor. 

In the Google experiment they used a fixed 2-qubit Sycamore gate with canonical parameters $\ax = 1, \ay = 1, \az = 1/6$, whereas for 1-qubit they randomly sampled from a set of square-roots of Pauli matrices~\cite{Google}. For the sake of simplicity, and to stay within the framework used in our paper, we take the 1-qubit gates to be random and uniformly distributed on U(2). The geometry used in the Sycamore processor and that we analyze is schematically represented in Fig.~\ref{fig:Google_geometry}. The protocols that we study are composed of commuting groups of 2-qubit gates denoted by 4 letters A,B,C and D, see Fig.~\ref{fig:Google_geometry}. For instance, the protocol used in the Google's supremacy experiment is ABCDCDAB. Due to size limitation we of course can not simulate a circuit with $n=53$ qubits, so we focus on a half-sized patch-circuit with $n=27$ used by Google. We shall in addition show results for two other protocol steps (ABCD and ABCDCADB) that turn out to be optimal at smaller sizes.

\begin{table*}[ht!]
	\begin{ruledtabular}
		\begin{tabular}{lcccccccccccc}
			Configuration &  \multicolumn{12}{c}{$\lambda$}\\
	
			
			& & \multicolumn{3}{c}{XY} & & \multicolumn{3}{c}{Sycamore} & & \multicolumn{1}{c}{CNOT} & & \multicolumn{1}{c}{random $\av$} \\
			
			\cmidrule(r){3-5}
			\cmidrule(r){7-9} 
			\cmidrule(r){11-11} 
			\cmidrule(r){13-13}
			
			& & n=9 & n=18 & n=27 & & n=9 & n=18 & n=27 & & n=27 & & n=27 \\
			
			\midrule
			
			ABCDCDAB	& & 0.319 & 0.161 & 0.141 & & 0.336 & 0.188 & 0.169 & & 0.548 & & 0.427 \\
			ABCDCADB	& & 0.192 & 0.128 & 0.143 & & 0.236 & 0.160 & 0.176 & & 0.547 & & 0.428	\\
			ABCD		& & 0.114 & 0.174 & 0.208 & & 0.133 & 0.196 & 0.229 & & 0.494 & & 0.358	
			
		\end{tabular}
		\caption{\label{tab:gaps_table} The effective eigenvalue $\lambda$ for different configurations, gates (XY, Sycamore, CNOT, and random), and system sizes $n$ in a 2D geometry (Fig.~\ref{fig:Google_geometry}). With random $\av$ we denoted the gate with randomly generated canonical parameters $\ax = 0.8501,\ay= 0.4628,\az= 0.1204$. The effective eigenvalue describes purity decay after the application of $n$ 2-qubit gates.}
	\end{ruledtabular}
\end{table*}
In Table~\ref{tab:gaps_table} we compare all three mentioned protocols, as well as different choices of a 2-qubit gate. In order to facilitate the comparison of entanglement generation rates between protocols with different number of gates we shall compare the effective eigenvalue $\lambda$. The effective eigenvalue is equal to the true eigenvalue $|\lambda_2|$ re-normalized to a step with $n$ gates, that is, in a protocol step with $T$ local gates $M_{i,j}$ it is $\lambda=|\lambda_2|^{n/T}$. For example, protocol steps with 8 letters have $T=(10n-30)/3$, whereas the ABCD protocol has $T=(5n-15)/3$. Similarly as in 1D protocols, fixing a configuration the fastest gate is the XY gate. Comparing different configurations (for small $n$ we checked all possible permutations of 8-letter protocols) we find that while at smaller $n$ there are configurations that have smaller $\lambda$ than the Google's ABCDCDAB, at $n=27$ the Google's choice seems to be the optimal one. We can see that the optimal eigenvalue $\lambda=0.141$ for the XY gate is by $\approx 15\%$ larger than for the BW PBC $|\lambda_2|=1/9$ (\ref{eq:BW_PBC}), however, applying a BW protocol on the 2D geometry in Fig.~\ref{fig:Google_geometry} would necessitate additional long-range 2-qubit gates. Similar gain ($\approx 10\%$) would be obtained also for the Sycamore gate. For non-optimal 2-qubit gates, like the CNOT, the optimal protocol can be different than ABCDCDAD even at $n=27$, see Table~\ref{tab:gaps_table}. 

In 1D protocols studied in the rest of the paper we have seen that the entanglement generation rate $\rE$ was usually not given by the spectral gap. In Fig.~\ref{fig:Google_purity} we show numerically calculated purity for protocols discussed in this section. \new{The subsystem A for our bipartition is for even $m$ composed of $m/2$ bottom rows of qubits (see Fig.~\ref{fig:Google_geometry}), while for odd $m$ we take the bottom $(m-1)/2$ rows plus an additional leftmost qubit in the next row.} We can see that for all cases the purity rate agrees with $-\log{|\lambda_2|}$. It is now evident that taking a CNOT gate instead of the XY or the Sycamore gate would results in a significantly slower entanglement generation. It is an important and nontrivial task to find the optimal configuration and choose the best gate for each topology at hand.

\section{Discussion}

Studying purity entanglement generation speed in quantum circuits with random single-qubit unitaries and a given \new{nearest-neighbor} 2-qubit gate we have identified circuits with the fastest entanglement production. They are the fastest possible and by a significant factor faster than previously studied cases, for instance, circuits with CNOT or random U(4) gates. These fastest scramblers are circuits with a brick-wall pattern of applied gates, with either open or periodic boundary conditions, and the XXZ 2-qubit gate, a special case of which is the XY gate.

Such random circuits should be of interest in experiments \new{where} they can reduce the running time and therefore increase the fidelity. They are simple examples of ``fastest'' chaotic many-body systems, see also Ref.~\cite{prosen19} for a Floquet system with such property. Interestingly, relaxation towards the asymptotic random-state entanglement in optimal protocols, as well as other, does not proceed with a single relaxation rate but rather exhibits a phase transition: relaxation (or, equivalently, the entanglement generation rate) is \new{faster} until the thermalization time (\new{thermalization time is the time} when the entanglement closes in on its volume-law saturation value $\sim \nA$), but then discontinuously transitions into a smaller rate. This happens because the initial relaxation rate is not given by the transfer matrix gap, as one would naively expect. On a mathematical level it arises due to non-symmetric Markovian transfer matrix even though the action of each individual gate is described by a symmetric matrix. \new{This is so} because the expansion coefficients in the spectral decomposition can get exponentially large in the size of a many-body system. As a consequence a subleading eigenvalue, even if it is gapped away from the leading one, can give the dominant relaxation rate. Even stranger is the observation that for the same non-Hermitian reasons it can happen that such exponentially large terms result in relaxation with a rate that is smaller that the one given by the leading eigenvalue. It is as if one would have an additional 'phantom' eigenvalue in the spectrum that is larger than any actual eigenvalue. 

\new{
  We did demonstrate that the same phenomenology occurs also for non-extremal fixed 2-qubit gates (non dual-unitary), as well as for the case of Haar-random 2-qubit gates. For the latter several conjectured exact results about the largest eigenvalue and the purity decay are left as open problems. We also show that spatial as well as temporal randomness of 1-site unitaries is not necessary. This suggests that the same phenomenology of a multistage relaxation should arise also in many other situations, including non-random systems. While we discussed the average entanglement, for large system sizes explicit averaging over different circuit realizations is not necessary -- a single circuit realization will already show the phenomenon.

  The phenomena therefore seem to be rather general, a common point being non-Hermitian many-body matrices. One wonders in how many other situations with non-Hermitian operators such a physics can arise (e.g., dissipative systems, scattering problems, etc.). A promising approach would be to find a continuous-time system with the same phenomenology. Alternatively, starting with a Hamiltonian with noise (instead of with a Floquet system -- a quantum circuit -- with random unitaries) one can arrive at the Lindblad master equation after averaging over noise. In some cases or limits the resulting~\cite{Austen18} non-Hermitian generator is equal to some well studied models, simplifying the insight. It has been shown~\cite{Austen18} that some features in such a setting are the same as in random circuits.

We stress that one needs sufficiently large systems to observed the effects. We are therefore dealing with genuine many-body physics. Convergence with $n$ is namely in some cases quite slow and one needs good numerical methods to access the required sizes.}

How do other quantities besides purity behave in such random circuits is also an interesting open problem. This includes other R\' enyi entropies as well as operators, correlation functions etc.. \new{Von Neumann entropy for instance does show multistage thermalization. Preliminary results show~\cite{tobe} that OTOCs also do exhibit effects of a phantom eigenvalue.} Because the extremal random circuits belong to a class of dual-unitary circuits non-Hermitian effects that we identify could be at play in such situations, and perhaps even more generally in some chaotic Floquet systems~\cite{prosen19,adamPRX20,Austen20}.

\new{Do Markov chains describing random circuits exhibit a cutoff phenomenon known to happen in some Markovian chains~\cite{Persi96,Kastoryano12}, and which has been speculated~\cite{Znidaric_2007} to actually occur in certain random circuits, and if yes, are phenomena we observe in any way related to it? In the cutoff phenomenon relaxation towards the invariant state as measured by the total variational distance is sudden -- upto the cutoff time there can be `memory' of the initial state and no relaxation, which is then suddenly 'forgotten' at the cutoff (mixing) time. Based on purity decay, which is always exponential but with a ``wrong'' rate that is only by a factor different than the inverse gap, it would be tempting to conclude that there is no cutoff. Namely, a necessary condition for a cutoff~\cite{cutoff} is that the mixing time is parametrically larger than the inverse gap (their ratio should diverge in the TDL). However, to really probe the cutoff one should study the full measure relaxation as quantified by the total variational distance which is in particular concerned with a worst-case relaxation scenario. We were on the other hand studying initial states that correspond to a valid purity vector (purities for different bipartitions can not be arbitrary) and relevant bipartitions with extensive $\nA$.}

While we explained the behavior in terms of non-orthogonality and the way expansion coefficients behave, the underlying physics remains to be analytically understood. When and why do such phase transitions in the rate happen? Is it associated with some change in the entanglement properties of the underlying state, like e.g. their Schmidt spectrum being different than that of random states~\cite{jpa07}; features like that have for instance been identified in either appropriate random ensembles~\cite{Anto07} or under random evolutions~\cite{Vinayak12,Jed19}. It has been observed that around thermalization time fluctuations qualitatively change in random circuits~\cite{Cotler20}.

\new{We focused on qubits and the case of nearest-neighbor protocols, however other topologies are also of interest, including systems in higher dimensions as well as systems with larger local Hilbert space (qudits)}. For instance, for the all-to-all coupling we have calculated the exact asymptotic expression for the spectral gap for any gate (Appendix~\ref{app:random_couplings}), however we did not explore the entanglement generation.


We would like to thank A.~Nahum, \new{S.~Gopalakrishnan, A.~G.~Green and A.~Lamacraft} for discussion. Support by Grants No.~J1-1698 and No.~P1-0402 from the Slovenian Research Agency is acknowledged.

\newpage
\appendix

\new{
  \section{Brief derivation of transfer matrix}
  \label{app:derivationM}

In this section we present a short derivation of equation (\ref{eq:M'}); the original derivation with a more detailed explanation can be found in Ref.~\cite{metoda_redukcija}.

In order to sketch the derivation of the Markov chain description, we will define a new set of operators on a doubled Hilbert space. These operators will be used to obtain a compact expression for purity for every possible bipartition of our system. Averaging over all independent one-site Haar random gates will results in the Markov chain description in equation (\ref{eq:M'}). 

Let us write down the purity of the state $\Psi$ for every possible bipartition of our set of qubits. We will encode a bipartition in a vector $\mathbf{s}=(s_1,s_2,\dots,s_n)$, where $s_i \in \{ \uparrow,\downarrow \}$, depending on whether the $i$-th qubit is in the subsystem A ($s_i = \uparrow$) or B ($s_i = \downarrow$) of a given bipartition. $\Psi$ is a physical state of a chain of $n$ qubits on a Hilbert space $\mathcal{H} = \mathbb{C}^{2}_1\otimes \mathbb{C}^{2}_2\otimes \dots \otimes \mathbb{C}^{2}_n$. In order to write down purity we will define two local operators $\chi_{\downarrow_i}$ and $\chi_{\uparrow_i}$ acting nontrivially on a site $i$ of the duplicated system $\mathcal{H}^{\otimes 2}$, namely

\begin{equation}
	\chi_{s_i} = 
	\begin{cases}
		\1_i = \sum_{\alpha,\beta = 0}^{1} |\alpha \beta \rangle_i \langle \alpha \beta|_i, 	\quad &s_i = \downarrow	 \\
		\mathrm{SWAP}_i = \sum_{\alpha,\beta = 0}^{1} |\beta \alpha \rangle_i \langle \alpha \beta|_i, 	\quad &s_i = \uparrow,
	\end{cases}
	\label{chi}
\end{equation}
where $|\alpha\rangle$,$|\beta\rangle$ are basis vectors $\{\ket{0},\ket{1} \}$ of the local Hilbert space $\mathbb{C}^{2}_i$. The index $s_i$ labels the local part of the bipartition and the operator $\chi_{s_i}$ thus depends on the choice of a bipartition. Defining $\chi_{\mathbf{s}}$ as the tensor product of all local operators $\chi_{s_i}$, we can neatly express purity of a state $\Psi$ for a given bipartition as

\begin{equation}
	I_{\mathbf{s}}(\Psi) = \mathrm{tr}\left[ \chi_{\mathbf{s}} \left(|\Psi\rangle\langle\Psi|\right)^{\otimes 2} \right],
\end{equation}
where the trace is evaluated on the duplicated system $\mathcal{H}^{\otimes 2}$. 

We are interested in the average evolution of purity under the application of one elementary step $U_{i,j} = W_{i,j} V_i V_j$. Averaging over one-site Haar random unitaries $V_i$ and $V_j$ \cite{metoda_redukcija} one gets average purity of the state $U_{i,j} \Psi$,

\begin{equation}
  \underset{V_i,V_j \in \text{Haar}}{\mathbb{E}} \left[ I_{\mathbf{s}}(U_{i,j} \Psi ) \right] = \sum_{\mathbf{t},\mathbf{t'}}  (W_{i,j})'_{\mathbf{s},\mathbf{t}'} O_{\mathbf{t}',\mathbf{t}} \,I_{\mathbf{t}}(\Psi),
  \label{eq:sum}
\end{equation}
where scalars $(W_{i,j})'_{\mathbf{s},\mathbf{t}'}$ are

\begin{equation}
	(W_{i,j})'_{\mathbf{s},\mathbf{t}'} = \mathrm{tr} \left[ \chi_{\mathbf{s}} W_{i,j}^{\otimes 2} \chi_{\mathbf{t}'} W_{i,j}^{\dagger} {}^{\otimes 2} \right]
\end{equation}
while scalars $O_{\mathbf{t}',\mathbf{t}}$ are matrix elements of the operator $O=O_1\otimes O_2 \otimes \dots \otimes O_n$ defined on $\mathcal{H}$, with 
\begin{equation}
	O_i = \frac{1}{3} |\uparrow\rangle\langle\uparrow| - \frac{1}{6}|\uparrow\rangle\langle\downarrow| - \frac{1}{6}|\downarrow\rangle\langle\uparrow|+\frac{1}{3}|\downarrow\rangle\langle\downarrow|.
\end{equation}

Summing over $\mathbf{t}'$ in Eq.~(\ref{eq:sum}) and defining matrix elements of $M'_{i,j}$ as $[M'_{i,j}]_{\mathbf{s},\mathbf{t}} = \sum_{\mathbf{t}'}(W_{i,j})'_{\mathbf{s},\mathbf{t}'} O_{\mathbf{t}',\mathbf{t}}$, we see that we get a simple Markovian mapping of average purities after one gate as written in Eq.~(\ref{eq:M'}).

}

\section{Calculation of the steady state purity}
\label{app:even_parity}

In this appendix we derive the eigenvector $\mathbf{\Phi}_{\infty}$ of $M$ with eigenvalue $\lambda = 1$. The steady state of $M$ will be of great importance in numerical gap calculations (see Appendix \ref{Appendix_power_method}). With the help of $\mathbf{\Phi}_{\infty}$ we are able to calculate the purity of random states (\ref{eq:I_inf}). 

Keeping in mind that we must consider only even-parity vectors in the basis of $M$, we can try to calculate the purity to which we will converge after infinite time. To do so, let us first derive the even-parity eigenvector $\mathbf{\Phi}_{\infty}$ with eigenvalue $1$. The eigenvector $\mathbf{\Phi}_{\infty}$ can be constructed by demanding that for every $i\in\{1,\dots,n\}$ we have $M_{i,i+1} \mathbf{\Phi}_{\infty} = \mathbf{\Phi}_{\infty}$. From equations (\ref{eq:v}) we see that the coefficients before basis vectors with the same number of up spins must be equal, and that the coefficients before vectors with $k$ up spins must be $3$ times greater than those with $(k-2)$ up spins. The eigenvector $\mathbf{\Phi}_{\infty}$ is thus

\begin{equation}
\mathbf{\Phi}_{\infty} = \sum_{\mathbf{e}_k \in \{\uparrow,\downarrow\}^{\otimes n}} 3^{n-w} \mathbf{e}_k,
\label{eigen}
\end{equation}
where the sum runs over all basis vectors $\mathbf{e}_k$ of $M$ with an even number of down spins and where $w=w(\mathbf{s})$ is the Hamming weight of the $n$-bit string $\mathbf{e}_k$, i.e., the number of digits equal to $\downarrow$ in $\mathbf{e}_k$. When $n=4$ the vector $\mathbf{\Phi}_{\infty}$ takes the form

\begin{align}
\Phi_{\infty} =& 9 |\uparrow \uparrow \uparrow \uparrow\rangle + \nonumber \\
&3 (|\uparrow \uparrow \downarrow \downarrow\rangle + |\uparrow \downarrow \uparrow \downarrow\rangle + |\downarrow \uparrow \uparrow \downarrow\rangle)+ \nonumber \\
&3(|\uparrow \downarrow \downarrow \uparrow\rangle +|\downarrow \uparrow \downarrow \uparrow\rangle + |\downarrow \downarrow \uparrow \uparrow\rangle) + \nonumber \\
&1 |\downarrow \downarrow \downarrow \downarrow\rangle.
\end{align}
We also present an alternative construction of the eigenvector $\mathbf{\Phi}_{\infty}$, which will be especially useful in the derivation of $I_\infty$ that will follow. The new way to obtain $\mathbf{\Phi}_{\infty}$ is by projecting the vector

\begin{equation}
	\bm{\phi} = \bigotimes_{i=1}^{n} (\sqrt{3},1)
\end{equation}
onto the subspace of even parity, that is

\begin{equation}
	\mathbf{\Phi}_{\infty} = \frac{1}{2}(\1+Z) \bm{\phi} = \frac{1}{2} (\1 + \prod_i \sz_i) \bm{\phi}.
\end{equation}

After infinite time we will converge to the eigenvector $\mathbf{\Phi}_{\infty}$. If we want to read the purities from the coefficients of our final vector, we must convert $\mathbf{\Phi}_{\infty}$ back to the basis of $M'$, i.e. we must obtain the vector $\mathbf{\Phi'}_{\infty}=A \mathbf{\Phi}_{\infty}$. We get

\begin{align}
	\mathbf{\Phi'}_{\infty} &= A \frac{1}{2}(\1+Z) \bm{\phi} = \prod_i A_i \frac{1}{2} (\1+\prod_i\sz_i) \bm{\phi} \nonumber \\
	   &= (\frac{1}{\sqrt{3}})^n \left[\bigotimes_{i=1}^{n} (1,2) + \bigotimes_{i=1}^{n} (2,1)\right]. 
	\label{purity_inf_izpeljava}
\end{align}
The first term at the rightmost position of equation (\ref{purity_inf_izpeljava}) will be, up to the normalization factor $(\frac{1}{\sqrt{3}})^n$, equal to $2^{\nA}$, and the second term will be $2^{\nB}$. We normalize the coefficients of $\mathbf{\Phi'}_{\infty}$ by demanding $I_{\emptyset}(\infty) = 1$, so 

\begin{equation}
	I_\infty = \frac{2^{\nA}+2^{\nB}}{1+2^n},
\end{equation}
which agrees with equation (\ref{eq:I_inf}).

\section{Matrix product state form of the steady state}
\label{app:MPS}

The steady state in the even sector $\ket{\Phi'_\infty}$ (\ref{eq:Phiinf}) is an (unnormalized) eigenvector of $M'$ with eigenvalue $1$. It has Schmidt rank $2$ for any bipartition and can be therefore written in the matrix-product-state (MPS) ansatz with matrices of size $2$.

Let us define two non-normalized orthogonal vectors $\ket{x^{(p)}_+}$ and $\ket{x^{(p)}_-}$ on $p$ qubits,
\begin{equation}
  \ket{x^{(p)}_\pm} = \sum_{\mathbf{s}\in \{0,1\}^{\otimes p}} \frac{2^{p-w} \pm 2^w}{2^p \pm 1} \ket{\mathbf{s}},
\end{equation}
where $w=w(\mathbf{s})$ is the Hamming weight of a $p$-bit string $\mathbf{s}$, i.e., the number of $s_j$ equal to $\downarrow$ in the string $\mathbf{s}$. The norm of those vectors is
\begin{equation}
  |x^{(p)}_\pm|^2 = \frac{2(5^p \pm 4^p)}{(2^p \pm 1)^2}.
\end{equation}
Observe that $\ket{\Phi'_\infty}=\ket{x^{(n)}_+}$.

The Schmidt decomposition of an $n$-qubit $\ket{x^{(n)}_+}$ for a bipartition into first $r$ (subsystems A) plus last $n-r$ qubits (subsystem B) is
\begin{equation}
  \ket{x^{(n)}_+}=\mu_+(n,r) \frac{\ket{x^{(r)}_+}}{|x^{(r)}_+|} \frac{\ket{x^{(n-r)}_+}}{|x^{(n-r)}_+|} + \mu_-(n,r) \frac{\ket{x^{(r)}_-}}{|x^{(r)}_-|} \frac{\ket{x^{(n-r)}_-}}{|x^{(n-r)}_-|},
\end{equation}
where the two Schmidt coefficients are
\begin{equation}
  \mu_\pm(n,r) = \sqrt{\frac{(5^r \pm 4^r)(5^n 4^r \pm 5^r 4^n)}{5^r}}\frac{1}{2^r(2^n+1)}. 
\end{equation}
The Schmidt decomposition of the 2nd vector is on the other hand
\begin{equation}
  \ket{x^{(n)}_-}=\nu_+(n,r) \frac{\ket{x^{(r)}_+}}{|x^{(r)}_+|} \frac{\ket{x^{(n-r)}_-}}{|x^{(n-r)}_-|} + \nu_-(n,r) \frac{\ket{x^{(r)}_-}}{|x^{(r)}_-|} \frac{\ket{x^{(n-r)}_+}}{|x^{(n-r)}_+|},
\end{equation}
with Schmidt coefficients $\nu_\pm(n,r)$
\begin{equation}
  \nu_\pm(n,r) = \sqrt{\frac{(5^r \pm 4^r)(5^n 4^r \mp 5^r 4^n)}{5^r}}\frac{1}{2^r(2^n-1)}.
\end{equation}
This recursive two-state structure of the Schmidt decompositions guarantees MPS representation with matrices of size $2$. More explicitly, writing
\begin{equation}
  I_\infty(\mathbf{s})= \langle y | A^{(s_1)}_1 A^{(s_2)}_2\cdots A^{(s_n)}_n| y\rangle
\end{equation}
where $y=(1,0)$, the matrices are
\begin{align}
  A^{(\uparrow)}_1 &=
  \frac{1}{\sqrt{2}}
  \begin{pmatrix}
    \mu_+(n,1) & -\mu_-(n,1) \\
    0 & 0 
  \end{pmatrix}\nonumber \\
  A^{(\downarrow)}_1 &= \frac{1}{\sqrt{2}}
  \begin{pmatrix}
    \mu_+(n,1) & \mu_-(n,1) \\
    0 & 0
  \end{pmatrix}
\end{align}
on the first site,
\begin{align}
  A^{(\uparrow)}_n &=
  \frac{1}{\sqrt{2}}
  \begin{pmatrix}
    1 & 0 \\
    -1 & 0 
  \end{pmatrix}\nonumber \\
  A^{(\downarrow)}_n &= \frac{1}{\sqrt{2}}
  \begin{pmatrix}
    1 & 0 \\
    1 & 0
  \end{pmatrix}
\end{align}
on the last, and
\begin{align}
  A^{(\uparrow)}_r &=
  \frac{1}{\sqrt{2}}
  \begin{pmatrix}
    \frac{\mu_+(n-r,1)}{|x^{(n-r)}_+|} & -\frac{\mu_-(n-r,1)}{|x^{(n-r)}_+|} \\
    -\frac{\nu_-(n-r,1)}{|x^{(n-r)}_-|} & \frac{\nu_+(n-r,1)}{|x^{(n-r)}_-|} \\
  \end{pmatrix} \nonumber \\
    A^{(\downarrow)}_r &= \frac{1}{\sqrt{2}}
  \begin{pmatrix}
    \frac{\mu_+(n-r,1)}{|x^{(n-r)}_+|} & \frac{\mu_-(n-r,1)}{|x^{(n-r)}_+|} \\
    \frac{\nu_-(n-r,1)}{|x^{(n-r)}_-|} & \frac{\nu_+(n-r,1)}{|x^{(n-r)}_-|} \\
  \end{pmatrix}
\end{align}
on sites $r=2,\ldots,n-1$.

\new{\section{Spectrum analysis}}

Let us first remind ourselves of an elementary fact about matrices that we shall repeatedly use. Let $A$ and $B$ be two finite-dimensional square matrices. Then the product $AB$ has the same spectrum as $BA$, $AB \simeq BA$. One way to see this is to write down the characteristic polynomial $p({\lambda})$ whose expansion coefficients can all be expressed in terms of invariants, for instances the traces $\tr{(AB)^r}$. Because $\tr{(AB)^r}=\tr{(BA)^r}$ the above equivalence immediately follows. A corollary of $AB \simeq BA$ is also the spectral equivalence under cyclic permutations that we shall use, e.g., $ABC \simeq CAB$. The other property that we will need is the spectral equivalence under transposition, $A^{\rm T} \simeq A$.

Physical initial states always have positive (even) parity $Z$ therefore only even eigenvalues of $M$ matter for the purity decay. Because all elementary gates $M_{i,j}$ preserve the parity, the parity of eigenvectors for spectrally equivalent configurations $A \simeq B$ corresponding to the same eigenvalue $\lambda$ is the same. More explicitly, for cyclic permutations, if one has a nonzero $\lambda$ and the associated eigenvector $\mathbf{x}$ of $ABC$, then $CABC\mathbf{x}=C\lambda\mathbf{x}=CAB\mathbf{y}=\lambda \mathbf{y}$, where $\mathbf{y}=C\mathbf{x}$. Because all matrices conserve parity, the parity of $\mathbf{x}$ is the same as that of $\mathbf{y}$. The same holds for the transposition, parities of eigenvectors $\mathbf{x}$ and $\mathbf{y}$ with eigenvalue $\lambda$ corresponding to $A$ and $A^{\rm T}$, respectively, is the same (in a basis with good parity $A$ and $A^{\rm T}$ have a block structure). In the two theorems we are going to prove we therefore do not have to keep track of the parity -- an eigenvalue with good parity will have the same parity in all members of the equivalence class.\\

\subsection{Proof of Theorem \ref{thm:OBC}}
\label{app:obc}

\begin{proof}
Let us begin with an arbitrary permutation of $n-1$ gates $M$ (an example for $n=4$ is $M=M_{3,4}M_{1,2}M_{2,3}$). Our first step is to put the matrix $M_{1,2}$ at the rightmost position in the product (the 1st gate) using cyclic permutations. We thus have $M \simeq A M_{1,2}=A \Ms{1}{1}$, where $A$ is a products of matrices that do not contain $M_{1,2}$. We next increase the length of $\Ms{1}{1}$ by one using the following step.

	Suppose we have
	\begin{equation}
	R M_{i,i+1} A \Ms{i-1}{1},
	\label{general_configuration_OBC}
	\end{equation}
	where $R$ and $A$ is a product of gates that does not include $M_{i,i+1}$ nor any of the gates in $\Ms{i-1}{1}=M_{i-1,i}\cdots M_{1,2}$. Note that the only gate with which the gates in $\Ms{i-1}{1}$ do not commute is $M_{i,i+1}$, and therefore one always has $[\Ms{i-1}{1},A] = 0$. Therefore, $R M_{i,i+1} A \Ms{i-1}{1} \simeq R M_{i,i+1} \Ms{i-1}{1} A \simeq A R M_{i,i+1}\Ms{i-1}{1} = A R \Ms{i}{1}$. We have increased the length of $\Ms{i-1}{1}$ by one. Iterating this step we end up with $M \simeq \Ms{n-1}{1}$. This concludes the proof.
\end{proof}	

\subsection{Proof of Theorem \ref{thm:PBC}}
\label{app:pbc}

\begin{proof}
In the first step we bring, by cyclic permutations, the gate $M_{1,2}$ to the rightmost position. Then in Step 1 (described below), we try to construct $\B{3}{1}$ ($\B{3}{1}$ is the first nontrivial brick-wall because $\B{1}{1}$ is equal to $\Ms{1}{1}=M_{1,2}$). Step 1 will either succeed in constructing $\B{3}{1}$, or, we will end up in the canonical form with $p=1$ (see Fig.~\ref{fig:PBC_equivalent}).
	
	{\em Step 1.--} First, we bring $M_{2,3}$ to the position of a 2nd gate. Writing $R M_{2,3} A M_{1,2}$, where $R$ and $A$ are arbitrary products excluding $M_{1,2}$ and $M_{2,3}$, we have two possibilities ($M_{2,3}$ does not commute with only two gates, $M_{1,2}$ and $M_{3,4}$; for $n=3$, when $M_{3,4}$ is in fact equal to $M_{3,1}$, the presented argument still works):
	\begin{enumerate}[label=(\alph*)]
		\item if $M_{3,4}\notin A$, so that $[M_{2,3},A]=0$, we have $R M_{2,3} A M_{1,2} \simeq R A M_{2,3} M_{1,2}$.
		\item if $M_{3,4}\in A$ we use $[M_{2,3},R]=0$ and cyclic permutations to get $R M_{2,3} A M_{1,2} \simeq M_{1,2} M_{2,3} R A$. Using the fact that all $M_{i,i+1}$ are real and symmetric, and that $M^{\rm T}$ and $M$ have the same spectra, we write $M_{1,2} M_{2,3} R A \simeq  A^{\rm T} R^{\rm T} M_{2,3} M_{1,2}$.
	\end{enumerate}
	Having a form $RM_{3,4}AM_{2,3}M_{1,2}$, we now try to bring $M_{3,4}$ to the rightmost position. There are again two possibilities ($n=4$ again doesn't influence the argument):
	\begin{enumerate}[label=(\roman*)]
		\item if $M_{4,5} \notin A$ we have $RM_{3,4}AM_{2,3}M_{1,2} \simeq R A M_{3,4} M_{2,3} M_{1,2} = RA \Ms{3}{1}$.
		\item if $M_{4,5} \in A$ we use $[M_{3,4},R]$ and a cyclic permutation to write $RM_{3,4}AM_{2,3}M_{1,2} \simeq R A M_{2,3} M_{1,2} M_{3,4}=RA \B{3}{1}$
	\end{enumerate}
	At this point we therefore either have $RA \B{3}{1}$, and we have succeeded (finished the Step 1), or $RA \Ms{3}{1}$ and we continue by trying to include $M_{4,5}$ into $\Ms{3}{1}$.
	
	Procedure is by now familiar; writing $R M_{4,5} A \Ms{3}{1}$ we have either $M_{5,6} \in A$, in which case $R M_{4,5} A \Ms{3}{1} \simeq R A \Ms{4}{1}$, or $M_{5,6} \notin A$, in which case $R M_{4,5} A \Ms{3}{1} \simeq R A \Ms{3}{1} M_{4,5} \simeq R A  M_{3,4} M_{4,5} M_{2,3}M_{1,2} \simeq R A M_{4,5} \B{3}{1}$, where in the last equivalence we used that $M_{3,4}$ commutes with all the gates in $A$ and $R$. We have therefore either increased the number of gates in $\Ms{3}{1}$ by one, or obtained the desired $\B{3}{1}$.
	
	We are now in a position to write the generic iteration step: in $R M_{i+1,i+2}A\Ms{i}{1}$ one has either (i) $M_{i+2,i+3} \notin A$, or (ii) $M_{i+2,i+3} \in A$. In case (i) we immediately get $\simeq R A \Ms{i+1}{1}$. In case (ii) we have $R M_{i+1,i+2}A\Ms{i}{1} \simeq RA\Ms{i}{1} M_{i+1,i+2} \simeq R A M_{i,i+1} M_{i+1,i+2} M_{i-1,i}\cdots M_{3,4}M_{2,3}M_{1,2}$. Using the fact that gates $M_{3,4}$ till $M_{i,i+1}$ commute with all gates in $R$ and $A$, we can use cyclic permutations to bring first $M_{i,i+1}$ to the rightmost position, then $M_{i-1,i}$, and so on, till we bring $M_{3,4}$ to the 1st position, resulting in $\B{3}{1}$. An example of such a transformation can be seen in Fig.~\ref{fig:PBC_StoB}.	
	\begin{figure}[h]
		\begin{center}
			\includegraphics[width=80mm]{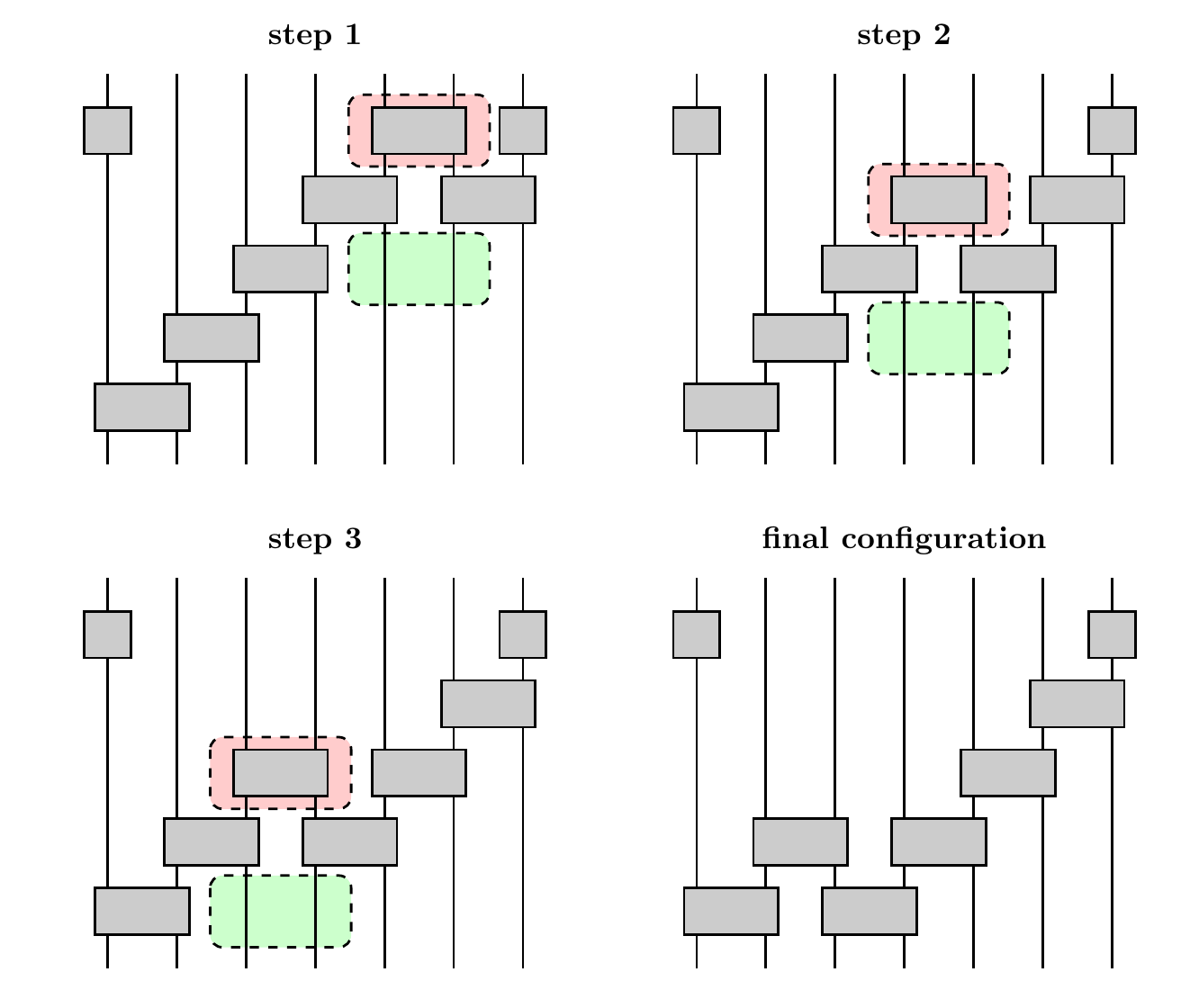}
			\caption{An example of a transformation from a configuration $M = R \Ms{i}{1} M_{i,i+1}$ to the configuration $M = R \B{3}{1}$. Using commutation relations and cyclic permutations we must transform the operator in the red box to the position denoted by the green box.}
			\label{fig:PBC_StoB}
		\end{center}
	\end{figure}

We see that repeating this procedure eventually either produces $\B{3}{1}$ at the position of the first three gates, or, we end up with $\Ms{n}{1}$, i.e., the canonical configuration with $p=1$. If we have the situation $R \B{3}{1}$ we continue with Step 2 which describes a generic step starting with $R \B{i}{1}$. It either increases the number of brick-wall gates, bringing us to $A \B{i+2}{1}$, or it will end up in $\Ms{n-i}{i+1}\B{i}{1}$.
	
	{\em Step 2.--} 	
	Starting with $R \B{i}{1}$ (with odd $i$) we see that $\B{i}{1}$, which acts on sites $1,\ldots,i+1$, does not commute only with two gates, $M_{n,1}$ and $M_{i+1,i+2}$. It behaves in exactly the same way as the gate $M_{1,2}$ in the Step 1, because none of the gates in $R$ acts on the inner qubits $2,\ldots,i$ of the block $\B{i}{1}$. Following exactly the same steps as in the Step 1 we try to add two gates to $\B{i}{1}$ in order to obtain $M_{i+1,i+2}\B{i}{1}M_{i+2,i+3}=\B{i+2}{1}$. In the Step 1(b) we used that $M_{1,2}^{\rm T}=M_{1,2}$, which is not true for $\B{i}{1}$, however, the transpose of $\B{i}{1}$ can always be transformed into $\B{i}{1}$ by using cyclic permutations and the fact that all the ``inner'' gates in $\B{i}{1}$ acting on qubits $2,\ldots,i$ commute with all the gates in $R$. The procedure analogous to the one in the Step 1 will therefore either result in $R \B{i}{1} \simeq A \B{i+2}{1}$, or in $R \B{i}{1} \simeq \Ms{n-i}{i+1}\B{i}{1}$ which is the canonical configuration with $p=(i+1)/2$. This concludes the proof.
\end{proof}

\subsection{Proof of eigenvalue $1/9$}
\label{app:9}

\new{We can also prove that $\lambda=1/9$ is an eigenvalue of $M$ for the brick-wall configuration with PBC and the XY gate.}\\

\begin{proof}
Let us for simplicity focus on even $n$. Taking PBC BW $M=\B{n}{1}$ with the XY gate we can explicitly construct an eigenvector $\mathbf{v}$ of $M$ with eigenvalue $1/9$. The ansatz for the eigenvector is
\begin{equation}
\mathbf{v} = \sum_{i=1}^{n/2} \mathcal{T}^{2i} (\mathbf{v}_\alpha \otimes \mathbf{v}_\alpha \otimes \dots \otimes \mathbf{v}_{3}),
\label{ansatz_lastni_PBC}
\end{equation}
where $\mathbf{v}_\alpha = \mathbf{v}_1 + \alpha \mathbf{v}_2$, where $\mathbf{v}_{1,2,3}$ are the eigenvectors of a 2-qubit $M_{i,i+1}$ (\ref{eq:v}) and $\mathcal{T}$ is the translation operator by one site on a circle, e.g. $\mathcal{T}^2 \mathbf{a}\otimes \mathbf{b} \otimes \mathbf{c} = \mathbf{c}\otimes \mathbf{a} \otimes \mathbf{b}$. Let us write the brick-wall step $M$ as $M=M_{\rm e}M_{\rm o}$, where $M_{\rm o}$ is the first layer (half-step) of the BW, i.e. it contains all $M_{i,i+1}$ with odd $i$, and $M_{\rm e}$ is the 2nd layer ($M_{i,i+1}$ with even $i$). The ansatz vector $\mathbf{v}$ is for an arbitrary $\alpha$ already the eigenvector of $M_{\rm o}$ with $\lambda=-1/3$. If we want it to be the eigenvector of the whole $M$, we must fix the constant $\alpha$ so that it is also the eigenvector of $M_{\rm e}$. Let us focus on a pair of qubits $(i,i+1)$ with even $i$. We can expand $\mathbf{v}$ as a linear combination of $n 4^{\frac{n}{2}-1}$ basis vectors $\mathbf{e}_k$ (we choose the same basis as in equation (\ref{eq:M_cd})). Each basis vector can be further decomposed as $\mathbf{e}_k = \mathbf{w}^k_{1,i-1} \otimes \mathbf{v}^k_{i,i+1} \otimes \mathbf{w}^k_{i+2,n}$, where the subscripts denote the range of qubits which these decomposition vectors describe. For every $\mathbf{v}^{k}_{i,i+1}$ we want either $M_{i,i+1} \mathbf{v}^{k}_{i,i+1} = \mathbf{v}^{k}_{i,i+1}$ or $M_{i,i+1} \mathbf{v}^{k}_{i,i+1} = -\frac{1}{3} \mathbf{v}^{k}_{i,i+1}$, so $\mathbf{v}^{k}_{i,i+1}$ must be either $(3,0,0,1)$ or $(0,\pm1,1,0)$. The vectors $\mathbf{v}^k_{i,i+1}$ depend on the neighboring pairs $(i-1,i)$ and $(i+2)$, which can be $\mathbf{v}_{\alpha} \otimes \mathbf{v}_{\alpha}$, $\mathbf{v}_{\alpha} \otimes \mathbf{v}_{3}$ or $\mathbf{v}_{3} \otimes \mathbf{v}_{\alpha}$ (\ref{ansatz_lastni_PBC}). We have 32 possible combinations on qubits $(i,i+1)$, but for now let us focus on 4 instances: $(9,0,0,\alpha^2)$, $(\alpha^2,0,0,1)$, $(0,3,\alpha^2,0)$ and $(0,\alpha^2,3,0)$. We obtain the desired vector if we fix $\alpha = \sqrt{3}$. For all other $28$ possibilities our demands are automatically fulfilled without the specification of $\alpha$; one can check this by writing down all 32 possibilities that come from terms $\mathbf{v}_{\alpha} \otimes \mathbf{v}_{\alpha}$, $\mathbf{v}_{\alpha} \otimes \mathbf{v}_{3}$ and $\mathbf{v}_{3} \otimes \mathbf{v}_{\alpha}$. For small systems ($n=4,6,8$) we numerically checked the correctness of our guess. The eigenvector $\mathbf{v}$ does not have a good parity and can be decomposed as $\mathbf{v} = \mathbf{v}_{\rm e} + \mathbf{v}_{\rm o}$, into an even and odd parity parts, both of which are, due to parity conservation of $M$, still eigenvectors. This concludes the proof.
\end{proof}

\subsection{Numerical data for $\lambda_2$}
\label{app:numerics1}

\begin{figure}[h]
	\begin{center}
		\includegraphics[width=100mm]{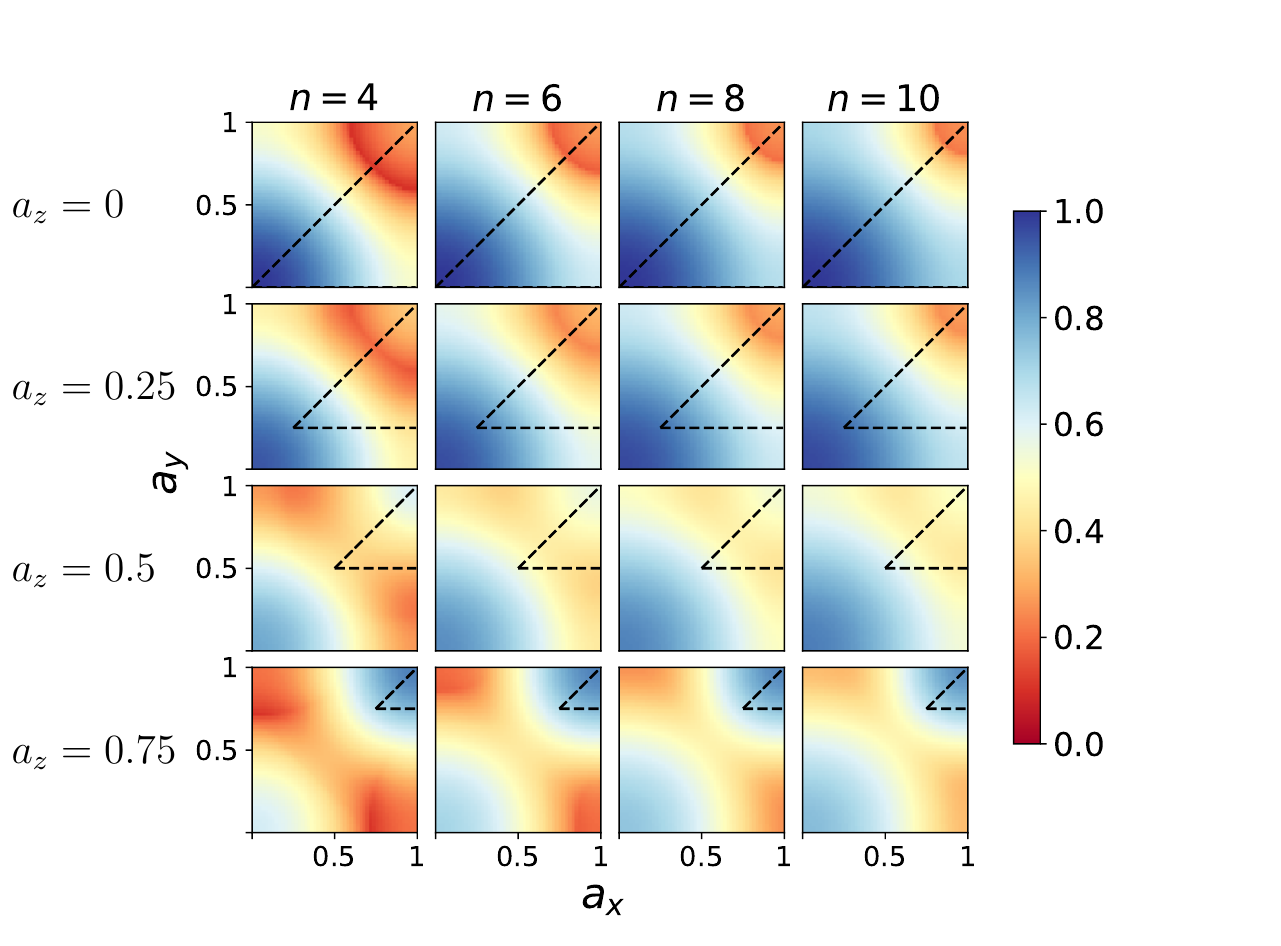}
		\caption{2nd largest eigenvalue $|\lambda_2|$ of $M$ for the OBC S protocol and different canonical gate parameters $\av$ (\ref{canonical_form}). The columns represent a fixed number of qubits and the rows are for different $\az$. The dotted triangles indicate the relevant set of irreducible parameters $\ax,\ay$.}
		\label{fig:OBC_gaps}
	\end{center}
\end{figure}

\begin{figure}[t!]
	\begin{center}
		\includegraphics[width=95mm]{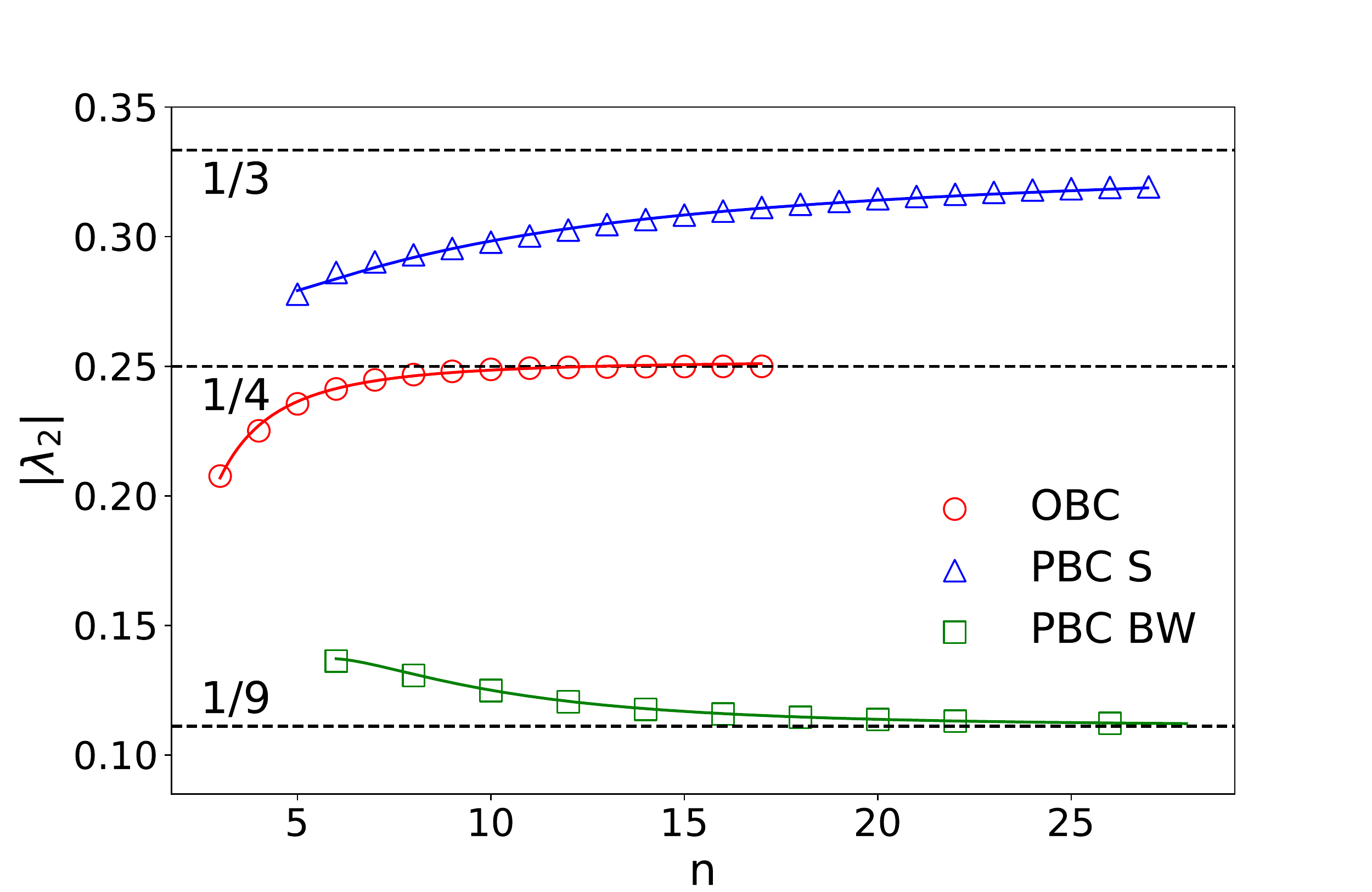}
		\caption{Eigenvalue $|\lambda_2|$ for OBC, and PBC configurations with S configuration (the equivalence class $p=1$) and BW configuration ($p=n/2$) for the two-qubit XY gate. The extrapolated values in the TDL (lines) are obtained from the fits (full curves) to numerical data (symbols). The fits are $|\lambda_2(n)| = 0.252 + 0.008 /n - 0.43 /n^2$ (OBC), $|\lambda_2(n)| = 0.334 - 0.44 /n + 0.83/n^2$ (PBC, $p=1$), and $|\lambda_2(n)| = 0.114 -0.20/n + 4.7/n^2 -16/n^3$ (PBC, $p=n/2$).}
		\label{fig:OBC_in_PBC_tdl}
	\end{center}
\end{figure}

\new{
In this appendix we present numerical results used to determine the fastest asymptotic scrambler for PBC and OBC protocols. To calculate $\lambda_2$ we used either exact diagonalization or the power method as described in Appendix~\ref{Appendix_power_method}.

Fig.~\ref{fig:OBC_gaps} shows color-coded plots of $|\lambda_2|$ for the S configuration with OBC for different values of all three canonical parameters $a_j$. Looking for the smallest $|\lambda_2|$, and how things change with increasing $n$, a promising candidate for the smallest $|\lambda_2|$ is $\mathbf{a}=(1,1,0)$, i.e., the XY gate. Studying more closely how $|\lambda_2|$ depends on $n$ for this XY gate, Fig.~\ref{fig:OBC_in_PBC_tdl}, we can conjecture that in the TDL one has $|\lambda_2|=1/4$ for any protocol with OBC (remember that for OBC the spectrum does not depend on the configuration).
}

\begin{figure}[h]
	\begin{center}
		\includegraphics[width=100mm]{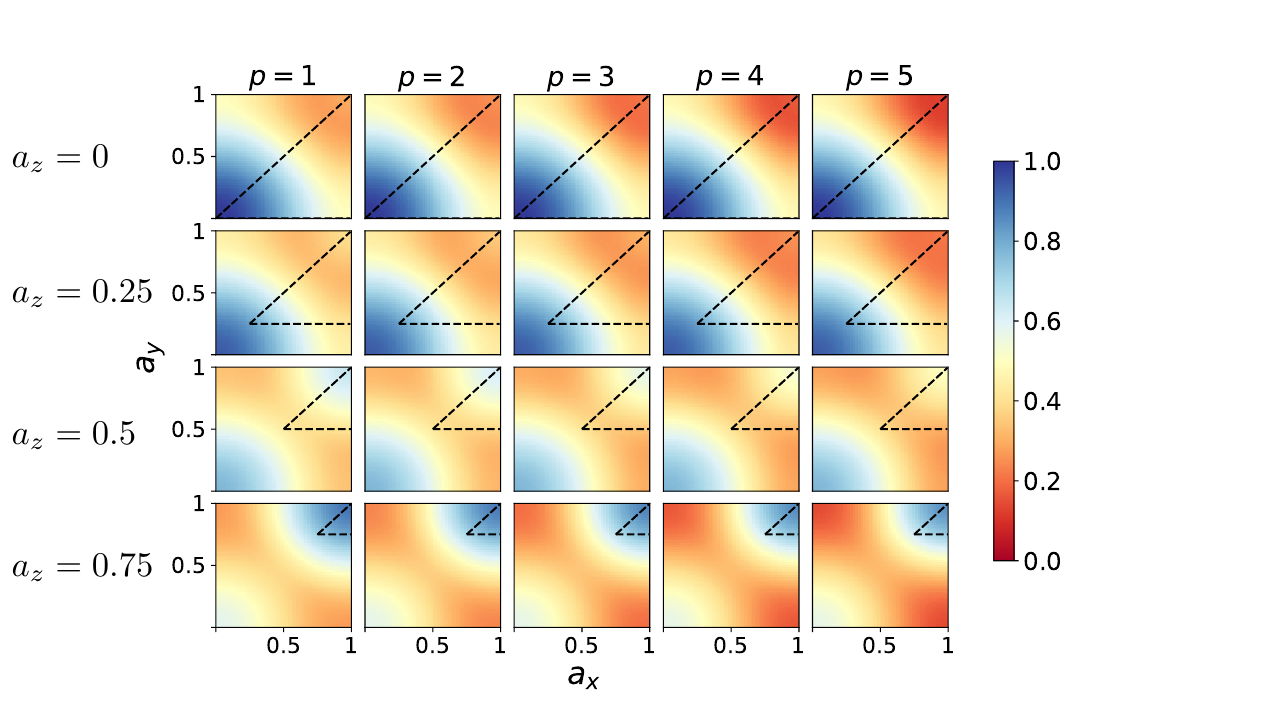}
		\caption{$|\lambda_2|$ for the PBC case at $n=10$. The columns represent a fixed configuration and the rows represent a fixed value for the parameter $\az$.}
		\label{fig:PBC_gaps_n10}
	\end{center}
\end{figure}

\begin{figure}[h]
	\begin{center}
		\includegraphics[width=100mm]{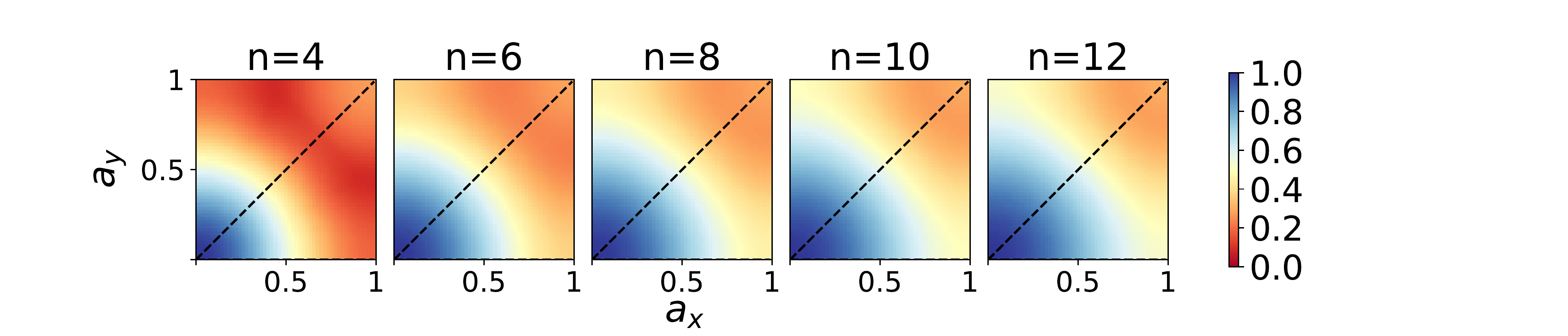}
		\caption{$|\lambda_2|$ of configurations $p=1$ for different values of parameters $\ax,\ay$ at $\az=0$.}
		\label{fig:PBC_p1}
	\end{center}
\end{figure}

\begin{figure}[h]
	\begin{center}
		\includegraphics[width=100mm]{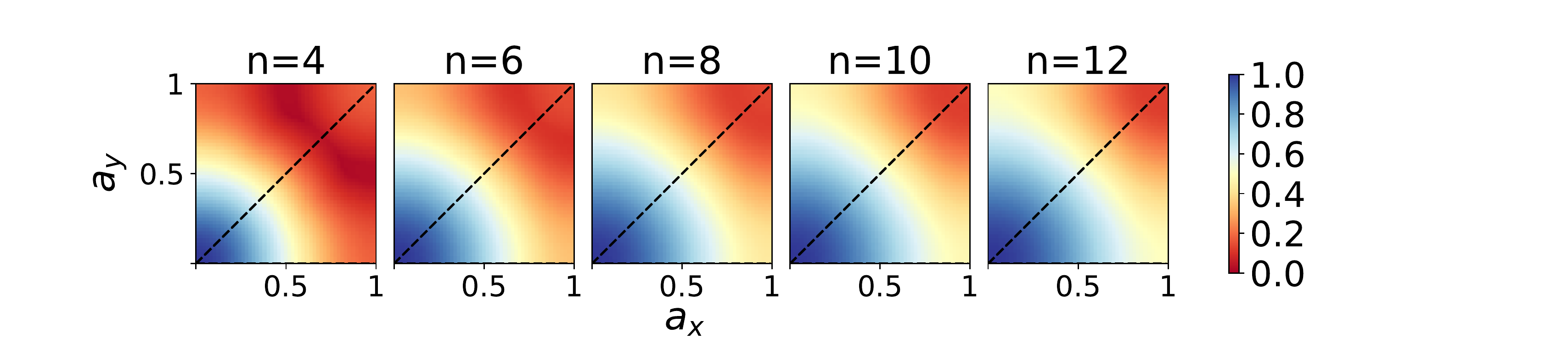}
		\caption{$|\lambda_2|$ of configurations $p=\lfloor n/2 \rfloor$ for different values of parameters $\ax,\ay$ at $\az=0$.}
		\label{fig:PBC_pn2}
	\end{center}
\end{figure}

Fig.~\ref{fig:PBC_gaps_n10} shows the values of $|\lambda_2|$ for different two-qubit gates (parameters $\ax,\ay,\az$) \new{with PBC} and different configurations (parameter $p$) for fixed $n=10$. From Fig.~\ref{fig:PBC_gaps_n10} we learn that $|\lambda_2|$ monotonically decreases as we increase $p$. Moreover the fastest entanglement generation comes from the region $\az=0$. Figs.~\ref{fig:PBC_p1},~\ref{fig:PBC_pn2} show the dependence of $|\lambda_2|$ on $\ax,\ay,\az=0$ and $n$ for $p=1$ (slowest scrambler) and $p=\n2$ (fastest scrambler) respectively. Contrary to OBC protocols, where the fastest asymptotic scrambler is obtained for all $\ax=\ay=1,\az\in [0,\ac\approx 0.32]$, the smallest $|\lambda_2|$ for PBC protocols comes from a single choice of two-qubit gates, namely $\ax=\ay=1,\az=0$ (XY gates).


\section{Numerical calculation of spectral gaps for large $n$}
\label{Appendix_power_method}

Whenever exact numerical diagonalization was too demanding in terms of memory, we calculated the dominant eigenvalue $|\lambda_2|$ using the power method. Briefly, the power method consists of iterations of form
\begin{align}
	\mathbf{u}_{i+1} &= M \mathbf{w}_{i}, \\
	\mathbf{w}_{i+1} &= \mathbf{u}_{i+1}/||\mathbf{u}_{i+1}||
\end{align} 
where $M$ is the matrix of which we want $|\lambda_2|$ and $\mathbf{w}_{k}$ is the vector we iterate. For almost all initial vectors $\mathbf{w}_0$ the power method converges to the vector with the greatest eigenvalue in absolute value. 

In our case, the matrix $M$ has one even-parity eigenvector $\mathbf{\Phi}_{\infty}$ with eigenvalue equal to $1$ (for details see Appendix \ref{app:even_parity}). If we wish to get $|\lambda_2|$ from our power iteration, we must subtract the component corresponding to $\mathbf{\Phi}_{\infty}$ from $\mathbf{w}_i$ at every step $i$. We iterate
\begin{align}
\mathbf{u}_{i+1} &= M \mathbf{w}_{i} - \mathbf{\Phi}_{\infty} \langle \mathbf{\Phi}_{\infty},\mathbf{w}_i \rangle, \\
\mathbf{w}_{i+1} &= \mathbf{u}_{i+1}/||\mathbf{u}_{i+1}||,
\end{align} 
where $\langle * , *\rangle$ is the standard inner product.

If $|\lambda_2|$ is non-degenerate, the power iteration converges to the corresponding eigenvector. In general $\lambda_2 \in \mathbb{C}$. Because the product $M$ is real, then if $\lambda_2 \in \mathbb{C}$ there is also a $\bar{\lambda}_2$ in the spectrum of $M$. Suppose we want to extract the value $|\lambda_2|$ from our power iteration. We begin with a vector $\mathbf{w}_0$ and denote with $\mathbf{w}_{\infty}$ the eigenvector of $M$ with eigenvalue $\lambda_2$. After $k \gg 1$ iterations we can write

\begin{equation}
	M^k \mathbf{w}_0 = \lambda_2^k c \mathbf{w}_{\infty} + \bar{\lambda}_3^k \bar{c} \bar{\mathbf{w}}_{\infty},
	\label{Mk_iteration}
\end{equation}
where $c = \langle \mathbf{w}_{\infty}, \mathbf{w}_0\rangle$. In equation (\ref{Mk_iteration}) we use the fact that at every step we subtract the component resulting from eigenvector with $\lambda = 1$; we can equivalently start with a vector orthogonal to $\mathbf{\Phi}_{\infty}$: $\langle \mathbf{w}_0, \mathbf{\Phi}_{\infty} \rangle = 0$.

We are interested only in the value $|\lambda_2|$, so we can compute the norm of $M^k \mathbf{w}_0$ and compare it with $M^{k+1} \mathbf{w}_0$. Writing $\lambda_2 = |\lambda_2| \mathrm{e}^{\ii \phi}$, $c = |c| \mathrm{e}^{\ii \psi_c}$ and $\langle \mathbf{w}_{\infty}, \bar{\mathbf{w}}_{\infty} \rangle = V \mathrm{e}^{\ii \psi_v}$ we get

\begin{equation}
	||M^k \mathbf{w}_0|| = 2 |\lambda_2|^{2k} |c|^2 \left[ 1 + V \cos(\tilde{k} \phi) \right].
\end{equation}
For sake of simplicity, we defined $\tilde{k} = k + \frac{\psi_c}{2 \psi} + \frac{\psi_v}{2 \psi} = k + \mathrm{const.}$. 

An arbitrary quotient of $(k+1)$-th and $k$-th step can be written as

\begin{equation}
	\frac{||M^{k+1} \mathbf{w}_0||}{||M^k \mathbf{w}_0||} = |\lambda_2|^2 \frac{1 + V \cos((\tilde{k}+1) \phi)}{1 + V \cos(\tilde{k} \phi)} = |\lambda_2|^2 \frac{v_{k+1}}{v_k},
\end{equation} 
where we used the notation $v_k = 1 + V \cos(\tilde{k} \phi)$. Now suppose $\phi = \frac{l}{m} 2\pi$ for integers $l,m$, then $v_{k+m} = v_{k}$ for every $k$. Taking the geometric mean of norms $||M^k \mathbf{w}_0||$ of $m$ subsequent iterations we get

\begin{equation}
	\sqrt[m]{\frac{v_{k+1}}{v_k} \frac{v_{k+2}}{v_k+1} \dots \frac{v_{k+m}}{v_k+m-1}} = |\lambda_2|^2.
\end{equation}
We found that for most cases $\phi \approx \frac{l}{m} 2\pi$ holds, hence a good estimation of $|\lambda_2|$ was possible.

\section{Further \new{entanglement} data}
\label{app:numerics}

\subsection{von Neumann entropy}
\label{app:S}

\begin{figure}[h]
  \centerline{\includegraphics[width=2.8in]{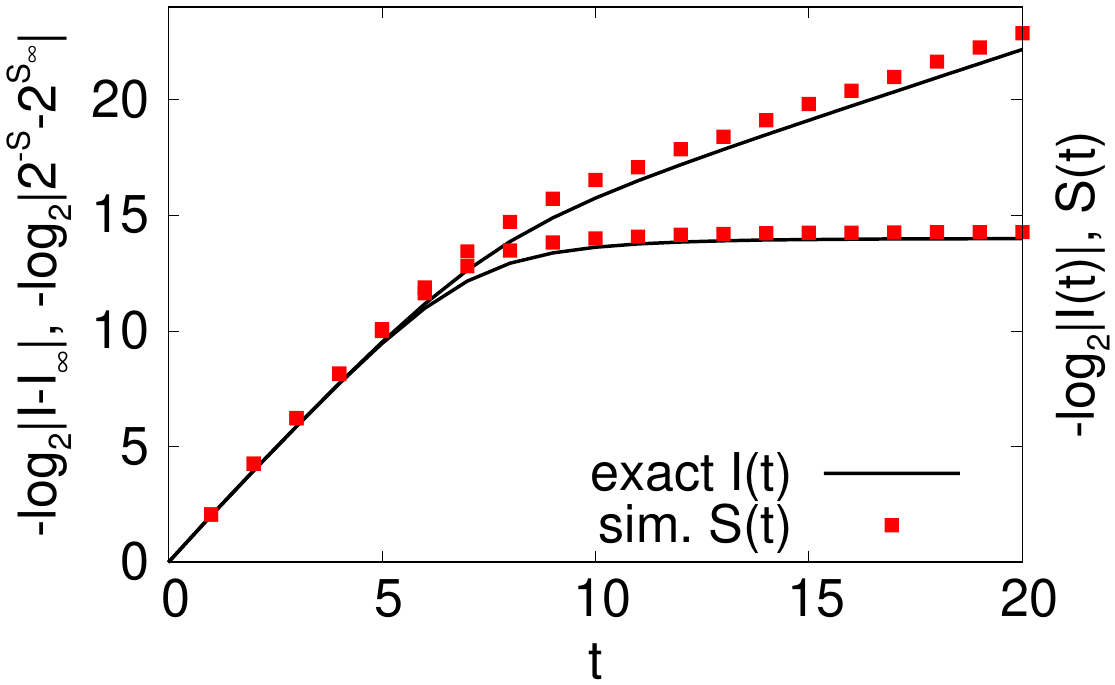}}

\caption{Exact average results for purity (black curves) are compared with the von Neumann entropy $S(t)$ (red squares). In the upper data the thermal value is subtracted (left label), while for the lower two saturating sets it is not (right label). All is for $n=30$ and S configuration with PBC and the XXZ gate with $\az=0.5$ (the same parameters as in Fig.~\ref{fig:crtapbc}(c,d)) for which the rate is $\rE=2\ln{2}$.}
\label{fig:S}
\end{figure}

In Fig.~\ref{fig:S} we compare the von Neumann entropy with the logarithm of the average purity that we studied in the rest of the paper. As we can see von Neumann entropy behaves rather similar as purity, or more precisely as $-\log{\ave{I(t)}}$. In particular it also exhibits a phase transition in the local rate at $t \approx t_\infty$ ($t_\infty \approx 8$ in the figure). 

\new{
\subsection{Fluctuations and randomness}
\label{app:flukt}

For sufficiently large times the state reached under random-circuit evolution is close to a random state. Due to the measure concentration in a large Hilbert space one can get good self-averaging properties even for a single random circuit realization. In Fig.~\ref{fig:flukt} we numerically check such self-averaging property for the S configuration with PBC and two different system sizes in order to get insight on how fluctuations behave with $n$. First thing we notice is that a single realization with random Haar i.i.d. 1-qubit unitaries in both space and time (red curves labeled by ``diff.t,diff.x'') is for large $n$ almost on top of the average purity. Therefore, while an explicit averaging over single-site Haar random unitaries simplifies analytical treatment (results in a Markovian process) it is not necessary for the observed phenomena.
\begin{figure}[h]
  \centerline{\includegraphics[width=2.8in]{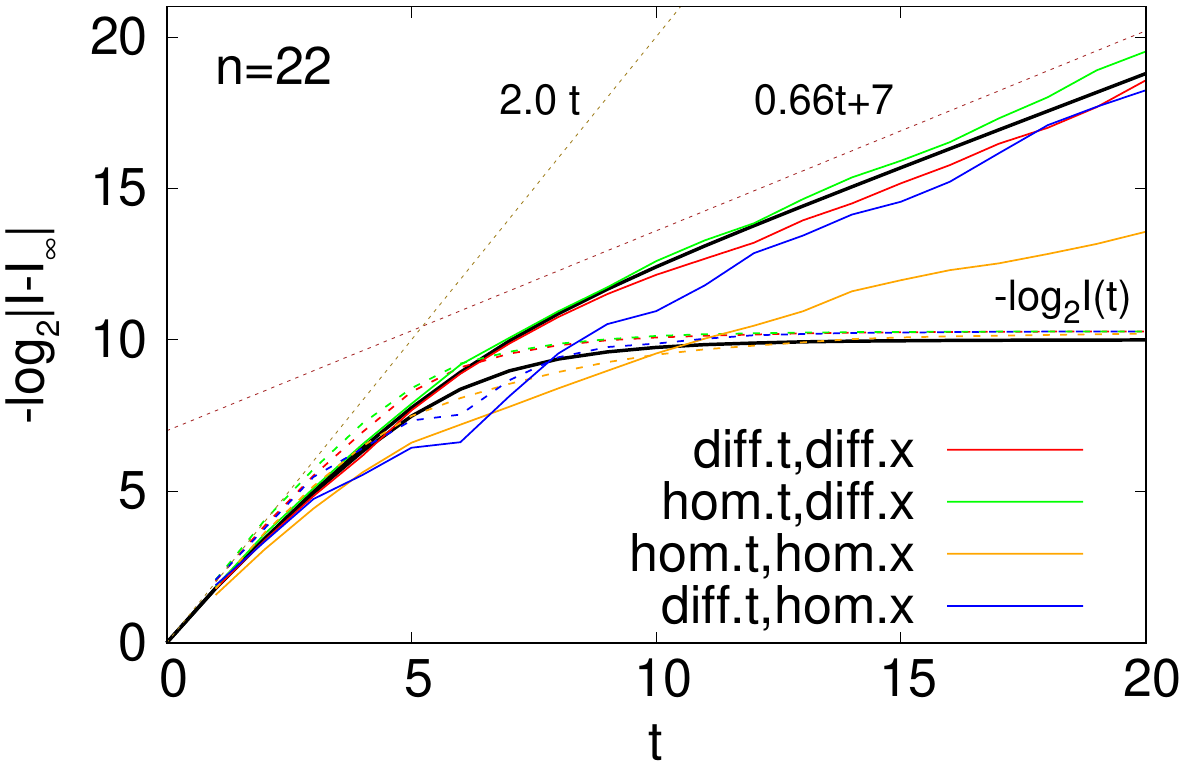}}
  \centerline{\includegraphics[width=2.8in]{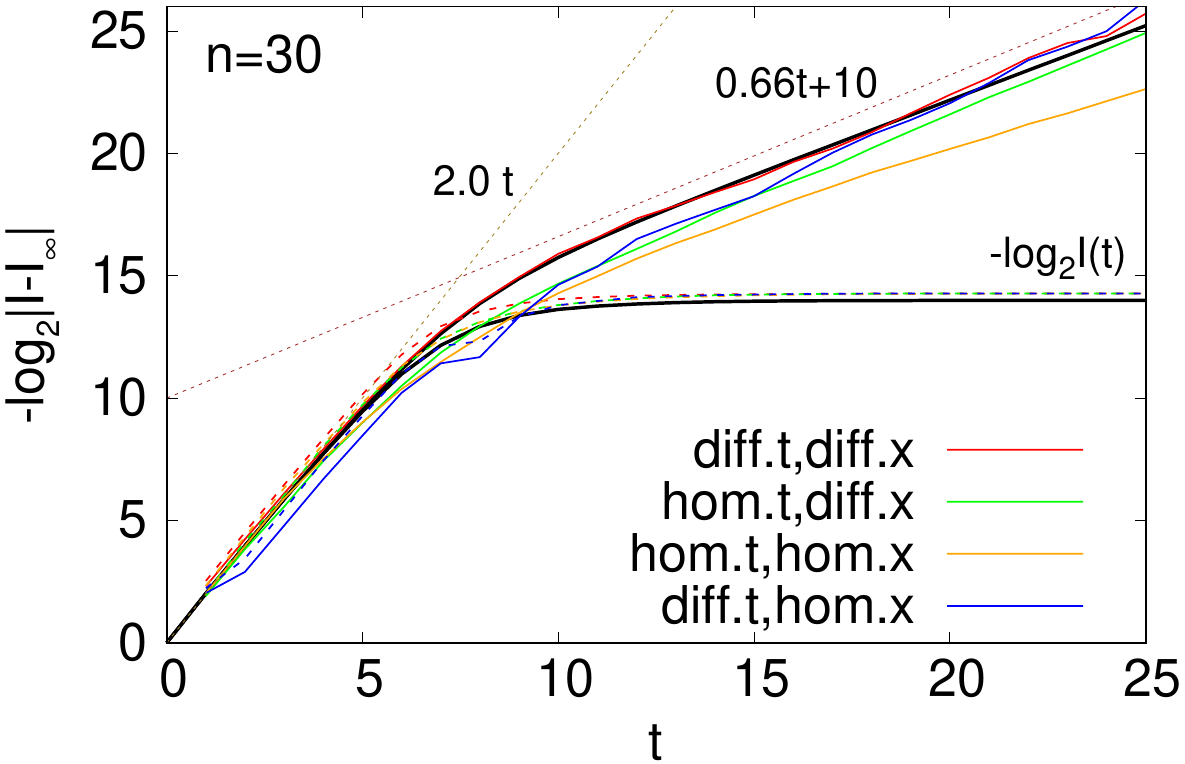}}
\caption{\new{(Color online) Single random circuit realization results for a single random product initial state (four colored curves with labels) and the average purity (smooth black full and dashed curves). All is for the S configuration with PBC and $\az=0.5$ (the same parameters as in Fig.~\ref{fig:crtapbc}(c,d)) and system size $n=22$ (top), and $n=30$ (bottom). Labels for colored curves denote whether 1-qubit Haar unitaries are different on each site (``diff.x'') or the same (``hom.x''), as well as whether they are the same at every timestep (``hom.t'') or different at every circuit layer (``diff.t'').}}
\label{fig:flukt}
\end{figure}

We also check how randomness in 1-site unitaries influences our results. To that end we compare our canonical case where 1-site rotations at each site and timestep are independent (``diff.t,diff.x'') with a situation when unitaries are the same at every site (``hom.x'') and/or the same at every timestep (``hom.t''). We interestingly see that if we use the same random 1-qubit unitary at every site, as well as at every time (``hom.x,hom.t''), i.e., the whole single relization of a circuits uses only one Haar 1-qubit unitary, one will get the same behavior in the thermodynamic limit. Based on the data we can conjecture that in the TDL explicit randomness in not necessary neither in space, nor in time. Comparing fluctuations between the 4 cases shown, they are expectedly the smallest for 1-qubit unitaries that are independently random in space and time, followed by the case of random in space and the same in time (every circuit layer uses the same 1-qubit unitaries), then the case with no spatial randomness but with new unitaries at every time, and lastly the largest fluctuations are observed for the circuits that is homogeneous in space and time. 
}

\section{Random all-to-all coupling}
\label{app:random_couplings}

Here we compute the spectral gap of the transfer matrix describing random quantum circuits where on every step we randomly choose two qubits $i,j$ on which we apply a 2-qubit gate. Opposed to the main text we therefore allow coupling between an arbitrary pair of qubits (not just n.n.). The average step can be written as
\begin{equation}
  \bar{M} = \frac{2}{n (n-1)} \sum_{i,j=1;i<j}^{n} M_{i,j},
  \label{eq:Mbar}
\end{equation}
where $M_{i,j}$ is the familiar matrix from Eq.(\ref{eq:M_cd}). We follow procedure from Ref.~\cite{PRA08} where the gap has been calculated for Clifford gates. Using Pauli notation (\ref{Mcd_Heisenberg}) and rewriting everything in terms of total spin operators $S_{\alpha} = 1/2 \sum_{i=1}^{n} \sigma_i^\alpha$ we get

\begin{align}
	\bar{M} = \frac{2 h}{n} S_z + \frac{4}{n(n-1)}\left(J_\mathrm{x} S_\mathrm{x}^2 + J_\mathrm{y} S_\mathrm{y}^2 + J_\mathrm{z} S_\mathrm{z}^2 \right)  \nonumber \\ + \left(d - \frac{J_\mathrm{x}+J_\mathrm{y}+J_\mathrm{z}}{n-1}\right) \1.
\end{align}
This is the Lipkin-Meshov-Glick (LMG) model~\cite{LMG model} for which one can calculate the spectral gap by taking the semiclassical limit $n \gg 1$ (i.e., spin size $S\to \infty$), and replacing spin operators with classical spins
\begin{align}
	S_\mathrm{x} &= S \cos\phi \sqrt{1-\mu^2},\\
	S_\mathrm{y} &= S \sin\phi \sqrt{1-\mu^2},\\
	S_\mathrm{z} &= S \mu.
\end{align} 
Expanding $\bar{M}$ around the energy maximum and quantizing the resulting harmonic oscillator we get the spectral gap
\begin{equation}
	1-\lambda_2 = \frac{3 h}{n} + \mathcal{O}(1/n^2),
	\label{delta_all-all}
\end{equation}
where $h = \frac{1}{9} \left(3-v\right)$, $v = \cos\left( \pi \ax \right)\cos\left( \pi \ay \right)+\cos\left( \pi \ax \right)\cos\left( \pi \az \right)+\cos\left( \pi \ay \right)\cos\left( \pi \az \right)$.
The gap is maximal and equal to $1-\lambda_2 = \frac{4}{3n}$ at $\ax=1,\az=0$ and an arbitrary $\ay$, which includes both XY ($\ay = 1$) and CNOT ($\ay = 0$). Those gates are therefore the fastest scrambler for the all-to-all coupling (numerically identified in Ref.~\cite{Znidaric_2007}). For the Clifford XY and CNOT gates the gap has been already calculated in Ref.~\cite{PRA08}, Eq.~(\ref{delta_all-all}) though extends it to any gate. Note that the normalization in Eq.~(\ref{eq:Mbar}) means that $\bar{M}$ represents the average action of a single 2-qubit gate. If we measure time in such units that $T=n$ gates are applied per unit of time, we will have $I(t)-I_\infty \asymp (1-3h/n)^{tn} \to \exp{(-3ht)}$, giving the purity rate $\rE=3h$, which is equal to $\rE=\frac{4}{3}$ for optimal gates.


\begin{thebibliography}{99}

\bibitem{book} B.~Zheng, X.~Chen, D.~L.~Zhou, and X.~G.~Wen, {\em Quantum Information Meet Quantum Matter} (Springer, 2019).

\bibitem{emerson03} J.~Emerson, Y.~S.~Weinstein, M.~Saraceno, S.~Lloyd, and D.~G.~Cory, \tit{Pseudo-random unitary operators for quantum information processing} Science {\bf 302}, 2098 (2003).

\bibitem{Google} F. Arute, K. Arya, R. Babbush et al., \tit{Quantum supremacy using a programmable superconducting processor} Nature {\bf 574}, 505 (2019).

  
\bibitem{Frank18} C.~W.~{von Keyserlingk}, T.~Rakovszky, F.~Pollmann, and S.~L.~Sondhi, \tit{Operator hydrodynamics, OTOCs, and entanglement growth in systems without conservation laws} Phys.~Rev.~X {\bf 8}, 0211013 (2018).

\bibitem{adam18} A.~Nahum, S.~Vijay, and J.~Haah, \tit{Operator spreading in random unitary circuits} Phys.~Rev.~X {\bf 8}, 021014 (2018).

\bibitem{vedika18} V.~Khemani, A.~Vishwanath, and D.~A.~Huse, \tit{Operator spreading and the emergence of dissipative hydrodynamics under unitary evolution with conservation laws} Phys.~Rev.~X {\bf 8}, 031057 (2018).

\bibitem{nick18}  N.~{Hunter-Jones}, \tit{Operator growth in random quantum circuits with symmetry} arXiv:1812.08291 (2018).

\bibitem{Chalker18} A.~Chan, A.~{De Luca}, and J.~T.~Chalker, \tit{Solution of a Minimal Model for Many-Body Quantum Chaos} Phys.~Rev.~X {\bf 8}, 041019 (2018).

\bibitem{Calabrese05} P.~Calabrese and J.~L.~Cardy, \tit{Evolution of entanglement entropy in one-dimensional systems} J.~Stat.~Mech. {\bf 2005}, P04010 (2005).

\bibitem{prosen19} B.~Bertini, P.~Kos, and T.~Prosen, \tit{Entanglement spreading in a minimal model of maximal many-body quantum chaos} Phys.~Rev.~X {\bf 9}, 021033 (2019).

  
\bibitem{Hayden07} P.~Hayden and J.~Preskill, \tit{Black holes as mirrors: quantum information in random subsystems} JHEP {\bf 2007}, 120 (2007).

  
\bibitem{Susskind08} Y.~Sekino and L.~Susskind, \tit{Fast scramblers} JHEP {\bf 2008}, 065 (2008).

\bibitem{Lorenzo20} L.~Piroli, C.~{S\" underhauf}, and X.-L.~Qi, \tit{A random unitary circuit model for black hole evaporation} JHEP {\bf 2020}, 63 (2020).
  
\bibitem{Suh14} H.~Liu and S.~J.~Suh, \tit{Entanglement tsunami: universal scaling in holographic thermalization} Phys.~Rev.~Lett. {\bf 112}, 011601 (2014).

\bibitem{Mezei16} H.~Casini, H.~Liub and M.~Mezei, \tit{Spread of entanglement and causality} J.~High Energy Phys. {\bf 2016}, 77 (2016).

 
\bibitem{Mezei17} M.~Mezei and D.~Standford, \tit{On entanglement spreading in chaotic systems} J.~High Energ.~Phys. {\bf 2017}, 65 (2017).

  
\bibitem{Fawzi12} W.~Brown and O.~Fawzi, \tit{Decoupling with random quantum circuits} Commun.~Math.~ Phys. {\bf 340}, 867 (2015).

\bibitem{thermalization} L.~{D'Alessio}, Y.~Kafri, A.~Polkovnikov, and M.~Rigol, \tit{From quantum chaos and eigenstate thermalization to statistical mechanics and thermodynamics} Adv.~Phys. {\bf 65}, 239 (2016).

  \new{
  \bibitem{tobe} J.~Bensa and M.~\v Znidari\v c, {\em in preparaton}.
    }
  
	 
\bibitem{Mori20} T.~Mori and T.~Shirai, \tit{Resolving a discrepancy between Liouvillian gap and relaxation time in boundary-dissipated quantum many-body systems} Phys.~Rev.~Lett. {\bf 125}, 230604 (2020).

\bibitem{PRE15} M.~\v Znidari\v c, \tit{Relaxation times of dissipative many-body quantum systems} Phys.~Rev.~E {\bf 92}, 042143 (2015).

  
\bibitem{emerson05} J.~Emerson, E.~Livine, and S.~Lloyd, \tit{Convergence conditions for random quantum circuits} Phys.~Rev.~A {\bf 72}, 060302 (2005).


\bibitem{oliveira07} R.~Oliveira, O.~C.~O.~Dahlsten, and M.~B.~Plenio, \tit{Generic entanglement can be generated efficiently} Phys.~Rev.~Lett. {\bf 98}, 130502 (2007). O.~C.~O.~Dahlsten, R.~Oliveira, and M.~B.~Plenio, \tit{The emergence of typical entanglement in two-party random processes} J.~Phys.~A {\bf 40}, 8081 (2007).


\bibitem{PRA08} M.~\v Znidari\v c, \tit{Exact convergence times for generation of random bipartite entanglement} Phys.~Rev.~A {\bf 78}, 032324 (2008).


\bibitem{Viola10} W.~G.~Brown and L.~Viola, \tit{Convergence rates for arbitrary statistical moments of random quantum circuits} Phys.~Rev.~Lett. {\bf 104}, 250501 (2010).

\bibitem{cwiklinski13} P.~\' Cwiklin\' nski, M.~Horodecki, M.~Mozrzymas, \L.~Pankowski, and M.~Studzi\' nski, \tit{Local random quantum circuits are approximate polynomial-designs: numerical results} J.~Phys.~A {\bf 46}, 305301 (2013).

\bibitem{metoda_redukcija} W.-T.~Kuo, A. A. Akhtar, D. P. Arovas, and Y.-Z.~You, \tit{Markovian Entanglement Dynamics under Locally Scrambled Quantum Evolution} Phys. Rev. B {\bf 101}, 224202 (2020). 

\bibitem{Swingle20} R.~Belyansky, P.~Bienias, Y.~A.~Kharkov, A.~V.~Gorshkov, and B.~Swingle, \tit{Minimal model for fast scrambling} Phys.~Rev.~Lett. {\bf 125}, 130601 (2020).


\bibitem{gross07} D.~Gross, K.~Audenaert, and J.~Eisert, \tit{Evenly distributed unitaries: on the structure of unitary designs} J.~Math.~Phys. {\bf 48}, 052104 (2007).
  
\bibitem{Harrow09} A.~W.~Harrow and R.~Low, \tit{Random quantum circuits are approximate 2-designs} Commun.~Math.~Phys. {\bf 291}, 257 (2009).
 

\bibitem{brandao16} F.~G.~S.~L.~Brandao, A.~W.~Harrow, and M.~Horodecki, \tit{Local random quantum circuits are approximate polyomial designs} Commun.~Math.~Phys. {\bf 346}, 397 (2016).

\bibitem{brandao16b} F.~G.~S.~L.~Brandao, A.~W.~Harrow, and M.~Horodecki, \tit{Efficient quantum pseudorandomness} Phys.~Rev.~Lett. {\bf 116}, 170502 (2016).

\bibitem{Hunter20} J.~Haferkamp and N.~{Hunter-Jones}, \tit{Improved spectral gaps for random quantum circuits: large local dimensions and all-to-all interactions} arXiv:2012.05259 (2020).


\bibitem{Zanardi12} A.~Hamma, S.~Santra, and P.~Zanardi, \tit{Quantum entanglement in random physical states} Phys.~Rev.~Lett. {\bf 109}, 040502 (2012).


\bibitem{AdamPRB19} T.~Zhou and A.~Nahum, \tit{Emergent statistical mechanics of entanglement in random unitary circuits} Phys.~Rev.~B {\bf 99}, 174205 (2019).

\bibitem{Adam17} A.~Nahum, J.~Ruhman, S.~Vijay, and J.~Haah, \tit{Quantum entanglement growth under random unitary dynamics} Phys.~Rev.~X {\bf 7}, 031016 (2017).

\bibitem{Braun08} L.~Arnaud and D.~Braun, \tit{Efficiency of producing random unitary matrices with quantum circuits} Phys.~Rev.~A {\bf 78}, 062329 (2008). 

  \bibitem{Znidaric_2007} M.~\v Znidari\v c, \tit{Optimal two-qubit gate for generation of random bipartite entanglement} Phys.~Rev.~A {\bf 76}, 012318 (2007).
  
\bibitem{foot3} There is an interesting observation about $\rE$ (or eqivalently $\vE$) between the U(4) gate in the BW and random n.n. configurations. The gap of the Markovian matrix is in the random n.n. protocol equal to a sum of 2-site matrices, whereas it is equal to a product of the same 2-site matrices in the BW case. Based on exact diagonalization results it was guessed~\cite{CP20} that the largest nontrivial eigenvalue for the BW case is $\lambda_2=(\frac{4}{5}\cos{\frac{\pi}{n}})^2$ for the OBC, and $\lambda_2=(\frac{4}{5}\cos{\frac{\pi}{n}})^4$ for the PBC (expressions are exact for any $n$ and agree with the asymptotic results for $\rE$ in Table~\ref{fig:pregled}). For the random n.n. protocol one on the other hand has~\cite{PRA08} $\lambda_2=[1-\frac{1}{n-1}(1-\frac{4}{5}\cos{\frac{\pi}{n}})]^{n-1}$ for the OBC, and $\lambda_2=[1-\frac{2}{n}(1-\frac{4}{5}\cos{\frac{\pi}{n}})]^{n}$ for the PBC.

  
\bibitem{CP20} M.~\v Znidari\v c, \tit{Entanglement growth in diffusive systems} Commun.~Phys. {\bf 3}, 100 (2020).

\bibitem{Arul20} S.~A.~Rather, S.~Aravinda, and A.~Lakshminarayan, \tit{Creating ensembles of dual unitary and maximally entangling quantum evolutions} Phys.~Rev.~Lett. {\bf 125}, 070501 (2020).

\bibitem{Karol15} D.~Goyeneche, D.~Alsina, J.~I.~Latorre, A.~Riera, and K.~\. Zyczkowski, \tit{Absolutely maximally entangled states, combinatorial designs, and multiunitary matrices} Phys.~Rev.~A {\bf 92}, 032316 (2015).

\bibitem{Yoshida15} F.~Pastawski, B.~Yoshida, D.~Harlow, and J.~Preskill, \tit{Holographic quantum error-correcting codes: toy models for the bulk/boundary correspondence} J.~Hep. {\bf 06}, 149 (2015).

\bibitem{brunoprb} L.~Piroli, B.~Bertini, J.~I.~Cirac, and T.~Prosen, \tit{Exact dynamics in dual-unitary quantum circuits} Phys~.Rev.~B {\bf 101}, 094304 (2020).

\bibitem{sarang19} S.~Gopalakrishnan and A.~Lamacraft, \tit{Unitary circuits of finite depth and infinite width from quantum channels} Phys.~Rev.~B {\bf 100}, 064309 (2019).

  
\bibitem{Bruno20} L.~Piroli and B.~Bertini, \tit{Scrambling in random unitary circuits: Exact results} Phys.~Rev.~B {\bf 102}, 064305 (2020).


\bibitem{Austen20} P.~W.~Claeys and A.~Lamacraft, \tit{Maximum velocity quantum circuits} Phys.~Rev.~Res. {\bf 2}, 033032 (2020).

  
\bibitem{foot2} Another choice of taking the thermodynamic limit would be to hold $\nA$ fixed and let $\nB \to \infty$. In such a case one would probe the local rather than the global thermalization, and would have strong kinematic effects. Namely, for a typical state in a large Hilbert space the reduced density operator $\rA$ will be very close to an identity~\cite{Karol}. For instance, tracing a random state over $\nB \gg \nA$ results in the spectrum of $\rA$ whose relative deviations from a flat one are negligible, \new{scaling as} $\sim q^{-(\nB-\nA)/2}$~\cite{jpa07}. Such measure concentration can swamp-out any dynamical effects (e.g., integrable vs. chaotic evolution). 

\bibitem{Karol} V.~M.~Kendon, K.~\. Zyczkowski, and W.~J.~Munro, \tit{Bounds on entanglement in qudit subsystems} Phys.~Rev.~A {\bf 66}, 062310 (2002).
  
\bibitem{jpa07} M.~\v Znidari\v c, \tit{Entanglement of random vectors} J.~Phys.~A {\bf 40}, F105 (2007).

\bibitem{foot1} Often it is incorrectly stated that a one 2-qubit gate can increase entanglement of a separable state by $\ln{q}$ (i.e., by one ebit). That is not true -- one 2-qudit gate acting on a bipartite separable state can produce maximal entanglement of $2\ln{q}$~\cite{Bennett02}. \new{In a special case of acting} on a separable state of 2-qubits the maximum is indeed $1$ ebit, however, if one has $2+2$ (or more qubits) the maximum is $2$ ebits. There are therefore three possible sources of confusion (sometimes misplaced in the literature): (i) a factor of $2$ in the number of cuts between the OBC (1 boundary link connecting subsystems A and B) and PBC (2 boundary links), (ii) how time is counted -- one full BW is counted as $t=1$ (i.e., $\sim n$ gates per unit of time; this is the units we use), or as $\tau=2$ ($\sim n/2$ gates per $\tau=1$), (iii) one two-qudit gate can increase entanglement of a bipartite separable state maximally by $2\ln{q}$.


\bibitem{Bennett02} C.~H.~Bennett, A.~W.~Harrow, D.~W.~Leung, and J.~A.~Smolin, \tit{On the capacities of bipartite Hamiltonians and unitary gates} IEEE Trans.~Inf.~Theory {\bf 49}, 1895 (2003).


\bibitem{BrunoSci} B.~Bertini, P.~Kos, and T.~Prosen, \tit{Operator entanglement in local quantum circuits I: chaotic dual-unitary circuits} SciPost Phys.~ {\bf 8}, 067 (2020).

  
  
	\bibitem{dekompozicija_1} N. Khaneja, R. Brockett, and S. J. Glaser, \tit{Time optimal control in spin systems} Phys. Rev. A {\bf 63}, 032308 (2001).
	
	\bibitem{dekompozicija_2_in_simetrije} B. Kraus, and J. I. Cirac, \tit{Optimal creation of entanglement using a two-qubit gate} Phys. Rev. A {\bf 63}, 062309 (2001).
	
	\bibitem{dekompozicija_recept} M. Blaauboer, and R. L. de Visser, \tit{An analytical decomposition protocol for optimal implementation of two-qubit entangling gates} J. Phys. A: Math. Theor. {\bf 41}, 39 (2008).
	 
	 
	 \bibitem{Lubkin} E. Lubkin \tit{Entropy of an $n$-system from its correlation with a $k$-reservoir} J. Math. Phys. {\bf 19}, 1028 (1978).

  
  
\bibitem{Paolo00} P.~Zanardi, C.~Zalka, and L.~Faoro, \tit{Entangling power of quantum evolutions} Phys.~Rev.~A {\bf 62}, 030301(R) (2000).

  \new{
  \bibitem{Makhlin00} Y.~Makhlin, \tit{Nonlocal properties of two-qubit gates and mixed states, and the optimization of quantum computations} Quant.~Info.~Proc. {\bf 1}, 243 (2002).

  \bibitem{Whaley03} J.~Zhang, J.~Vala, S.~Sastry, and K.~B.~Whaley, \tit{Geometric theory of nonlocal two-qubit operations} Phys.~Rev.~A {\bf 67}, 042313 (2003).

  \bibitem{footeU} The bipartite entangling power $e(W)$ is defined~\cite{Paolo00} as $e(W)=1-\ave{I(W\ket{\psi_{\rm A}}\ket{\psi_{\rm B}})}$ where the averaging is over the Haar measure on ${\rm U}(2^{\nA})\otimes {\rm U}(2^{\nB})$ of initial bipartite product states, see e.g. Ref.~\cite{arul20} and references therein for studies of $e(U)$. In random circuits one is though typically interested in fully factorized initial product states (${\rm U(2)}\otimes {\rm U(2)} \otimes \cdots \otimes {\rm U(2)}$) and not bipartite product initial states. While one can generalize the entangling power with respect to multipartite product initial states, e.g. Ref.~\cite{Scott04}, in random circuits things are further complicated by having random 1-site unitaries at each step of a protocol (not just at the beginning), and the fact that the unitary propagator for the whole circuit depends in a nontrivial way on the number of layers (time) and the 2-qubit gate $W$. While the entangling power has been of use in small systems, e.g., 2 qubits, where exact results are simple to obtain, it is not clear how to make use of it (go beyond Markovian techniques that we use) in the TDL and a multipartite setting.

\bibitem{arul20} B.~Jonnadula, P.~Mandayam, K.~\. Zyczkowski, and A.~Lakshminarayan, \tit{Entanglement measures of bipartite quantum gates and their thermalization under arbitrary interaction strength} Phys.~Rev.~Research {\bf 2}, 043126 (2020).
    
    \bibitem{Scott04} A.~J.~Scott, \tit{Multipartite entanglement, quantum-error-correcting codes, and entangling power of quantum evolutions} Phys.~Rev.~A {\bf 69}, 052330 (2004).
}


\bibitem{andrew21} \new{{S.-H.}~Lin, R.~Dilip, A.~G.~Green, A.~Smith, and F.~Pollmann, \tit{Real- and imaginary-time evolution with compressed quantum circuits} PRX Quantum {\bf 2}, 010342 (2021).}

  \new{
  \bibitem{sarang20} S.~Gopalakrishnan and M.~J.~Gullans, \tit{Entanglement and purification transitions in non-Hermitian quantum mechanics} arXiv:2012.01435 (2020).
    }

  \new{
  \bibitem{cutoffASEP} C.~Labbe and H.~Lacoin, \tit{Cutoff phenomenon for the asymmetric simple exclusion process and the biased card shuffling} Ann.~Probab. {\bf 47}, 1541 (2019).
    }

  \new{
  \bibitem{footBWS} Stacking odd-$n$ consecutive $S$ configurations we can write it in terms of a BW configuration as $[S_1^{n-1}]^n=L [B_1^{n-1}]^{(n+3)/2} U$, where $L$ is a 'lower triangle' block of gates and $U$ an 'upper triangle' of gates, each containing $(n-1)(n-3)/2$ gates; disregarding $L$ and $U$ would imply in the TDL $\lambda_2({\rm S})=\sqrt{\lambda_2({\rm BW})}$.}
  
\bibitem{Vidal} G.~Vidal, \tit{Efficient classical simulation of slightly entangled quantum computations} Phys.~Rev.~Lett. {\bf 91}, 147902 (2003).
  
           \new{
\bibitem{footBW} We conjecture that in the TDL and Haar-random U(4) gates in BW configuration with OBC and half-half symmetric bipartition purity at any finite fixed $t$ is equal to $I(t)=(16/25)^t$. This agrees with the asymptotic decay $(4/5)^{2t}$ obtained in Refs.~\cite{Frank18,adam18} (the explicit finite-$n$ expression in Eq.(25) of Ref.~\cite{Frank18} however is not equal to this result because one has to correctly account for boundary conditions).
             }

\new{\bibitem{Austen18} D.~A.~Rowlands and A.~Lamacraft, \tit{Noisy coupled qubits: Operator spreading and the Fredrickson-Andersen model} Phys.~Rev.~B {\bf 98}, 195125 (2018).
}

         
\bibitem{adamPRX20} T.~Zhou and A.~Nahum, \tit{Entanglement membrane in chaotic many-body systems} Phys.~Rev.~X {\bf 10}, 031066 (2020).

           
\bibitem{Persi96} P.~Diaconis, \tit{The cutoff phenomenon in finite Markov chains} Proc.~Natl.~Acad.~Sci.~U.S.A. {\bf 93}, 1659 (1996).

\bibitem{Kastoryano12} M.~J.~Kastoryano, D.~Reeb, and M.~M.~Wolf, \tit{A cutoff phenomenon for quantum Markov chains}  J.~Phys.~A {\bf 45}, 075307 (2012).

  \new{
  \bibitem{cutoff} D.~A.~Levin, Y.~Peres, and E.~L.~Wilmer, {\em Markov Chains and Mixing Time}, (AMS, 2017).
    }

\bibitem{Anto07} P.~Facchi, U.~Marzolino, G.~Parisi, S.~Pascazio, and A.~Scardicchio, \tit{Phase transitions of bipartite entanglement} Phys.~Rev.~Lett. {\bf 101}, 050502 (2008).
           
\bibitem{Vinayak12} Vinayak and M.~\v Znidari\v c, \tit{Subsystem dynamics under random Hamiltonian evolution} J.~Phys.~A {\bf 45}, 125204 (2012).

\bibitem{Jed19} P.-Y.~Chang, X.~Chen, S.~Gopalakrishnan, and J.~H.~Pixley, \tit{Evolution of entanglement spectra under generic quantum dynamics} Phys.~Rev.~Lett. {\bf 123}, 190602 (2019).
           
\bibitem{Cotler20} J.~Cotler, N.~{Hunter-Jones}, and D.~Ranard, \tit{Fluctuations of subsystem entropies at late times} arXiv:2010.11922 (2020).


	 \bibitem{LMG model} H. J. Lipkin, N. Meshkov, and A. J. Glick,  \tit{Validity of many-body approximation methods for a solvable model: Exact solutions and perturbation theory} Nuclear Physics {\bf 62}, 188-198 (1965).


  
\end{thebibliography}
\end{document}